\def\msun{\ensuremath{{\rm\,M_\odot}}}
\def\Msun{\msun}

\renewcommand{\vec}[1]{\mbox{\boldmath $#1$}}

\documentclass[preprint2]{pp7}

\usepackage[dvipsnames]{xcolor}
\usepackage{rotating}
\usepackage{multirow}
\usepackage{todonotes}
\usepackage[normalem]{ulem} 

\usepackage{savesym}
\savesymbol{captionbox}
\usepackage{caption}
\restoresymbol{CAPTIONBOX}{captionbox}

\usepackage{ulem}

\newcommand{\ppvi}{{\it Protostars \& Planets VI\/}}

\input{pp7.h}

\begin{document}

\title{\textbf{\LARGE The Origin and Evolution of Multiple Star Systems }}

\author {\textbf{\large Stella S. R. Offner}$^{1}$, \textbf{Maxwell Moe}$^{2}$,  \textbf{Kaitlin M. Kratter}$^{2}$, \textbf{Sarah I. Sadavoy}$^{3}$, \textbf{Eric L. N. Jensen}$^{4}$,
\textbf{John J. Tobin}$^{5}$ }
\affil{$^{1}$\small\it The University of Texas at Austin, $^2$University of Arizona, $^3$Queen's University, $^{4}$Swarthmore College, $^5$National Radio Astronomy Observatory }

\begin{abstract}
\baselineskip = 11pt
\leftskip = 1.5cm 
\rightskip = 1.5cm
\parindent=1pc
{\small Observational advances over the last decade have enabled high-resolution, interferometric studies of forming multiple systems, statistical surveys of multiplicity in star-forming regions, and new insights into disk evolution and planetary architectures in these systems. In this review, we compile the results of observational and theoretical studies of stellar multiplicity. We summarize the population statistics spanning system evolution from the protostellar phase through the main-sequence phase and evaluate the influence of the local environment. In short, most stars are born in multiple stellar systems, most main sequence stars are members of multiple systems, but most star systems are single. We describe current models for the origin of stellar multiplicity and review the landscape of numerical simulations and evaluate their consistency with observations. We review the properties of disks and discuss the impact of multiplicity on planet formation and system architectures. Finally, we summarize open questions
and discuss the technical requirements for future observational and theoretical progress.
 \\~\\~\\~}
\end{abstract}  


\section{\textbf{INTRODUCTION}}

The formation of multiple star systems -- systems of two or more gravitationally bound stars with separations $\lesssim 0.1$~pc -- takes place during the earliest phases of star formation. The majority of such systems form and evolve to their final configuration during the time period spanned by the collapse of dense cores through the end of mass accretion. This review summarizes the current statistics of multiple stars, describes theoretical models and observational evidence for different modes of multiple star formation, and discusses the implications for planet formation and system evolution.

Studying multiple formation has historically been challenging from several perspectives. Dust in star-forming regions absorbs short-wavelengths of light and re-emits at longer wavelengths, requiring high-resolution, radio observations to detect forming stars.
Theoretical models, meanwhile, require both large dynamic range ($\sim$1~au -10~pc) and treatment of non-linear, complex physical processes, including turbulence, radiative feedback, and magnetic effects.
Consequently, when the last Protostars and Planets conference occurred in 2013, the mechanisms for multiple formation were largely theoretical and the properties of young multiple systems were poorly constrained. 

At that time a major expansion of the Very Large Array (VLA) had just concluded, and  
only the first cycle of observations by the Atacama Large Millimeter/submillimeter Array (ALMA) had been taken. 
Since then, the power of these telescopes has revolutionized our understanding of young {\it forming} multiples,
and stunningly detailed images of disks have provided a comprehensive understanding of the impact of binaries on disks, both in a statistical sense 
and within individual multiple systems. 
 Meanwhile large-scale, uniform surveys of star-forming regions, clusters, and field stars have generated new insights into the statistics of stellar multiplicity and its dependence on mass, age, metallicity, and cluster density. For already-formed planetary systems, radial velocity (RV) and high-resolution imaging surveys of {\it Kepler} and {\it TESS} planet hosts have recently revealed that binaries substantially sculpt planet formation, shedding new light on planet formation and migration models.  Computational advances in speed and methodology have enabled studies of multiple star formation that span an unprecedented dynamic range, model different star cluster environments, and explore the role of multiple physical effects acting in concert. There is now consensus that {\it most} stars both form and reside on the main sequence with at least one stellar companion: single stars are the exception not the rule. 

 This review begins by introducing key definitions and summarizing the statistics of observed multiple star systems (\S2), with a special focus on their configuration and properties at the end of the star formation process. We then proceed with an overview of theoretical models describing the origin and early evolution of multiple star systems in \S3. Using the framework of these models, in \S4 we examine protostellar multiple systems, assessing how well they confirm or refute theoretical expectations. In \S5 we discuss the impact of stellar multiplicity on planet formation and system architectures. \S6 summarizes remaining open questions and discusses observational and theoretical prospects for enhancing the understanding of multiple systems over the coming decade. We conclude in \S7.

\section{\textbf{OBSERVED STELLAR MULTIPLICITY}}
\label{sec:ObsStellarMultiplicity}

\index{Binaries!statistics|} Observations show that stellar multiplicity evolves substantially during formation and changes more slowly thereafter. In order to distinguish between primordial multiplicity, the results of dynamical evolution, and the final system configurations for main-sequence (MS) field stars, we separately consider the multiplicity of very young stars, i.e., protostars, and older sources. We adopt the standard convention of grouping young stellar objects (YSOs) according to their spectral energy distribution, i.e., by ``Class" as defined by \citet{LadaWilking1984} and \citet{AdamsLada1987}, whereby younger, more embedded sources have more reddening from an envelope and disk. We caution, however, that the inclination of the source outflow and disk with respect to the line of sight affect the degree of extinction and may shift an older source from a later to an earlier Class or vice versa \citep{RobitailleWhitney2007,OffnerRobitaille2012}. Thus, while Class provides a rough, if imperfect, measure of sample age it is an unreliable age indicator for individual sources.

In this section, we compile and compare the multiplicity statistics of MS and non-embedded pre-MS stars (mostly Class II/III T Tauri stars) from several surveys (mostly optical and near-IR).  These statistics represent the final outcome of the star formation process, providing insight into how stars and planets form and must subsequently evolve, and serve as a critical benchmark for simulations and models. A detailed analysis of embedded Class~0/I binaries observed by the VLA and ALMA is discussed in \S\ref{sec:ProtoStatistics}.

\index{Binaries!separation|(} 
Many multiplicity properties also vary as a function of the physical distance between sources. In the literature, pair separations are often characterized as ``close" or ``wide," however, these terms have no fixed definition. Throughout this review, we define {\it close}, {\it intermediate}, and {\it wide} separations  as $\lesssim$\,10\,au, $\sim$\,10\,-\,300\,au and $\gtrsim$\,300\,au, respectively, unless otherwise specified.

\subsection{\textbf{Detection Methods and Definitions}}

\index{Binaries!fraction|(} 

 No single observational technique provides a complete picture of multiplicity. For MS and T Tauri stars, spectroscopic\index{Binaries!spectroscopic} and eclipsing\index{Binaries!eclipsing} binaries probe close separations within $a$~$<$~10~au \citep{AbtGomez1990,Latham2002,Melo2003,Sanade-Mink2012,MoeKratter2019,KounkelCovey2019}. Interferometry, sparse aperture masking, {\it Hubble Space Telescope} ({\it HST}) imaging, and astrometry reveal binaries across close to intermediate separations of $a$~=~1\,-\,300~au \citep{Reid2001,Gizis2002,KrausIreland2008b,RaghavanMcAlister2010a,DieterichHenry2012,RizzutoIreland2013,SanaLe-Bouquin2014,DeFurio2019}. Speckle imaging, lucky imaging, and adaptive optics (AO) resolve binaries across intermediate to wide separations of $a$~=~10\,-\,1,000~au \citep{ShatskyTokovinin2002,Close2003,LawHodgkin2008,BergforsBrandner2010,JansonHormuth2012,De-RosaPatience2014,Ward-DuongPatience2015,TokovininBriceno2020}. Common proper motion confirms that wide binaries with $a$~$=$~300~\,-\,20,000~au are in fact gravitationally bound  \citep{LepineBongiorno2007,El-BadryRix2018,WintersHenry2019,Hartman2020}.  Other techniques can identify unresolved binaries, e.g., excess luminosity method or deblending of spectroscopic binaries, but they cannot determine their orbital separations \citep{Hubrig2001,BardalezGagliuffi2014,GulliksonKraus2016,Widmark2018}. For  volume-limited populations of M-dwarfs \citep{WintersHenry2019} and solar-type MS stars \citep{RaghavanMcAlister2010a,Tokovinin2014}, the combination of these overlapping techniques provides a $>$90\% complete census of MS companions orbiting the primary. 

We first define the {\it multiplicity fraction} (sometimes called the binary fraction) as the fraction of primaries with at least one companion:

\begin{equation}
 {\rm MF} = \frac{\rm B+T+Q+...}{\rm S+B+T+Q+...},
 \label{MFdef}
\end{equation}

\noindent where {\rm S}, {\rm B}, {\rm T}, and {\rm Q} are the number of single stars, binaries, triples, and quadruples, respectively. Higher ordered multiples are rare but exist, including the two known septuples, AR~Cas and $\nu$~Sco, both of which have early-B primaries \citep{EggletonTokovinin2008}. Similarly, the {\it triple/high-order fraction} is:

\begin{equation}
{\rm THF} = \frac{\rm T+Q+...}{\rm S+B+T+Q+...}.
 \label{TFdef}
\end{equation}
\index{Multiple Systems!higher order}

\noindent Meanwhile, the {\it companion frequency} is the average frequency of companions per primary:

\begin{equation}
  {\rm CF} = \frac{\rm B+2T+3Q+...}{\rm S+B+T+Q+...} = \int f_{\rm log\,a}\,d{\rm log}a,
  \label{CFdef}
\end{equation}

\noindent where $f_{\rm log\,a}$ is the frequency of companions per decade of orbital separation.   

\begin{figure*}[h!]
\centerline{
\includegraphics[trim=0.6cm 3.0cm 0.2cm 2.6cm, clip=true, width=6.7in]{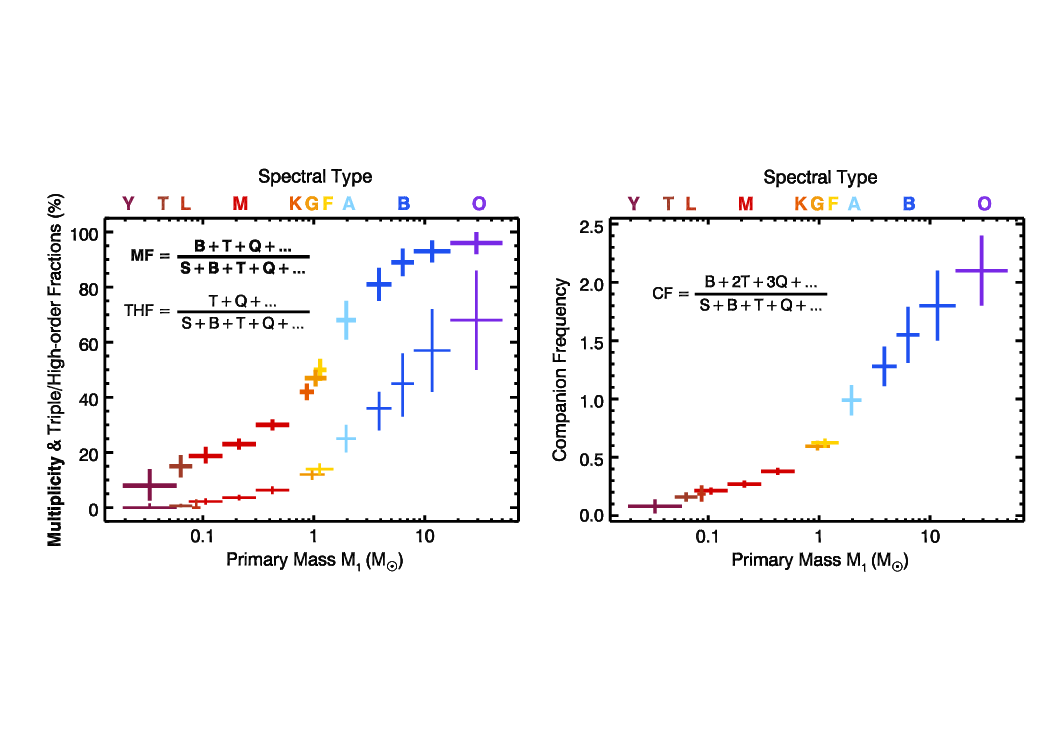}}
\vspace{-0.05in}
\caption{\small Bias-corrected multiplicity fraction (left; thick), triple/high-order fraction (left; thin), and companion frequency (right) of BDs and MS stars. All three quantities increase monotonically with primary mass.  The indicated spectral types at the top roughly correspond to the mean primary masses of field dwarfs. See Table \ref{table:MultStat} for detailed references.} 
\label{fig:MultFrac}
\end{figure*}

Measurement uncertainties in the multiplicity fraction and triple/high-order fraction follow binomial statistics, while Poisson statistics govern the companion frequency, which can exceed unity \citep{BurgasserKirkpatrick2003, SanaLe-Bouquin2014,MoeDi-Stefano2017}. Here we report multiplicity statistics with respect to volume-limited samples. For magnitude-limited samples, corrections for Malmquist bias, which is sometimes called the Branch bias in the context of binary stars \citep{Branch1976}, must be applied since unresolved twin binaries are $\Delta$m~=~0.75~mag brighter than their single-star counterparts. Systematic uncertainties, such as those associated with correcting for incompleteness, generally dominate the overall error for samples exceeding a few hundred systems. In the following subsections, we compute bias-corrected MF, THF, and CF from various surveys and present the results in Table~\ref{table:MultStat} and Fig.~\ref{fig:MultFrac}.

\begin{figure*}[t]
\centerline{
\includegraphics[trim=0.6cm 2.6cm 0.6cm 3.2cm, clip=true, width=6.4in]{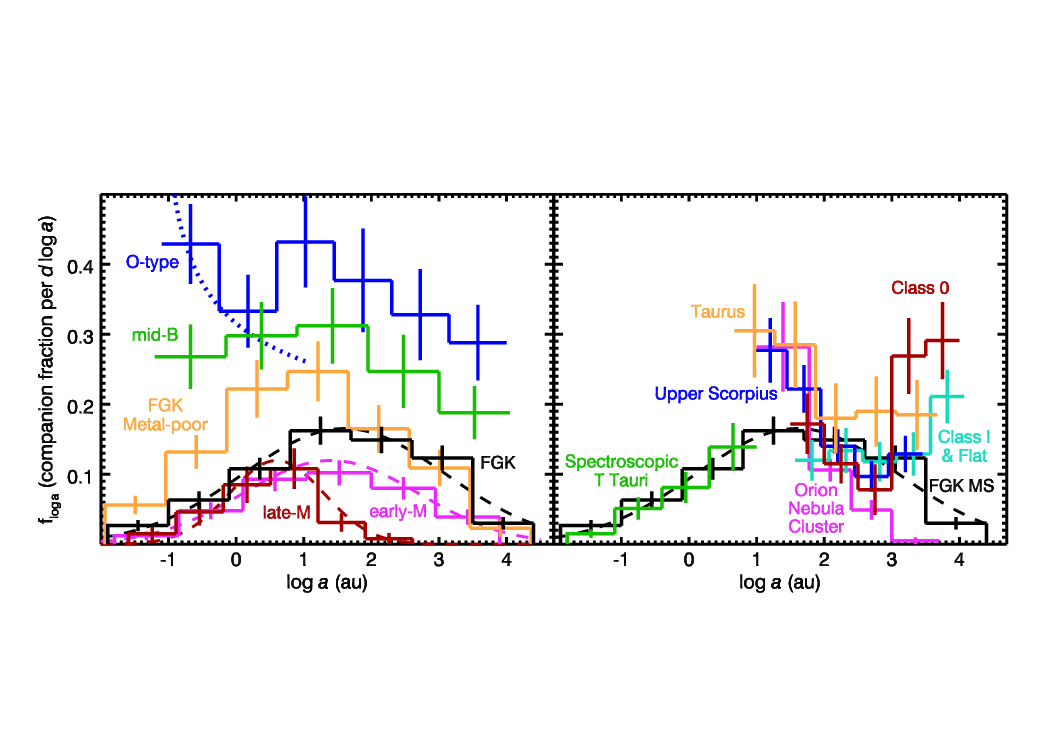}}
\caption{{\small Frequency $f_{\rm log\,a}$ of companions per decade of orbital separation. {\bf Left:} the separation distribution of MS binaries as a function of spectral type and metallicity. Late-M \citep[red;][]{WintersHenry2019}, early-M \citep[magenta;][]{WintersHenry2019}, and FGK \citep[black;][]{RaghavanMcAlister2010a} binaries follow lognormal distributions (dashed fits - see parameters in Table~2). The distribution of all companions to mid-B \citep[green;][]{MoeDi-Stefano2017} and O \citep[blue;][]{Sanade-Mink2012,SanaLe-Bouquin2014} stars roughly follow \"{O}pik's law. The dotted blue line shows the fit to inner binary companions to O stars, which skew significantly toward shorter separations. The close binary fraction of FGK dwarfs increases significantly at low metallicity \citep[orange;][]{MoeKratter2019}. {\bf Right:} separation distribution of young solar-type binaries across different star-forming environments and classes, where we display the FGK MS distribution (black) for reference. The MS distribution is well matched by the bias-corrected close binary fraction within $a$ $<$ 10\,au of Class II/III T Tauri stars inferred from spectroscopy \citep[green;][]{KounkelCovey2019}. However, imaging surveys reveal an excess of companions to pre-MS stars across $a$~=~10\,-\,50~au in most environments, in conflict with the MS and \citet{KounkelCovey2019} results. Across $a$ = 100\,-\,5,000 au, the dense Orion Nebula Cluster exhibits a deficit of wide companions \citep[magenta;][]{ReipurthGuimaraes2007,Duchene2018}, intermediate density regions like Upper Scorpius are consistent with the field population \citep[blue;][]{TokovininBriceno2020}, and sparse, low-density Taurus shows a slight excess \citep[orange;][]{KrausIreland2011c}. Class~I and flat-spectrum protostars (cyan) and especially Class~0 protostars (red) in Perseus and Orion exhibit a substantial excess of companions across $a$ = 1,000\,-\,10,000~au \citep{TobinOffner2021}.}  Note that the two panels depict populations with very different ages, underscoring the evolution in multiplicity that occurs during formation.\index{Class 0}\index{Class I}\index{T Tauri}}
\label{fig:Floga}
\end{figure*}

\index{Binaries!mass ratio|)}
We display the separation distribution $f_{\rm log\,a}$ of all companions for different stellar populations in Fig.~\ref{fig:Floga}.  To distinguish the effect of hierarchical triples on the overall separation distribution, we define $\widetilde a_{\rm all}$ and $\widetilde a_{\rm in}$ as the median separations of all companions and of inner binaries only (excluding outer tertiaries), respectively. Close binaries exhibit different trends with respect to primary mass, mass ratio, metallicity, and environment compared to intermediate and wide separation binaries (see \S\ref{sec:MultTrends}). We therefore define the {\it close binary fraction} CBF and {\it wide binary fraction} WBF as the fraction of primaries with at least one companion below $a$~$<$~10~au and across $a$~$=$~100\,-\,10,000\,au, respectively. We display both $\widetilde a_{\rm all}$ and $\widetilde a_{\rm in}$ in Fig.~\ref{fig:Mediana} and both CBF and WBF in Fig.~\ref{fig:CloseWide} as a function of {\it primary mass $M_1$} and other stellar parameters.

The {\it minimum companion mass} $M_{\rm comp}$ or {\it mass ratio} $q$ = $M_{\rm comp}$/$M_1$ included in reported multiplicity statistics varies from survey to survey, but typically brown dwarf (BD) primaries include all BD companions, FGKM-dwarfs include only MS companions above $M_{\rm comp}$ $>$ 0.075$\Msun$, and OBA stars include MS companions above $q$~$>$~0.1. About 20\% of companions within $a$~$<$~10~au of AFG primaries are white dwarfs (WDs; \citealt{MoeDi-Stefano2017,MurphyMoe2018}), and nearly 30\% of OB stars are the products of binary evolution \citep{deMink2013}. In order to compare field MS multiplicity statistics to their pre-MS counterparts, we exclude evolved stars or systems with WD companions when possible.


\setlength{\tabcolsep}{5.0pt}
\begin{deluxetable}{lcccccccccccc}
\tabletypesize{\tiny}

 \tablecaption{Multiplicity Statistics of Brown Dwarfs and Main-Sequence Stars}

\startdata
 & & & & &  & CBF (\%) & WBF (\%) & & & $\gamma_{\rm trunc}$ & $\gamma_{\rm trunc}$ & $\gamma_{\rm trunc}$ \\
 Survey & $M_1$\,(\Msun) & N & 
 MF (\%) & THF (\%) & CF & 
$a$ $<$ 10\,au & $a$ = 10$^2$\,-\,10$^4$\,au &
$\widetilde a_{\rm all}$\,(au) & $\widetilde a_{\rm in}$\,(au) &
$a$ $<$ 1\,au & 
$a$ = 1\,-\,10$^2$\,au &
$a$ $>$ 10$^2$\,au \\
\hline
\citet{Fontanive2018} & 0.019\,-\,0.058 & 47 & 
  8\,$\pm$\,6 & $<$\,2 & 0.08\,$\pm$\,0.06 & 
 8\,$\pm$\,6 & $<$\,2 & 
 2.9\,$\pm$\,1.1 & - & 
 - & ~~~~4.8\,$\pm$\,2.2 & - \\
\citet{Burgasser2007}& 0.05\,-\,0.08 & 162 & 
  15\,$\pm$\,4$^{\rm a}$ & 0.6\,$\pm$\,0.3 & 0.16\,$\pm$\,0.04$^{\rm a}$ & 
 - & - & 
-& - & 
 - & - & - \\
 \citet{Close2003} & 0.080\,-\,0.095 & 39 & 
 19\,$\pm$\,7$^{\rm a}$ & $<$\,3 & 0.19\,$\pm$\,0.07$^{\rm a}$ & 
 16\,$\pm$\,6 & $<$\,3 & 
 3.7\,$\pm$\,1.3 & - & 
 - & ~~~~3.3\,$\pm$\,1.2 & - \\
\citet{Allen2007}& 0.06\,-\,0.15 & 361 & 
  20\,$\pm$\,4 & $<$\,1 & 0.20\,$\pm$\,0.04 & 
  14\,$\pm$\,3 & $<$\,0.4 & 
 6.9\,$\pm$\,1.4 & - & 
 - & ~~~~1.7\,$\pm$\,0.5 & - \\
 \citet{WintersHenry2019}: 20 pc & 0.075\,-\,0.15 & 185 & 
 19\,$\pm$\,3 & 2.2\,$\pm$\,1.1 & 0.21\,$\pm$\,0.03 & 
 16\,$\pm$\,3 & 0.5\,$\pm$\,0.5 & 
 3.9\,$\pm$\,1.2 & 3.1\,$\pm$\,1.1 & 
 - & - & - \\
  \citet{WintersHenry2019}: 20 pc & 0.15\,-\,0.30 & 336 & 
 23\,$\pm$\,2 & 3.6\,$\pm$\,1.0 & 0.27\,$\pm$\,0.03 & 
 14\,$\pm$\,2 & 4.8\,$\pm$\,1.2 & 
 10\,$\pm$\,3 & 6\,$\pm$\,2 & 
 - & ~~~~0.7\,$\pm$\,0.5 & - \\
  \citet{WintersHenry2019}: 20 pc & 0.3\,-\,0.6 & 350 & 
 30\,$\pm$\,2 & 6.3\,$\pm$\,1.4 & 0.38\,$\pm$\,0.03 & 
 15\,$\pm$\,2 & 12\,$\pm$\,2 & 
 26\,$\pm$\,4 & 14\,$\pm$\,3 & 
 - & ~~~~0.1\,$\pm$\,0.4 & - \\
\citet{El-BadryRix2019b}: 200 pc & 0.1\,-\,0.4 & 5,220 &
- & - & - & - & - & - & - & - & - & ~~~~0.4\,$\pm$\,0.3 \\
~~~~~~~~'' & 0.4\,-\,0.6 & 12,717 &
- & - & - & - & - & - & - & - & - & $-$0.2\,$\pm$\,0.3 \\
~~~~~~~~'' & 0.6\,-\,0.8 & 9,542 &
- & - & - & - & - & - & - & - & - & $-$0.9\,$\pm$\,0.2 \\
~~~~~~~~'' & 0.8\,-\,1.2 & 11,588 &
- & - & - & - & - & - & - & - & - & $-$1.2\,$\pm$\,0.2 \\
~~~~~~~~'' & 1.2\,-\,2.5 & 3,238 &
- & - & - & - & - & - & - & - & - & $-$1.3\,$\pm$\,0.3 \\
 \citet{RaghavanMcAlister2010a}: 25 pc & 0.75\,-\,1.25 & 454 & 
46\,$\pm$\,3 & 12\,$\pm$\,2 & 0.60\,$\pm$\,0.04 & 
20\,$\pm$\,2 & 22\,$\pm$\,2 & 
49\,$\pm$\,6 & 29\,$\pm$\,4 & 
 ~~~~1.5\,$\pm$\,0.5 & ~~~~0.2\,$\pm$\,0.4 & $-$1.0\,$\pm$\,0.4 \\
 ~~~~~~~~~~'' & 0.75\,-\,1.00 & 323 & 42\,$\pm$\,3 & - & - & - & - & - & - & 
 - & - & - \\
 ~~~~~~~~~~'' & 1.00\,-\,1.25 & 131 & 50\,$\pm$\,4 & - & - & - & - & - & - & 
 - & - & - \\
 \citet{Tokovinin2014}: 67 pc & 0.85\,-\,1.5 & 4,847 & 
47\,$\pm$\,3 & 14\,$\pm$\,2 & 0.62\,$\pm$\,0.04 & 
24\,$\pm$\,2 & 19\,$\pm$\,2 & 
31\,$\pm$\,5 & 22\,$\pm$\,4 & 
 ~~~~1.7\,$\pm$\,0.5 & ~~~~0.4\,$\pm$\,0.4 & $-$0.7\,$\pm$\,0.4 \\
\citet{De-RosaPatience2014}: 75 pc & 1.6\,-\,2.4 & 435 &
- & - & - &
- & 28\,$\pm$\,4 & 
- & - & 
- & $-$1.3\,$\pm$\,0.4 & $-$2.2\,$\pm$\,0.4 \\
\citet{MurphyMoe2018} & 1.4\,-\,2.3 & 2,224 &
- & - & - & - & - & - & - & - & $-$1.1\,$\pm$\,0.3 & - \\
\citet{MoeKratter2021} & 1.6\,-\,2.4 & - &
68\,$\pm$\,7 & 25\,$\pm$\,5 & 0.99\,$\pm$\,0.13 & 
37\,$\pm$\,5 & - & 
32\,$\pm$\,8 & 13\,$\pm$\,3 &
- & - & - \\
\citet{MoeDi-Stefano2017}$^{\rm b}$ & 3\,-\,5 & - &
81\,$\pm$\,6 & 36\,$\pm$\,8 & 1.28\,$\pm$\,0.17 & 
46\,$\pm$\,7 & 40\,$\pm$\,6 & 
28\,$\pm$\,7 & 8\,$\pm$\,2 &
~~~~0.5\,$\pm$\,0.6 & $-$1.0\,$\pm$\,0.5 & $-$1.3\,$\pm$\,0.5 \\
\citet{MoeDi-Stefano2017}$^{\rm b}$ & 5\,-\,8 & - &
89\,$\pm$\,5 & 45\,$\pm$\,11 & 1.55\,$\pm$\,0.24 & 
54\,$\pm$\,8 & 49\,$\pm$\,7 & 
25\,$\pm$\,7 & 6.3\,$\pm$\,1.5 &
~~~~0.3\,$\pm$\,0.5 & $-$1.7\,$\pm$\,0.5 & $-$2.2\,$\pm$\,0.6 \\
\citet{MoeDi-Stefano2017}$^{\rm b}$ & 8\,-\,17 & - &
93\,$\pm$\,4 & 57\,$\pm$\,15 & 1.8\,$\pm$\,0.3 & 
61\,$\pm$\,10 & 55\,$\pm$\,8 & 
23\,$\pm$\,7 & 4.2\,$\pm$\,1.1 &
~~~~0.1\,$\pm$\,0.4 & $-$1.6\,$\pm$\,0.5 & $-$1.9\,$\pm$\,0.5 \\
Sana et al.$^{\rm c}$ & 17\,-\,50 & - &
96\,$\pm$\,4 & 68\,$\pm$\,18 & 2.1\,$\pm$\,0.3 & 
70\,$\pm$\,11 & 62\,$\pm$\,9 & 
19\,$\pm$\,6 & 1.7\,$\pm$\,0.5 &
$-$0.1\,$\pm$\,0.5 & $-$1.4\,$\pm$\,0.4 & $-$2.1\,$\pm$\,0.5 \\
\enddata
 \vspace{-0.5cm}
 \tablecomments{\scriptsize All statistics are computed after correcting for incompleteness and removing systems with WD companions. (a): To measure MF and CF from the BD imaging surveys, we add 4\% to their reported resolved binary fractions to account for unresolved spectroscopic binaries. (b): A compilation of B-type multiplicity surveys, including \citet{AbtGomez1990}, \citet{ShatskyTokovinin2002}, and \citet{RizzutoIreland2013}. (c): From the combined \citet{Sanade-Mink2012} and \citet{SanaLe-Bouquin2014} O~star surveys. }
 \label{table:MultStat}
\end{deluxetable}
\setlength{\tabcolsep}{6pt}

Historically, the mass-ratio distribution has been approximated by a single power-law $f_q$~$\propto$~$q^{\gamma}$ across 0~$<$~$q$~$<$~1, but larger samples indicate that up to three parameters may be required to accurately fit the distributions \citep{DucheneKraus2013d,MoeDi-Stefano2017,El-BadryRix2019b}. For this review, we wish to easily compare trends, even if the samples are small or incomplete toward small mass ratios. We therefore fit a simplified power-law slope $\gamma_{\rm trunc}$ across the truncated interval 0.4~$<$~$q$~$<$~1 where all the compiled surveys are complete (see Table~\ref{table:MultStat}).  

\subsection{\textbf{Multiplicity Statistics versus Primary Mass}}
\label{sec:statisticsMS}
\subsubsection{\textbf{FGK dwarfs}}
\label{sec:FGKstars}\index{Stars!FGK}

{\it Frequencies:} About half of field FGK primaries have MS companions, which follow a broad lognormal separation distribution peaking near $\widetilde a_{\rm all}$ = 40~au \citep[][ see Figs.~\ref{fig:MultFrac}\,-\,\ref{fig:CloseWide}]{DuquennoyMayor1991a,RaghavanMcAlister2010a,Tokovinin2014a}.  In addition to plotting the separation distribution directly from the \citet{RaghavanMcAlister2010a} data in Fig.~\ref{fig:Floga}, we also display a lognormal fit to $f_{\rm log\,a}$, where the fit parameters are listed in Table~2.


Across close and intermediate separations, the binary fraction of FGK MS stars in young open clusters 
are consistent with their field MS counterparts \citep{Bouvier1997,Patience2002,Elliott2014,DeaconKraus2020,Torres2021}. 
The CBFs measured in old open clusters also match the field values, albeit the cluster cores exhibit a substantial excess of binaries due to mass segregation \citep{GellerMathieu2012,Geller2021}. Meanwhile,  {\it Gaia} observations reveal a deficit of wide binaries in open clusters, especially those with higher stellar densities, which is likely due to dynamical disruptions \citep{DeaconKraus2020,Niu2020}.

{\it Mass Ratios:} The overall mass-ratio distribution of solar-type binaries is roughly uniform, but closer solar-type binaries systematically favor larger mass ratios \citep{DuquennoyMayor1991a,RaghavanMcAlister2010a,Tokovinin2014,MoeDi-Stefano2017}.  Specifically, FGK binaries across close and intermediate separations exhibit a 30\% and 10\% excess fraction of twins with $q$ $>$ 0.95, respectively, relative to the underlying uniform distribution \citep{LucyRicco1979,Tokovinin2000,MoeDi-Stefano2017}. These observations suggest that solar-type binaries within $a$ $<$ 200~au coevolved in a shared mass reservoir, perhaps a circumbinary disk (see \S\ref{sec:diskfrag}). \citet{El-BadryRix2019b} also found that wide binaries exhibit a very small but statistically significant excess twin fraction, suggesting some wide companions originate from intermediate separations but then subsequently widen dynamically. Aside from the excess twins, solar-type binaries beyond $a$ $>$ 200~au are skewed toward smaller mass ratios, but their distribution is still top-heavy compared to random pairings drawn from the IMF \citep {LepineBongiorno2007,MoeDi-Stefano2017,El-BadryRix2019b}.

There is a dearth of BD companions within $a$ $<$ 0.5~au to solar-type primaries, which is commonly called the BD desert \citep{GretherLineweaver2006,CsizmadiaHatzes2015,ShahafMazeh2019}. 
Meanwhile, across intermediate separations, the roughly uniform mass-ratio distribution of solar-type binaries extends down to the BD-planet boundary near $M_{\rm comp}$ = 13\,M$_{\rm J}$ \citep{WagnerApai2019,NielsenDe-Rosa2019}. 
While the multiplicity fractions of MS stars reported in Table~\ref{table:MultStat} exclude BD companions, only $\approx$\,4\% of solar-type stars have BD companions, affecting the integrated multiplicity statistics very little.

{\it Eccentricities:} Eccentricity distributions are typically referenced to a thermal distribution defined as $p_e =2e$, where $p_e$ is the probability of given eccentricity $e$ \citep{Ambartsumian1937,Heggie1975}. The average $e$ of solar-type binaries increases with orbital separation, transitioning from a sub-thermal (smaller eccentricities) to thermal distribution near 200\,au and to super-thermal beyond $>$\,1,000\,au \citep{TokovininKiyaeva2016,Tokovinin2020}. The closest binaries are impacted by tides. For MS stars with convective envelopes below the Kraft break ($T_{\rm eff}$ $<$ 6,200\,K; $M_1$ $<$ 1.2\,\Msun), binaries are tidally circularized below $P$ $\lesssim$ 8~days; above the Kraft break, only binaries within $P$ $\lesssim$ 2~days are fully circularized \citep{MeibomMathieu2005,Abt2006,MoeDi-Stefano2017,Geller2021,ZanazziWu2021,Torres2021}. 


{\it Triple Statistics:} \index{Multiple Systems!higher order}
 About 14\% of solar-type primaries are in triples and higher order multiples \citep[][see~Fig.~\ref{fig:MultFrac}]{Tokovinin2014}. The closest binaries within $P<3$ days are found almost exclusively ($96\%$) in triples, whereas only a third of binaries with $P$ $\approx$ 20 days have tertiary companions \citep{TokovininThomas2006,LaosStassun2020}. Beyond these periods,  \citet{Tokovinin2014} 
concluded that the joint distribution $f(P_{\rm in},P_{\rm out})$ of inner periods $P_{\rm in}$ and outer periods $P_{\rm out}$ is consistent with both companions being independently drawn from the same overall lognormal period distribution for solar-type binaries, but with the added constraint of dynamical stability. Various theoretical studies have proposed slightly different criteria for dynamical stability in triples \citep{MardlingAarseth2001a,Georgakarakos2008}. Based on their sample of solar-type triples with reliable orbital solutions, \citet{Tokovinin2004} measured an empirical relation:

\begin{equation}
    P_{\rm out}/P_{\rm in} > \frac{5}{(1-e_{\rm out})^3},
\end{equation}

\noindent where $e_{\rm out}$ is the eccentricity of the outer tertiary. 


\citet{Tokovinin2017b} examined the relative orbital orientations of visually resolved triples. For outer tertiaries beyond $a_{\rm out}$~$>$~1,000~au, they found an equal number of prograde and retrograde orbits with respect to the inner binaries, suggesting isotropic orientations. Meanwhile, the majority of compact solar-type triples within $a_{\rm out}$~$<$~50~au have prograde configurations, demonstrating a high degree of alignment between the inner and outer orbits. \citet{Tokovinin2017b} also found tentative evidence that triples become more misaligned with increasing primary mass. Finally, \citet{BorkovitsHajdu2016} identified a large population of very compact triples with $a_{\rm out}$~$<$~10~au. All have mutual inclinations within $I_{\rm mutual}$ $<$ 60$^{\circ}$, about half of which are nearly coplanar with $I_{\rm mutual}$ $<$ 20$^{\circ}$. We discuss the special implications of triple star statistics for formation models in \S\ref{sec:secular}.


\subsubsection{\textbf{M-dwarfs}}
\label{sec:Mdwarfs}

\begin{figure}[h]
\centerline{
\includegraphics[trim=2.0cm 1.0cm 3.0cm 1.0cm, clip=true, width=3.1in]{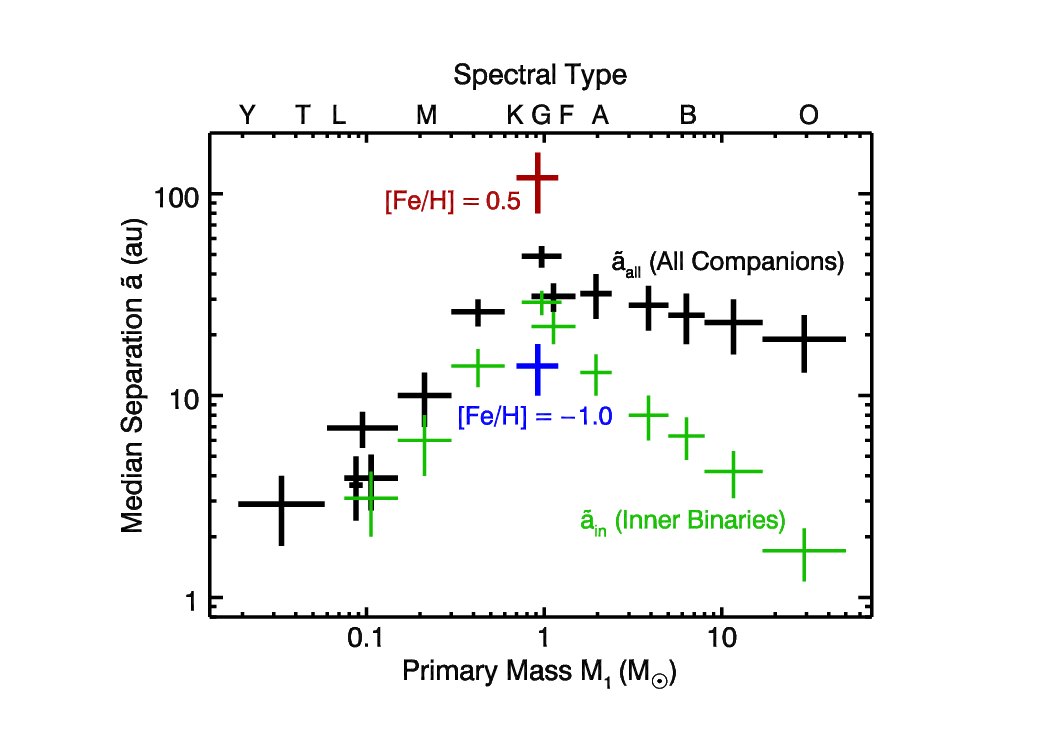}}
\vspace{-0.1in}
\caption{{\small Median separations of all companions (thick black) and of inner binaries only (thin green) as a function of primary mass. Metal-poor solar-type binaries are skewed toward smaller separations (blue) while metal-rich solar-type binaries favor longer periods (red). A significant fraction of massive stars are in triples, so their inner binary distributions are substantially skewed toward shorter separations compared to all companions.} }
\label{fig:Mediana}
\end{figure}

\begin{figure}[t!]
\centerline{
\includegraphics[width=3.2in]{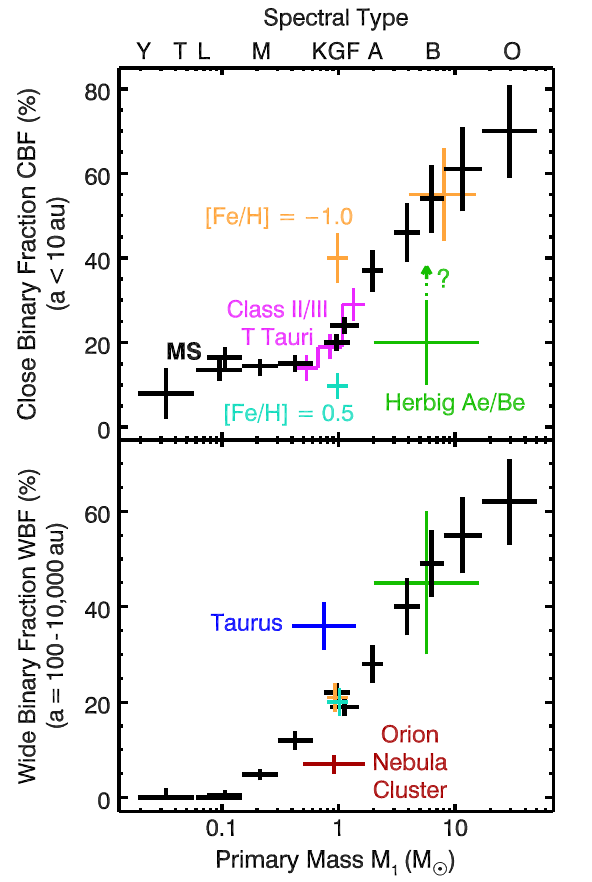}}
\vspace{-0.1in}
\caption{{\small The close ({\bf top}) and wide ({\bf bottom}) binary fractions of MS stars (thick black) both increase with primary mass but with markedly different trends.  Metal-poor (orange) solar-type stars have a larger CBF than their metal-rich (cyan) counterparts \citep{MoeKratter2019}, while the CBF of OB stars is constant with metallicity \citep{MoeDi-Stefano2013}. The WBF of solar-type stars is also metallicity invariant \citep{MoeKratter2019,El-BadryRix2019}.  The CBF of Class II/III T~Tauri stars (magenta) increases with mass like the field MS population \citep{KounkelCovey2019}. We highlight discrepancies in the WBF between the dense Orion Nebula Cluster (red; \citealt{Duchene2018}) and sparse Taurus star-forming region (blue; \citealt{KrausIreland2011c}), which exhibit a deficit and excess, respectively, compared to the field MS. The WBF of Herbig Ae/Be stars (green) is consistent with their field MS analogs  \citep{KouwenhovenBrown2005a,Baines2006}, but the CBF of Herbig Ae/Be stars appears lower \citep{CorporonLagrange1999,ApaiBik2007a,Sana2017}; ``?" indicates this could arise from a selection bias whereby close binaries shorten disk lifetimes. }}
\label{fig:CloseWide}
\end{figure}

{\it Frequencies}: Building on the historical work of \citet{FischerMarcy1992a}, many subsequent studies have employed a variety of detection techniques to fill in the multiplicity parameter space of the most common stars in the galaxy
\citep{Close2003,Gizis2003,Bouy2003,Daemgen2007,LawHodgkin2008,BergforsBrandner2010,JansonHormuth2012,DieterichHenry2012,Janson2014,Ward-DuongPatience2015}.
By combining previous samples with their own imaging survey, \citet{WintersHenry2019} provided the most complete census of the multiplicity statistics for a 25-pc volume-limited sample of M-dwarfs primaries. We split their sample into three mass intervals and focus on the 20-pc subset that is relatively complete (see their Figs.~19-20 and results in our Table~\ref{table:MultStat}). The multiplicity fraction, triple fraction, and median companion separation all increase with primary mass within the M-dwarf regime (see Figs.~\ref{fig:MultFrac}-\ref{fig:Mediana}). As shown in Fig.~\ref{fig:CloseWide}, the wide binary fraction dramatically increases from 1\% to 12\% across primary masses 0.1\,-\,0.5\,\Msun.  In Fig.~\ref{fig:Floga}, we plot the separation distribution of early-M ($M_1$ = 0.4\,-\,0.6\,\Msun) and late-M ($M_1$ = 0.075\,-\,0.15\,\Msun) binaries from the 20-pc subset of \citet{WintersHenry2019} along with their best-fit lognormal distributions (see Table~2 for parameters).


\citet{WintersHenry2019} excluded both BD and WD companions from their reported multiplicity statistics but kept records of such systems in their sample. The fraction of M-dwarfs within 20 pc that have observed BD companions is only 2\%, which increases slightly to 4\% after accounting for incompleteness. Whether or not BD companions are included, the multiplicity fraction of late-M dwarfs is lower than that of early-M dwarfs, which in turn is less than that of solar-type primaries.

{\it Mass ratios:} Across intermediate separations, early-M binaries follow a roughly uniform mass-ratio distribution, similar to that of  solar-type binaries. Meanwhile, mid-M and late-M binaries are skewed substantially toward equal masses, which is not a selection effect due to incompleteness or exclusion of low-mass companions \citep{Close2003,LawHodgkin2008,DieterichHenry2012,Janson2014}. Wide companions ($a$ $>$ 100 au) to M-dwarfs are skewed toward slightly smaller masses \citep{Ward-DuongPatience2015,El-BadryRix2019b}, similar to the separation trend observed in solar-type binaries.

\begin{table}[t!]
\begin{center}
\small
\caption{~~~~~~~~{\sc Lognormal Fit Parameters}}
\vspace{0.1in}
\small
\begin{tabular}{lccc}

$M_1$ (\Msun) & $\mu$ (au) & $\sigma_{\rm log\,a}$ & CF \\
\hline
0.075\,-\,0.15 & 4 & 0.7 & 0.21 \\
0.3\,-\,0.6 & 25 & 1.3 & 0.38 \\
0.75\,-\,1.25 & 40 & 1.5 & 0.60 \\
\end{tabular}
\end{center}
\vspace*{-0.2cm}
{\small NOTE - Mean $\mu$ and dispersion $\sigma_{\rm log\,a}$ of lognormal separation distributions $f_{\rm log\,a}$ scaled to the companion frequency CF integrated across $-$2.0 $<$ log\,$a$\,(au) $<$ 4.5 as defined in Eqn.~\ref{CFdef}}.

\end{table}



\subsubsection{\textbf{Brown dwarfs}}
\label{sec:BrownDwarfs}\index{Brown Dwarfs}

{\it Frequencies:} Although completeness corrections become more challenging with decreasing primary mass, surveys confirm that the MF continues to decline through the BD regime (see Table~\ref{table:MultStat} and Fig.~\ref{fig:MultFrac}). The period-integrated MF is roughly $\approx$\,20\% for L/early-T dwarfs and only $\approx$\,8\% for late-T/early-Y dwarfs  \citep{Reid2001,Close2002,BurgasserKirkpatrick2003,Bouy2003,Close2003,Gizis2003,Allen2007,Burgasser2007,Aberasturi2014,BasriReiners2006,Joergens2008,Blake2010,Hsu2021,Fontanive2018}. The vast majority of BD binaries have $a$~$=$~1\,-\,20~au. The binary fraction within $a<1$\,au is only $\approx$\,4\%, and only a handful of systems have been identified at wide separations \citep{Burgasser2007,Allen2007,Joergens2008,RadiganLafreniere2009,FahertyGoodman2020}. The lognormal separation distribution of BD binaries narrowly peaks near $\widetilde a_{\rm all}$ = 3\,au, similar to that of late-M binaries (see Fig.~\ref{fig:Mediana}). 



{\it Mass Ratios:} BD binaries are extremely skewed toward equal masses, i.e., $\gamma_{\rm trunc}$ = 2\,-\,3 for L/early-T binaries \citep{BurgasserKirkpatrick2003,Close2003,Allen2007} and possibly up to $\gamma_{\rm trunc}$ $\approx$ 5 for late-T/early-Y binaries \citep{Fontanive2018}. These measurements are not biased by incompleteness to smaller mass ratios, nor do they exclude objects below the deuterium burning limit.

\subsubsection{\textbf{A stars}}
\label{sec:Astars}
\index{Stars!A}
{\it Frequencies:}
Moving toward earlier spectral types, we find that the trend of an increasing MF with primary mass continues. The wide binary fraction of A-dwarfs is slightly larger than the solar-type value \citep{De-RosaPatience2014}. Early spectroscopic binary studies of A stars focused on their magnetic and chemical peculiar properties \citep{Abt1965,Abt1983}. There has been no recent systematic A-dwarf spectroscopic or eclipsing binary survey, but the upcoming {\it Gaia} full third data release 
will soon remedy this (\S\ref{sec:nextgen}). Instead, close binary properties of A stars have been inferred from examining {\it Kepler} phase modulations of $\delta$ Scuti stars, which are pulsating late-A/early-F MS stars \citep{MurphyMoe2018}. They found an A-dwarf binary fraction of 14\% across 0.6\,-\,3.5~au, double the G-dwarf value across the same separation interval. Since the data remain incomplete, we use an interpolation from \citet{MoeKratter2021} to derive the full bias-corrected multiplicity statistics for A-dwarfs across the entire separation range (see Table~\ref{table:MultStat} and Fig.~\ref{fig:MultFrac}).

{\it Mass ratios:}
Across intermediate separations, A-dwarf binaries peak near $q$ = 0.3 and exhibit a very small but statistically significant excess twin fraction \citep{De-RosaPatience2014,MoeDi-Stefano2017,MurphyMoe2018}. Meanwhile, wide A-type binaries are skewed toward even smaller mass ratios but are still slightly top-heavy compared to random pairings drawn from the IMF \citep{De-RosaPatience2014,MoeDi-Stefano2017}.

\subsubsection{\textbf{B stars}}
\label{sec:Bstars}
\index{Binaries!massive}
\index{Stars!B}

{\it Frequencies:} Building on the groundbreaking work by \citet{AbtGomez1990}, surveys of B-stars show binary fractions that continue to rise with primary mass across all orbital separations \citep{Duchene2001,ShatskyTokovinin2002,Roberts2007,RizzutoIreland2013,MoeDi-Stefano2015b,Caballero-Nieves2020,BanyardSana2022}. \citet{MoeDi-Stefano2017} combined the data from spectroscopy, eclipsing binaries, long-baseline interferometry, and adaptive optics to compute the statistics reported in Table~\ref{table:MultStat}. Both the masses and companion frequencies of B stars span a large interval, increasing from CF = 1.1 for 3\Msun\ primaries to 1.9 for 16\Msun\ primaries (see Fig.~\ref{fig:MultFrac}). 



The overall separation distribution of companions to mid-B stars roughly follows 
\"{O}pik's law, i.e., uniform in log\,$a$ \citep{Opik1924,KouwenhovenBrown2007,KobulnickyFryer2007,MoeDi-Stefano2017}. 
Fig.~\ref{fig:Floga} displays $f_{\rm log\,a}$ for $q$ $>$ 0.1 companions to 8\,\Msun\ mid-B primaries as reported in \citet{MoeDi-Stefano2017}. Early-B inner binaries favor shorter separations with a median of $\widetilde a_{\rm in}$ $\approx$ 6 au \citep[][see Fig.~\ref{fig:Mediana}]{RizzutoIreland2013,MoeDi-Stefano2013}. 

{\it Mass ratios:} B-type binaries within $a$ $<$ 1 au follow a uniform mass-ratio distribution \citep{AbtGomez1990,Kobulnicky2014}. Companions to B-type stars across intermediate separations of $a$~=~1\,-\,100~au peak near $q$ $\approx$ 0.3  \citep{RizzutoIreland2013,GulliksonKraus2016,MoeDi-Stefano2017}, similar to their A-type counterparts. Wide companions to B-type stars are substantially skewed toward small mass ratios ($\gamma_{\rm trunc}$ $\approx$ $-$2) but with a slight flattening below $q$ $\lesssim$ 0.3 \citep{AbtGomez1990,ShatskyTokovinin2002,MoeDi-Stefano2017}.

\subsubsection{\textbf{O stars}}
\index{Stars!O}
\label{sec:Ostars}
{\it Frequencies:} The total multiplicity fraction of O stars in clusters is MF $>$\,90\% and the companion frequency is CF = 2.1\,$\pm$\,0.3 above $q$ $>$ 0.1 (see Figs.~\ref{fig:MultFrac}\,-\,\ref{fig:CloseWide}). Hence, the majority of O stars are in triples and higher ordered multiples (THF = 68\%\,$\pm$\,18\%). As with solar-type stars, these statistics derive from a wide range of survey methods: spectroscopy, interferometry, sparse aperture masking, AO, lucky imaging, and common proper motion \citep{Turner2008,MasonHartkopf2009a,Sanade-Mink2012,Chini2012,Kobulnicky2014,Sota2014,SanaLe-Bouquin2014,Aldoretta2015,MoeDi-Stefano2017,MaizApellaniz2019,Caballero-Nieves2020}. Given the high fraction of triples, we display both the total and inner-binary separation distributions in Fig.~\ref{fig:Floga}. 



The statistics above describe O stars in young clusters. In contrast, the $\approx$\,20\% of O stars that are runaway and/or in the field exhibit a lower binary fraction \citep{MasonHartkopf2009a,Chini2012,LambOey2016}. The multiplicity statistics of older field and runaway O stars have been altered by dynamical effects and/or binary evolution  \citep{Hoogerwerf2001}. Hence, we consider only cluster O stars, which have multiplicity properties more similar to their primordial distributions. 

{\it Mass Ratios:} Spectroscopic and eclipsing binary surveys of O stars reveal a fairly uniform mass-ratio distribution within $a$ $<$ 0.5~au but with a small excess twin fraction \citep{PinsonneaultStanek2006a,Sanade-Mink2012,Kobulnicky2014,MoeDi-Stefano2017,ShenarSana2022}. 
Nonetheless, the excess twin fraction among O-type binaries within $a$ $<$ 0.5 au is only 10\%, which is lower than the 30\% excess twin fraction measured among solar-type binaries across the same separation interval \citep{MoeDi-Stefano2017}. There is no excess twin fraction among O-type binaries beyond $a$ $>$ 0.5 au \citep{Sanade-Mink2012}. Similar to early-B binaries, the mass-ratio distribution of O-type binaries beyond $a$ $>$ 100~au is skewed steeply toward small mass ratios 
but with a flattening below $q$ $<$ 0.3 \citep{SanaLe-Bouquin2014,MoeDi-Stefano2017}.

\subsubsection{\textbf{Summary of observed MS trends}}
\label{sec:MultTrends}

The multiplicity fraction increases monotonically with primary mass from MF $\approx$ 20\% for BDs and late-M dwarfs to $\approx$\,50\% for solar-type stars to MF $>$ 90\% for OB stars (see Fig.~\ref{fig:MultFrac}). The triple fraction increases even more dramatically from THF $\approx$ 2\% for late-M dwarfs to 14\% for FGK dwarfs to nearly 70\% for O stars. Similarly, the companion frequency is CF = 0.2 for low-mass stars while OB stars have CF $\approx$ 2.0 companions per primary on average. 

Figs.~\ref{fig:Floga}-\ref{fig:CloseWide} illustrate the nuances of the separation distribution  with respect to primary mass. BD, M-dwarf, and FGK-dwarf binaries follow lognormal separation distributions, where the peak increases from $\widetilde a_{\rm all}$ = 3 au for BD and late-M dwarfs to $\widetilde a_{\rm all}$ = 40 au for solar-type stars. Meanwhile, companions to OBA stars roughly follow \"{O}pik's law, albeit inner binary companions to early-B and especially O stars are skewed toward shorter separations, e.g., $\widetilde a_{\rm in}$ $\approx$ 2~au. Interestingly, the close binary fraction is nearly constant at CBF = 15\% across $M_1$ = 0.05\,-\,0.8\,\Msun\, 
but then rapidly increases with primary mass, reaching 70\% for O stars. Meanwhile, the wide binary fraction increases dramatically from WBF = 1\% for late-M stars to 12\% for early-M stars, and then continues to moderately increase up to 60\% for OB stars. The different trends in the close versus wide binary fractions can help constrain the various binary formation processes that operate on different scales.

The binary mass-ratio distribution becomes skewed toward more unequal masses with increasing primary mass and increasing orbital separation. It is firmly established that BD and late-M binaries intrinsically favor equal masses. We note that some binary populations favor mass-ratio distributions that deviate from a power-law model: close and intermediate-period solar-type binaries follow uniform distributions with a 30\% and 10\% excess twin fraction, respectively (see \S\ref{sec:FGKstars}),  very close O-type binaries follow a uniform distribution with a 10\% excess twin fraction (\S\ref{sec:Ostars}), and intermediate-period companions to BA primaries peak near $q$ = 0.3 (\S\ref{sec:Astars}-\ref{sec:Bstars}). Finally, wide companions to OBA primaries match the slope $\gamma_{\rm trunc}$ = $-$2.35 expected from random pairings drawn from a Salpeter IMF, but only across the truncated interval $q$ = 0.4\,-\,1.0. The mass-ratio distribution flattens below $q$~$<$~0.3, so the overall distribution is slightly top-heavy compared to random pairings from the IMF (see  \S\ref{sec:Astars}-\ref{sec:Ostars}). Observed mass-ratio distributions demonstrate that the component masses in close and intermediate-period binaries coevolved during their prior fragmentation and accretion phases, while the components in wide binaries formed mostly, but not fully, independently (see \S\ref{sec:TheoreticalModels}). 

\subsection{\textbf{Metallicity Dependence}}
\label{sec:MSmetals}\index{Stars!metallicity}
Since {\it PPVI}, the field has seen a substantial change in our understanding of the metallicity dependence of binary statistics.
Early surveys of metal-poor halo stars measured a spectroscopic binary fraction that is consistent with the disk population \citep{Latham2002,Carney2005}. However, it is more difficult to detect spectroscopic binary companions to metal-poor stars, which have weaker absorption lines, and so the bias-corrected close binary fraction of halo stars is definitively larger \citep{MoeKratter2019}. There are now several surveys indicating that the close binary fraction of solar-type stars decreases significantly with metallicity \citep{GretherLineweaver2007,Rastegaev2010,RaghavanMcAlister2010a,BadenesMazzola2018,MoeKratter2019,Mazzola2020}. For example, both APOGEE spectroscopic binaries and {\it Kepler} eclipsing binaries reveal a close binary fraction that decreases by a factor of four across $-$1.0 $<$ [Fe/H] $<$ 0.5 \citep[][see  Fig.~\ref{fig:CloseWide}]{MoeKratter2019}. \citet{Mazzola2020} measured an even steeper anti-correlation between the close binary fraction and chemical abundances of $\alpha$ elements like C, O, Mg, and Si. 

Meanwhile, the wide binary fraction across $a$ = 200\,-\,1,000~au appears to be metallicity invariant \citep{MoeKratter2019,El-BadryRix2019}.  \citet{El-BadryRix2019} utilized their catalog of common-proper-motion binaries from {\it Gaia} to  demonstrate that a strong metallicity dependence emerges only within $a$ $<$ 200~au. They measured a factor of three decrease in the fraction of solar-type binaries near $a$ = 50 au across $-1.0$ $<$ [Fe/H] $<$ 0.5, nearly the factor of four variation observed within $a$ $<$ 10~au across the same metallicity interval.  Figs.~\ref{fig:Floga}~\&~\ref{fig:Mediana} show that metal-poor solar-type binaries skew toward shorter separations while metal-rich binaries favor long periods. These observations indicate metallicity plays a crucial role in binary fragmentation at high densities and small separations (see \S\ref{sec:MetalsTheory}).

Finally, the small population of very wide solar-type binaries across $a$ = 1,000\,-\,10,000 au exhibits a non-monotonic metallicity dependence in their companion frequency, peaking near [Fe/H] = 0.0 \citep{Hwang2021}.
The deficit of extremely wide companions ($a$ $>$ 5,000 au) to older, metal-poor halo stars is at least partially due to dynamical disruption via galactic tides and gravitational perturbations. However, \citet{Hwang2021} argued that disruptions alone cannot explain the metallicity dependence observed in binaries across $a$ = 1,000\,-\,5,000~au, which have larger binding energies. They instead concluded that metal-poor halo stars initially formed in denser clusters, resulting in a smaller very wide binary fraction (see also \S\ref{sec:statisticsPMS}). Meanwhile, \citet{Hwang2021} offered multiple theories for the smaller very wide binary fraction of [Fe/H] = 0.5 disk stars, including radial migration within the Milky Way and/or dynamical unfolding of initially unstable triples. 

Unlike solar-type stars, the close binary fraction of massive stars does not decrease with metallicity. Both eclipsing and spectroscopic binary surveys of OB stars in the metal-poor Magellanic Clouds reveal a close binary fraction that matches their Milky Way counterparts at solar-metallicity \citep[][see Fig.~\ref{fig:CloseWide}]{MoeDi-Stefano2013,Sana2013,Dunstall2015}.  Moreover, the period and mass-ratio distributions of close OB binaries in the Magellanic Clouds are consistent with the Milky Way distributions \citep{MoeDi-Stefano2013,AlmeidaSana2017,VillasenorTaylor2021}. Recent surveys of massive metal-poor stars in older environments indicate a slightly lower binary fraction \citep{Bodensteiner2021,Neugent2021}, but this may be due to binary evolution and/or other selection effects. 

\subsection{\textbf{Pre-MS Phase}}
\label{sec:statisticsPMS}
\index{Pre-Main Sequence}

\subsubsection{\textbf{T Tauri stars}}\label{statisticsTTS}
\index{T Tauri}

Spectroscopic surveys of Class II/III T Tauri stars reveal a close binary fraction that is consistent with their field MS counterparts \citep{Mathieu1994,Melo2003,Elliott2014,Prato2007,KounkelCovey2019}. Specifically, \citet{KounkelCovey2019} measured a bias-corrected close binary fraction that increases from CBF = 12\% for mid-M T Tauri stars to CBF = 33\% for early-F T Tauri stars (see Fig.~\ref{fig:CloseWide}). They also found that close T Tauri binaries follow the same short-period tail of the lognormal separation distribution as field MS binaries across $a$ = 0.1\,-\,10 au (see Fig.~\ref{fig:Floga}). These observations demonstrate that the primary mass dependence and separation distribution of close binaries are largely set by the beginning of the T Tauri stage. 

Two recent spectroscopic surveys suggest that the close binary fraction decreases with age across $\tau$ = 2\,-\,100~Myr \citep{Jaehnig2017,Zuniga-Fernandez2021}, but we argue these results are due to selection biases. Based on APOGEE observations, \citet{Jaehnig2017} reported completeness-corrected CBF = 12\%\,-\,20\% for $M_1$ $\approx$ 0.5\,\Msun\ pre-MS stars in five different star-forming groups, which is consistent with the early-M dwarf field value. In contrast, they measured only CBF $\approx$ 4\% for low-mass MS stars in the $\approx$\,100 Myr old Pleiades cluster. This discrepancy likely arises because the low-mass stars in the Pleiades observed by APOGEE are concentrated toward the periphery of the cluster, which has a lower binary fraction due to mass segregation \citep{RaboudMermillod1998}. Moreover, the close binary fraction of FGK MS stars in the Pleiades measured from more robust long-term radial velocity monitoring is CBF = 25\%\,$\pm$\,3\% \citep{Torres2021}, similar to the corresponding field value (see \S\ref{sec:FGKstars}). 
\citet{Zuniga-Fernandez2021} recently measured an even larger solar-type close binary fraction of CBF = 30\%\,$\pm$\,6\% averaged across three young moving groups with $\tau$ $<$ 20~Myr, including Beta Pic, and measured only CBF $\approx$ 10\%\,-\,15\% in slightly older associations with $\tau$ = 20\,-\,100~Myr. However, of their ten spectroscopic binaries in Beta Pic, six have slow rotation velocities and/or periods compared to their single star analogs, suggesting they are older non-members (only two have short orbital periods where tides could have reduced the spin). Moreover, four of the ten have $>$\,20\% probabilities of not being Beta Pic members, while only two of the remaining 32 members have such low reliabilities. After removing likely non-member spectroscopic binaries, the close binary fraction in young comoving groups matches the field value.

Meanwhile, AO and speckle imaging surveys of T~Tauri stars in star-forming environments like Taurus, Chamaeleon, Ophiuchus, and Scorpius reveal an excess of companions across $a$ = 10\,-\,200 au \citep{GhezNeugebauer1993,ReipurthZinnecker1993,Leinert1993,Ghez1997,KrausIreland2011c,TokovininBriceno2020}, which is discontinuous with the spectroscopic binary results (see Fig.~\ref{fig:Floga}). It was originally assumed that only low-density environments exhibited such an excess while high-density environments would have a deficit, and therefore the field population derived from a combination of these environments. However, \citet{Duchene2018} found tentative evidence for the same excess of companions across $a$ = 10\,-\,60 au in the dense Orion Nebula Cluster (ONC). 
Notably \citet{DeFurio2019} found no excess in this separation range for low-mass M-dwarfs in Orion.
Thus the discontinuity and excess near $a$ = 10~au 
in most star-forming regions poses a mystery and warrants further study.

Across wide separations of $a$ = 100\,-\,10,000 au, the binary fraction is sensitive to the stellar density of the environment.  Fig.~\ref{fig:Floga} shows that the dense Orion Nebula Cluster (ONC) exhibits a deficit of wide companions \citep{Scally1999,Kohler2006,ReipurthGuimaraes2007,Duchene2018,JerabkovaBeccari2019}. Environments with intermediate density such as Upper Scorpius and the Orion OB1 association are consistent with the field \citep{Bradner1996,KrausIreland2008b,KounkelMegeath2016,TokovininBriceno2020,TokovininPetr-Gotzens2020}.  Finally, low-density star-forming regions like Taurus and Chamealeon exhibit a slight excess of wide companions \citep{GhezNeugebauer1993,ReipurthZinnecker1993,Leinert1993,Ghez1997,KohlerLeinert1998,ConnelleyReipurth2008, KrausIreland2011c,Joncour2017}. These observations suggest that the majority of wide binaries are dynamically disrupted in extremely dense regions, and that the field MS population derives from a mixed composition of low, intermediate, and high-density environments. Specifically, for every solar-type primary born in a dense ONC environment, a counterpart must be born in a sparse Taurus-like association such that the average of the star-forming environments yields a wide binary fraction similar to Upper Scorpius and the field MS population. 

However, within the same star-forming region, there is tentative evidence for the opposite trend between the wide binary fraction and the immediately surrounding stellar density. 
For example, \citet{KounkelMegeath2016} imaged 129 Class I protostars and 197 Class II/III T Tauri stars across the Orion Molecular Cloud in the near-IR with {\it HST} and the InfraRed Telescope Facility. Across $a$ = 100\,-\,1,000 au, they found that protostars and T Tauri stars have very similar wide binary fractions, which are consistent with the field MS population. Surprisingly, \citet{KounkelMegeath2016} found that young stars in sub-regions of higher spatial stellar density exhibited a slightly larger wide binary fraction. Recent ALMA observations of Class 0/I protostars in Orion seem to confirm this result (\citealt{TobinOffner2021}; see \S\ref{sec:ProtostarMult}).\index{Class 0}\index{Class I}

There are noticeable differences between close versus wide T Tauri binaries. Similar to their MS counterparts, T Tauri binaries within $a$ $<$ 200~au follow a uniform mass-ratio distribution with a small excess twin fraction, while wider companions are skewed toward smaller masses \citep{KohlerLeinert1998,TokovininBriceno2020}. As discussed in \S\ref{sec:diskmasses}, binaries within $a$ $\lesssim$~50~au truncate the disk masses, radii, fluxes, and lifetimes compared to single stars and wider binaries \citep{JensenMathieu1996, HarrisAndrews2012a,KrausIreland2012,CheethamKraus2015}. Hence, a sample of T Tauri stars selected according to their large disk masses will be systematically biased against close binaries. For example, \citet{KounkelCovey2019} measured the fraction of Class II and III T Tauri stars that are in double-lined spectroscopic binaries (SB2s), which have large mass ratios $q$ $>$ 0.6. Compared to the field MS population, they found that Class II stars exhibit a deficit of SB2s while Class III stars have a surplus. \citet{KounkelCovey2019} concluded that close T Tauri binaries with large mass ratios quickly consume or disrupt their disks, thereby accelerating the Class II phase and extending the duration of the Class III phase compared to single stars and wider binaries.

\subsubsection{\textbf{Herbig Ae/Be stars and Massive YSOs}}
\index{Stars!Herbig Ae/Be}

According to various high-resolution imaging surveys, the wide binary fraction of Herbig Ae/Be stars and massive YSOs in young star-forming environments is WBF = 30\%\,-\,60\%, increasing with stellar mass \citep{KouwenhovenBrown2005a,Baines2006,Wheelwright2010,Gravity2018,Pomohaci2019}. These values are consistent with their MS analogs in the field and slightly older clusters (see Fig.~\ref{fig:CloseWide}).  Wide companions to OB stars in the Trapezium Cluster of the ONC skew toward small mass ratios $q$ $\approx$ 0.2, similar to their counterparts in slightly older clusters \citep{Gravity2018}. The samples of wide Herbig Ae/Be binaries are not sufficiently large to detect a dependence on surrounding stellar density as measured for T~Tauri stars.  

Meanwhile, the close binary fraction of Herbig Ae/Be stars and massive YSOs inferred from spectroscopy is only CBF = 10\%\,-\,30\% \citep{CorporonLagrange1999,ApaiBik2007a,Sana2017}, considerably lower than the corresponding MS fraction. 
One interpretation is that 1~Myr old massive stars do not have any close companions, and that the companions at intermediate separations harden on Myr timescales (see also \citealt{Ramirez-Tannus2021}). An equally compelling  explanation is that close companions to massive pre-MS stars shorten disk lifetimes, similar to their T~Tauri analogs, and therefore the subset of massive stars within a coeval cluster that retains a disk is less likely to have close companions. Resolving this ambiguity requires a systematic survey for close binary companions to both massive YSOs with disks and normal OBA stars without disks within the same cluster  \citep{BordierFrost2022}. Larger samples of spectroscopic and eclipsing OB binaries \index{Binaries!eclipsing}\index{Binaries!spectroscopic} in young star-forming environments within the Large Magellanic Cloud reveal a sizeable fraction with close companions \citep{Sana2013,MoeDi-Stefano2015a}. For example, \citet{MoeDi-Stefano2015a} discovered 18 early-B MS stars with 1\,-\,2\,\Msun\ pre-MS eclipsing binary companions across $P$ = 3\,-\,8~days, a few of which are as young as 0.6\,-\,1.0~Myr according to the measured temperatures and inflated radii of the pre-MS companions still on the Hyashi track.

\subsection{\textbf{Overall Multiplicity Statistics}}

Given the substantial excess of wide companions to YSOs compared to field stars (see \S\ref{sec:statisticsPMS} and \S\ref{sec:ObsFormMult}), most star systems are born as multiples. However, subsequent dynamical disruptions reduce the overall multiplicity fraction, especially for low-mass binaries that have lower binding energies \citep{Kroupa1995}. As emphasized in \citet{Lada2006}, most MS field stars are M-dwarfs and most M-dwarf field stars are single, and thus most MS star {\it systems} are single. By weighting our multiplicity statistics with the IMF of primary stars \citep{Kroupa2013}, we compute that only 35\% of zero-age MS star {\it systems} are multiple, consistent with the conclusion of \citet{Lada2006}. Nonetheless, most MS {\it stars} are members of multiples after considering the fact that multiples contain two or more stars. For example, given 100 M-dwarfs, only 49 are single, 16 are paired with 16 other M-dwarfs, 12 are binary companions to more massive BAFGK primaries, and 7 are tertiaries in triples and higher-ordered multiples. By weighting our statistics with the IMF, we conclude that 58\% of MS stars are members of multiples.

\index{Binaries!statistics|)} 
\index{Binaries!mass ratio|)} 
\index{Binaries!separation|)} 
\index{Binaries!fraction|)}




\begin{figure*}[h]
 \epsscale{1.75}
 \plotone{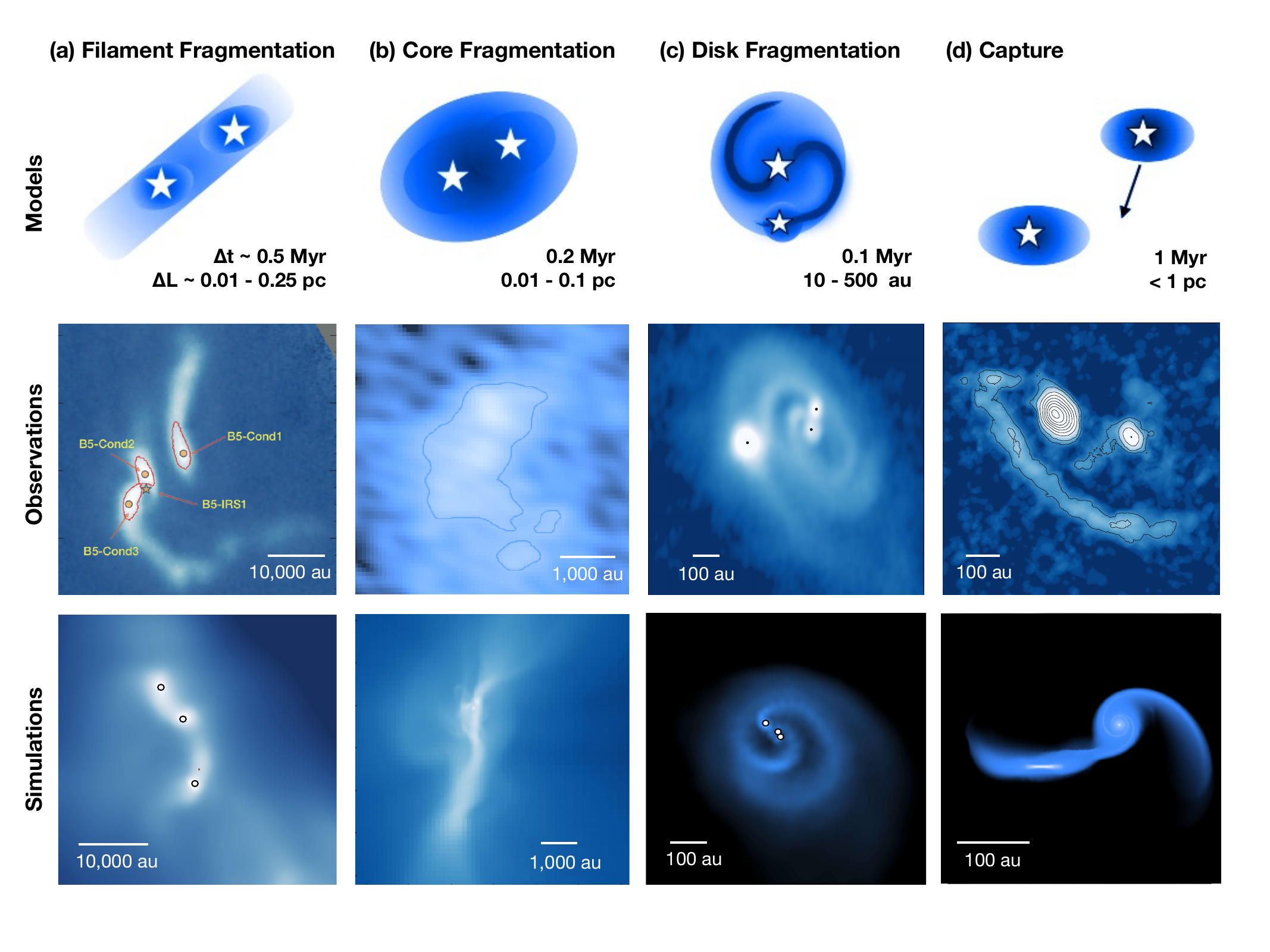}
 \vspace{-0.25in}
 \caption{\small Summary of mechanisms for multiple formation. {\bf Top:} Model and approximate range of time and length scales for each process. {\bf Middle:} Proposed observational examples. From left to right: B5 in Perseus \citep{PinedaOffner2015}, SM1N in Ophiuchus \citep{KirkDunham2017}, L1448 IRS3B in Perseus \citep{ReynoldsTobin2021} and RW Aur \citep{RodriguezLoomis2018}. {\bf Bottom:} Examples from numerical simulations. From left to right:
 \citet{GuszejnovGrudic2021}
 \citet{OffnerDunham2016},  \citet{Bate2018}, and \citet{MunozKratter2015}. }
 \label{schematic}
\end{figure*}

\section{\textbf{MODELS FOR MULTIPLE STAR FORMATION}}\label{sec:TheoreticalModels}


A successful theory for multiple formation must both reproduce observed multiplicity statistics (\S2) and model young, individual sources (\S\ref{sec:ObsFormMult} and \S\ref{sec:PlanetFormation}). To date, there is no comprehensive, predictive theory for multiple star formation. Models have instead tackled the origin of multiplicity at different spatial scales and explored the influence of different physical processes.

By definition, multiple formation requires two or more events of gravitational collapse\index{Gravitational Collapse} that ultimately form a bound stellar system. We define the process by which self-gravitating objects develop substructures that evolve and collapse independently as {\it fragmentation}\index{Fragmentation}.
\citet{Hoyle1953a} first proposed star formation through the hierarchical fragmentation of collapsing gas.  
However,
this ``classic" star formation paradigm,  described only by the interplay of gravity and thermal pressure, produces smooth, isotropic collapse \citep[e.g.,][]{Shu1977}. Some of the first numerical star-formation calculations demonstrated that multiple fragmentation events are not possible under these conditions alone \citep{Larson1972}. 

Observations and theoretical studies have reinforced that this simple picture is a poor representation of how most stars form. Instead, star formation is regulated not only by gravity and thermal pressure but also turbulence\index{Turbulence} and magnetic fields\index{Magnetic Fields}, which introduce rotation and asymmetry, crucial ingredients that facilitate additional collapse events within dense cores\index{Dense Cores}, filaments\index{Filaments} and accretion disks. The high degree of clustering in many star-forming regions also suggests that interactions between stars are important for the evolution of multiple systems.  

Current ideas for multiple formation can be divided into three main categories: theories in which multiples form via fragmentation of a core\index{Dense Cores!fragmentation} or filament\index{Filaments!fragmentation} (\S\ref{sec:corefrag}), via fragmentation of a massive accretion disk\index{Disks!fragmentation} (\S\ref{sec:diskfrag}) or through dynamical interactions (\S\ref{sec:capture}). This third mode can also rearrange the hierarchy and multiplicity of systems formed via the prior fragmentation channels. Figure \ref{schematic} summarizes these mechanisms and provides examples from observations and numerical simulations.
\S\ref{sec:enviornment} reviews the landscape of state-of-the-art calculations, approaches the origin of stellar multiplicity holistically and discusses   our current understanding of how physics and environment shape multiplicity statistics.

\subsection{\textbf{Core and Filament Fragmentation}} \label{sec:corefrag}

Stars form in dense cores, most of which are observed to be embedded within $\sim 0.1$\,pc filaments in nearby star-forming regions \citep{KonyvesAndre2015}. High-resolution observations have discovered even smaller filaments within cores that host forming stars \citep[][see Fig.~\ref{schematic}]{PinedaOffner2015}.  Core and filament fragmentation models posit that stellar multiples arise from over-densities that develop and collapse\index{Gravitational Collapse} within these parent structures, producing initially widely separated systems ($\gtrsim500$~au). 
Note that the terms {\it core} and {\it filament} represent morphological descriptions rather than physical definitions, and here, we define over-densities with small aspect ratios ($r_a/r_b < 3/1$) as cores and structures with higher aspect ratios as filaments. However, we stress that these terms likely represent two limits on a continuous spectrum of initial gas morphologies and physical conditions rather than wholly distinct classes of objects. Below we discuss different processes that regulate the fragmentation of these structures and the predicted signatures of their formation.

\subsubsection{\textbf{Modes of fragmentation}}\label{sec:modes} 

Two proposed drivers of core fragmentation are rotation\index{Rotational Fragmentation} and turbulence\index{Turbulent Fragmentation}, which produce density and velocity asymmetries. Velocity gradients observed in early observations of dense cores \citep[e.g.,][]{GoodmanBenson1993a} highlighted the importance of angular momentum\index{Dense Cores!angular momentum}\index{Angular Momentum|(} in core evolution. Early numerical simulations including solid body rotation demonstrated that rapidly rotating, collapsing cores are prone to fragmentation, thereby leading to binary formation \citep{Larson1972}.  \citet{InutsukaMiyama1992a} quantified the fragmentation criterion for rotating cores in terms of the initial ratio of thermal to gravitational energy, $\alpha_{\rm vir}$, and ratio of rotational to gravitational energy, $\beta_{\rm rot}$, where $\alpha_{\rm vir} \beta_{\rm rot} > 0.12$ produces fragmentation and
\medmuskip=-1mu
\thinmuskip=-1mu
\thickmuskip=0mu
\begin{equation}
\scalebox{0.95}{$\alpha_{\rm vir}\beta_{\rm rot} = 0.02 \left( \frac{T}{10{\rm\ K}} \right) \left( \frac{M}{1\ \msun} \right)^{-2} \left( \frac{R}{0.1{\rm\  pc}} \right)^{4} \left( \frac{\Omega}{10^{-14}{\rm\ s}^{-1}} \right)^{2}$}
\end{equation}
Fragmentation also requires $\alpha < 0.5$, 
since thermal pressure may otherwise prevent warm cores from fragmenting \citep{TsuribeInutsuka1999}.
These criteria are straightforward, but they fail to encapsulate the complexity of real cores and thus are not readily extended to construct a general theory of binary formation.

While recent high-resolution core observations validate the prevalence of velocity gradients,  current interpretation of their physical meaning is nuanced. Gradients may instead signify core formation via converging flows \citep{ChenMundy2020} or represent the largest turbulent fluctuation in the core \citep{GoodwinWhitworth2004a}. The latter explanation suggests that turbulence, either directly or indirectly, rather than rotation produces multiple star formation (see \S\ref{sec:angmom} for more discussion).

Most current star-formation theories appeal to turbulence to trigger the growth of structure, including the formation of cores and filaments, within molecular clouds. 
Several groups have developed rigorous, semi-analytic frameworks for star formation through turbulent fragmentation, in which the statistical properties of turbulence, such as the power spectrum and density probability distribution function, determine core masses and regulate gravitational collapse \citep[see review by][]{LeeOffner2020}. Only one model, which we discuss here, makes quantitative predictions for stellar multiplicity.
\citet{GuszejnovHopkins2015} use a grid of perturbations to hierarchically compute the fragmentation of a molecular cloud assuming that turbulence both seeds collapse and provides pressure support\index{Turbulent Fragmentation}.
 This method preserves the spatial relationships of collapsing objects, such that density fluctuations within cores naturally lead to fragmentation and multiple formation \citep{GuszejnovHopkins2017}. 
Sub-regions collapse when the local gravitational energy exceeds the combined turbulent and thermal energy. This may be phrased as a critical density threshold:
\begin{equation}
\scalebox{1.05}{
$\rho_{\rm crit} = \frac{\rho_0}{1+\mathcal{M}^2} \left( \frac{R}{R_0} \right)^{-2} \left[ 1 + \mathcal{M}^2 \left( \frac{R}{R_0}\right)^{p-1}  \right],$}
\end{equation}
where $\rho_0$, $R_0$ and $\mathcal{M}$ are the initial density, radius, and Mach number of the parent cloud, respectively. This formalism assumes turbulent energy scales as $E(R) \propto R^{p}$ and the cloud is  isothermal. For supersonic turbulence $p=2$ while for subsonic turbulence $p=5/3$\index{Turbulence}, where the transition occurs at the sonic scale, $R_s$. For a virialized molecular cloud, where $R_s \simeq 0.65\,$pc and $R_f< 0.1$\,pc is the fragment radius, we can write \citep{GuszejnovHopkins2015}:
\medmuskip=-2mu
\thinmuskip=-2mu
\thickmuskip=0mu
\arraycolsep=0pt
\begin{eqnarray}
\scalebox{1.0}{$n_{\rm crit}$} &=&\scalebox{1.0}{$\frac{n_0}{1+\mathcal{M}^2} \left( \frac{R_f}{R_0} \right)^{-2} \left[ 1 + \left( \frac{R_f}{R_s}\right)^{2/3}  \right]$}\\
 & \scalebox{1.0}{$\gtrsim$}& \scalebox{1.0}{$4.2 \times 10^6~{\rm cm}^{-3}~ \left( \frac{n_{H,0}}{10^2 {\rm cm}^{-3}} \right) \left( \frac{R_0}{10{\rm pc}} \right)^2 \left(\frac{R_f}{10^3{\rm au}}\right)^{-2}\left( \frac{\mathcal{M}}{10}\right)^{-2},$} \nonumber 
\end{eqnarray} 
where the mass of a $10^3$\,au radius fragment is $M_f = 4/3 \pi \rho_{\rm crit} R_f^3 \simeq 0.07~\Msun$ for typical cloud parameters.   
\citet{GuszejnovHopkins2017} used this formalism to directly predict multiplicity properties finding relatively good agreement with observed statistics despite their simplistic treatment of turbulence (see \S\ref{sec:enviornment} for more discussion).

Filamentary structure is a natural by-product of turbulence. Dense filaments are prone to sub-fragmentation at regular internals like ``beads on a string," which provides a mechanism to form widely-spaced multiples if filaments are sufficiently dense and compact (see Fig.~\ref{schematic})\index{Filaments!fragmentation}. An isothermal filament collapses if its mass per unit length, $M_{\rm l}$,  exceeds a critical value.
A filament threaded laterally by a magnetic field\index{Magnetic Fields}, $B_{\parallel}$, will become unstable when $M_{\rm l}$ exceeds   \citep{InutsukaMiyama1997,Tomisaka2014}: 
\begin{eqnarray}
\scalebox{1.0}{$M_{\rm l,crit}$} &\simeq &\scalebox{1.0}{ $\frac{2 c_s^2}{G} (1+ \beta_{\rm mag}^{-1})$} \\
&\simeq& \scalebox{1.0}{ $16.4 \left( \frac{T}{10 {\rm K}}\right) + 
7.8 \left(\frac{n_{\rm H}}{10^4 {\rm cm}^{-3}}\right)^{-1} \left( \frac{B}{10 \mu  {\rm G}} \right)^{2} M_\odot {\rm pc}^{-1},$} \nonumber
\end{eqnarray}
where $\beta_{\rm mag} = 8 \pi c_s^2 \rho  / B_{\parallel}^2$ is the ratio of thermal to magnetic pressure.  Several authors have developed semi-analytic models deriving the spectrum of fragment masses (i.e., the core mass function\index{Core Mass Function}) from filament properties \citep{Inutsuka2001,AndreArzoumanian2019}. We can apply these arguments to derive expected relationships for multiple formation from filament fragmentation.

In order for density perturbations to grow, analytic arguments suggest that substructures within filaments must have separations of at least 2.5 times the filament full-width half maximum, $W_{\rm fwhm}$  \citep{FischeraMartin2012}. This suggests that fragments that are sufficiently closely spaced to remain bound must form in filaments with widths of ~$\lesssim 0.1$\,pc$ / 2.5 \sim 9\times10^3$\,au.  Assuming turbulent support is negligible, a filament fragment will collapse if its mass exceeds the critical Bonnor-Ebert mass \citep{AndreArzoumanian2019}:
\medmuskip=0mu
\thinmuskip=0mu
\begin{eqnarray}
\scalebox{1.0}{$M_{\rm BE, th}$} &\simeq&\scalebox{1.0}{$ 1.3 \frac{c_s^4}{G^2 \Sigma_{\rm fil}} $} \label{filmass}\\
&\simeq& \scalebox{1.0}{$1.1 \left( \frac{T}{10 {\rm K}} \right)^2 \left( \frac{M_{\rm l,crit}}{40 M_\odot {\rm pc}} \right)^{-1} \left( \frac{W_{\rm fwhm}}{10^4~{\rm au}} \right) M_\odot$} \nonumber,
\end{eqnarray}
\medmuskip=2mu
\thinmuskip=2mu
\thickmuskip=2mu
where the filament surface density, $\Sigma_{\rm fil}$, is the ratio of the filament mass-per-length to the width of the filament $\Sigma_{\rm fil} = M_{\rm line}/W_{\rm fwhm}$. 
 Equation \ref{filmass} suggests that the formation of closely-spaced fragments with small masses requires a high filament mass-per-length.
This implies that strong magnetic fields facilitate the formation of dense, narrow filaments and thus play a critical role in multiples formed via filament fragmentation\index{Filament!fragmentation}.

While these expressions are relatively simple and provide intuition for the origin of multiplicity in different theoretical frameworks, they make a variety of simplifying assumptions. Consequently, using these to accurately {\it predict} future multiplicity from the properties of a particular observed filament or core is not straight-forward. Moreover, debate in the star-formation community continues on the relative roles of rotation, turbulence and magnetic fields on gas structure formation and evolution.  In \S\ref{sec:enviornment}, we examine the impact of non-linear physical effects in star-cluster formation calculations and  discuss the impact of different processes on multiple formation across all scales.

\subsubsection{\textbf{Predicted signatures}}\label{sec:signatures}

Multiple systems formed from core and filament fragmentation have various characteristics imprinted during formation that shape their statistical distributions and may point to their origins long after the natal gas is dispersed. 

Gravitational fragmentation\index{Gravitational Fragmentation} of an isothermal, turbulent cloud is scale free and thus requires additional physics to set the minimum fragmentation scale \citep{GuszejnovGrudic2020}. Current theoretical models appeal to either angular momentum, magnetic support or tidal forces to set this minimum scale \citep{GuszejnovHopkins2017,HaugbollePadoan2018,LeeHennebelle2018}; all of these effectively impose a lower limit of $\sim 10^2$~au. 
Fragments that collapse with separations $\gtrsim 0.1$~pc are unlikely to be initially bound and will naturally drift apart.
These two constraints limit the initial separations of multiples formed from turbulent fragmentation to $\sim 10^2$~au - 0.1~pc.  This separation range is consistent with those of multiplies that form  in magneto-hydrodynamic simulations of turbulent clouds and cores \citep{OffnerDunham2016,LeeOffner2019pub,KuffmeierCalcutt2019}.  Arc- and bridge-like gas structures, produced by formation in a turbulent environment, may connect the pairs and facilitate accretion \citep{KuffmeierCalcutt2019}.  Wide embedded and accreting, multiples exist at such separations for a relatively short time period, migrating to $\lesssim 10^2$~au within $\sim 10^5$ years \citep[][see \S \ref{sec:capture}]{OffnerDunham2016,LeeOffner2019pub}. 

Turbulent fragmentation\index{Turbulent Fragmentation} is only efficient when there is a significant reservoir of gas, i.e., during the embedded phase. Consequently, multiples in such systems, assuming they form from the same parent structure, are expected to have similar ages ($\Delta t \lesssim t_{\rm ff} \sim 3 \times 10^{5} (n_{\rm H_2}/10^4 \rm{cm}^{-3})^{-0.5}$\,yr). Once collapse commences, gravity curtails the growth of perturbations, limiting the number of objects that can form by this process. Consequently, turbulent fragmentation may only produce two or three members within a core  \citep{OffnerDunham2016,GuszejnovHopkins2017,LeeOffner2019pub}. Since collapse events happen independently and occur while there is a significant gas reservoir, the formation of very low-mass companions ($\lesssim 0.08 M_\odot$) 
and systems with high mass ratios are expected to be rare \citep{Fisher2004}.

Due to their wide initial separations, multiples formed from turbulent fragmentation accrete gas with different net angular momentum. This frequently produces misaligned stellar spins\index{Stars!angular momentum}, accretion disks\index{Disks!angular momentum} and protostellar outflows\index{Protostellar Outflows}  \citep{Bate2018, OffnerDunham2016,LeeOffner2019pub}. Stellar spins and outflows of wide-separation protostellar pairs formed in magneto-hydrodynamic (MHD) simulations are consistent with a random distribution  \citep{OffnerDunham2016,LeeOffner2019pub}. The misalignment persists to the end of the calculation, suggesting that misaligned spins are a signature of this formation mechanism that may be observable after the birth cloud disperses (see \S\ref{sec:Orientations}). The relative orientations of protostellar outflows, which are observable before significant dynamical evolution occurs, provide a means to directly probe the initial angular momentum direction (see \S \ref{sec:Orientations}).  However, outflow angles can be difficult to measure for close pairs and in clustered regions; selection effects due to the difficulty of distinguishing aligned protostellar outflows may cause wide pairs to appear preferentially anti-aligned rather than randomly distributed \citep{OffnerDunham2016}. 


\subsubsection{\textbf{Rotation, turbulence, and angular momentum inheritance}}\label{sec:angmom}

Several authors have drawn a direct line between the angular momentum generated by turbulence and multiplicity properties. \citet{Fisher2004} developed a semi-empirical model relating randomly drawn turbulent velocity fields to binary angular momentum, which in turn set binary periods and separations. This approach, which depends on several input parameters such as the core-to-star formation efficiency  
correctly predicts that more equal-mass binaries have smaller binary periods and that lower-mass systems have smaller typical binary separations. \citet{JumperFisher2013} extended this model to explore the origin of the ``brown dwarf desert," in which solar mass stars have relatively few brown dwarf\index{Brown Dwarfs} companions at short periods (see \S\ref{sec:FGKstars}). They showed that low-mass brown dwarfs have smaller mean separations, where the small fraction of observed wide-separation brown dwarf systems are explained by turbulent stochasticity.

Groups carrying out hydrodynamic calculations of turbulent molecular clouds\index{Molecular Clouds} both with and without magnetic fields have shown that turbulence naturally produces velocity gradients with the magnitude and distribution seen in observations \citep[e.g.,][]{ChenMundy2020}. However, numerical studies of fragmentation in collapsing, magnetized cores show poor correlation between the initial degree of rotation and later fragmentation \citep{KuznetsovaHartmann2020}. \citet{KuznetsovaHartmann2020} instead propose that dynamical torques are more directly responsible for core fragmentation. Ultimately, multi-physics simulations highlight the difficulty of applying simple theoretical models to predict the complex and nonlinear outcomes of star formation. 

Non-ideal magnetic effects, which influence angular momentum transport on both core and filament scales, further confuse the relationship between initial gas properties and fragmentation \citep{ZhaoTomida2020}. The origin and distribution of angular momentum is in turn crucial for the formation and properties of protostellar disks\index{Disks!protostellar}, which may also host binary formation as we discuss in the next section.  In \S\ref{sec:enviornment}, we present a more detailed discussion of multiplicity derived from multi-physics calculations of forming clusters, which include all four mechanisms for multiple formation.

\subsection{\textbf{Disk Fragmentation}}\label{sec:diskfrag}

\subsubsection{\textbf{Summary of the instability}}
Independent of multiple formation on the scales of cores and filaments, protostellar disks, which form around the individual stars, may also be prone to fragmentation\index{Disks!fragmentation} into binary or higher order multiples\index{Disks!in binary systems}. The notion that gravitational instability in a shearing disk might produce binaries dates back over 30 years \citep{AdamsRuden1989,ShuTremaine1990,Bonnell1994}. The physical mechanism at the heart of the instability is the same derived for the development of spiral arms in galaxies by \cite{Toomre1964}. The criterion for the growth of the axisymmetric, $(m=0)$ mode of the instability in razor-thin disks, where gravity overcomes support from thermal pressure on small scales and rotation on large scales is:
\begin{equation}
    Q =\frac{c_s \Omega}{\pi G \Sigma} = f \frac{M_*}{M_d}\frac{H}{r} \leq 1,
\end{equation}
where $c_s$ is the disk sound speed, $\Omega$ the disk orbital frequency, and $\Sigma$ the surface density.  The second form of the above equation relates the instability to the star-disk-mass ratio, $M_*/M_d$ and aspect ratio, $H/r$, where $H$ is the disk scale height, and the order unity scale factor $f$ depends on the assumed surface density power law. The Toomre Q\index{Toomre Q} criterion formally describes the onset of an instability rather than successful fragmentation of the disk into an independent bound object.

Numerical models of protostellar disk fragmentation often reference a secondary criterion for instability to lead to fragmentation, the so-called ``cooling criterion." An early version of this mandate for disks to cool rapidly was first described by \cite{ThompsonStevenson1988} but was brought into much clearer focus by \cite{Gammie2001a}. The constraint that disks radiate away heat on roughly the orbital timescale ($t_{\rm cool} < \sim 3-7\Omega^{-1}$) is critical for understanding the transition from the so-called gravito-turbulent state to fragmentation in viscously heated disks. For protostellar disks, the cooling constraint is often more easily satisfied than $Q\leq1$ \citep{KratterLodato2016}.  Moreover for thick, massive disks, the critical $Q$ value for fragmentation drops substantially below unity \citep{LauBertin1978} due to the extended vertical distribution of matter. Consequently, the canonical combination of $Q\sim 1$ and $t_{\rm cool} \Omega \sim O(1)$ may be insufficient for predicting the onset of fragmentation in irradiated, massive disks \citep{TsukamotoTakahashi2015,TakahashiTsukamoto2016}. 

\subsubsection{\textbf{The onset of instability}}
As the literature is replete with discussion of the nature of gravitational instability\index{Disks!gravitational instability} in general, we focus in this review on the likely pathway towards instability that is most relevant for multiple formation. Gravitational instability ( via modes $m\geq1$) is most likely to arise in the outer disk, as $Q(r) \propto r^{p-(q+3)/2}$, where $q$ is the power-law index of temperature $T \propto r^{-q}$, and $p$ is the power-law index for surface density $\Sigma \propto r^{-p}$. Except for pathological combinations of these indices, $Q$ declines with radius. At disk radii of several 10s to 100s of au, the disk temperature is primarily set by stellar irradiation, and thus somewhat insensitive to changes in disk properties \citep{Clarke2009, KratterMatzner2008a,RiceArmitage2011}. Consequently,  disks most likely become unstable through an increasing surface density, $\Sigma$, that is not matched by an increased temperature. This situation arises when the infall rate onto the disk 
exceeds the accretion rate through the disk and onto the star \citep{KratterMatzner2010a,HarsonoAlexander2011,ZhuHartmann2012}. Such high infall rates are likely during the Class 0 phase\index{Class 0}.    This scenario is particularly conducive to the formation of more equal mass binaries (rather than extreme mass-ratio BDs\index{Brown Dwarfs} or super Jupiters\index{Planets!super Jupiters}) because the requirement for rapid infall also provides ample mass supply for the newborn fragment. In fact, for a wide range of disk conditions, the secondary object will out-compete the primary in accretion, driving mass ratios towards unity \citep{BateBonnell1997,OchiSugimoto2005a,YoungClarke2015}. This behavior is consistent with the preference for more equal mass ratios among close binaries (see \S\ref{sec:statisticsMS}). Note that even at arbitrarily high infall rates, the disk mass will not exceed the stellar mass, as global $m=1$ mode instabilities arise and lead to binary formation \citep{AdamsRuden1989}. Numerical simulations that invoke disk masses comparable to stellar masses as initial conditions are therefore somewhat unphysical.

An alternate pathway occurs when the disk temperature declines faster than accretion can alter the surface density. A likely candidate for this scenario is stellar luminosity variation due to accretion changes driven by gravitational or other instabilities\index{Disks!gravitational instabilities}. A sudden drop in the central luminosity could also precipitate fragmentation, so long as the drop in temperature does not correspond to a rise in the cooling timescale that prohibits fragmentation \citep{DunhamVorobyov2014,KuffmeierFrimann2018}.

\subsubsection{\textbf{Fragment survival and subsequent evolution}}
In order for gravitational instability and fragmentation\index{Disks!fragmentation} to produce binary formation, several benchmarks must be met. Fragments must cool efficiently so that they collapse to sizes well below their Hill radius\index{Hill Radius}, $r_H = a(M_{\rm frag}/(3M_*))^{1/3}$, before being sheared apart by interactions with the disk or even other fragments. Soon after formation, they are also subject to inward migration through the disk\index{Disks!migration}, which can  lead to  tidal disruption as the Hill radius shrinks faster than the object \citep{BegelmanCioffi1989a,Gammie2001a,KratterMurray-Clay2011a,Nayakshin2010,BoleyHayfield2010}. Merger with the host star is also possible \citep{VorobyovBasu2005, ZhuHartmann2012}. We defer a discussion of disk-driven migration in general to \S\ref{sec:gasmig}. Full modeling of these migration dependencies is crucial for predicting the expected population of binaries (separation, mass ratio) that derive from disk fragmentation plus migration. For example, their analytic model of disk fragmentation followed by inward migration and preferential mass accretion onto the secondary, \citet{TokovininMoe2020} reproduced the observed excess twin fraction and BD desert among close solar-type binaries.\index{Brown Dwarfs}

\subsection{\textbf{Capture, Dynamical Evolution, and Migration}}\label{sec:capture}
The fragmentation modes discussed thus far provide excellent explanations for a large swath of observed binary systems, even if they are not fully predictive analytic models. One glaring omission in these models is the frequency and efficiency of orbital evolution post fragmentation.\index{Binaries!dynamical evolution} This omission is particularly acute for the closest binaries ($a \lesssim 1$ au), where in-situ fragmentation is all but prohibited by the size of the first hydrostatic core \citep{Larson1969}. To understand the origin of the closest binaries, as well as early evolution of orbital properties, we review the current models for capture and orbital evolution.\index{Binaries!separation}

\subsubsection{\textbf{N-body versus gas-mediated capture}}
As described in \S\ref{sec:corefrag}, multiple systems can form at large distances, nearly independently as single stars, but nevertheless be gravitationally bound to each other. 
Additionally, multiple stars in a star cluster may begin their lives unbound, but due to energy and angular momentum exchange with the surrounding gas cloud or individual circumstellar disks, subsequently become bound and remain so. Other objects (often hierarchical multiples) will alternate between weakly bound and unbound over the first free fall time of the cluster. We dub this process, seen in a range of scales in numerical simulations {\it gas-mediated capture} \citep{Ostriker1994,MoeckelBally2007,Bate2012,MunozKratter2015,Geller2021,Cournoyer2021}.
The net frequency of capture via these mechanisms is likely low; these systems would be indistinguishable observationally from those formed via fragmentation\index{Disks!fragmentation}, except possibly cases where two disks appear to be on a collision course (see panel d, Fig.~\ref{schematic}).

On the other hand, in regions of high stellar density, interactions between truly unbound stars in the absence of gas can lead to the formation of new binaries and partner exchange\index{Binaries!dynamical evolution}. The outcome of such interactions is fairly well understood and can be modelled statistically. For example the adage that hard binaries are hardened by such interactions while soft binaries become softer \citep{Heggie1975} suggests that very frequent interactions should tend to drive binary separations towards a bimodal rather than unimodal distribution -- notably inconsistent with most observations (see Fig.~\ref{fig:Floga}). More recent work has emphasized the important contributions of higher order multiples\index{Multiple Systems!higher order} to the overall interaction rates in clusters of moderate density \citep{GellerLeigh2015,HamersSamsing2020}. In particular binary-single and binary-binary interactions can produce a much more diverse set of outcomes, including the formation of very compact binaries \citep{DorvalBoily2017}. While relevant for dense open clusters and globular clusters, these processes are sub-dominant in gas-rich environments. Moreover, even when rapid dynamical instabilities occur that change orbital parameters, system multiplicity, or hierarchy, they rarely lead to the order of magnitude changes in separation that are required to explain very close binaries  \citep[e.g.,][]{Wall2019}. Orbital hardening is often driven by the ejection of a formerly bound companion; because the ejected object is typically the lowest mass, the total energy and angular momentum\index{Angular Momentum|)} removed from the system in such an interaction causes more moderate changes in semi-major axis \citep{ValtonenKarttunen2006,Kratter2011}.\index{Binaries!dynamical evolution}

Partner exchange and shifting hierarchies also occur frequently in gas-rich environments \citep{Bate2012,RyuLeigh2017,Cournoyer2021}. Such interactions are fundamental to shaping the final multiplicity distribution observed for pre-MS and field stars and may even explain some of the evolution between the Class 0 and I phase (see \S\ref{sec:ProtoStatistics}).  Crucial as it may be, there is little fundamental theory to describe this process. Although some theories of multiple formation argue that most if not all objects are born in binaries \citep{Kroupa1995, Kroupa2008}, modern numerical simulations produce some stars that are never bound to another cluster member \citep{Bate2012,LeeOffner2019pub}.

\subsubsection{\textbf{Gas-driven migration}}\label{sec:gasmig}
In addition to allowing for capture, gas-rich star-forming environments also facilitate dramatic orbital evolution of binaries from their birth separation.\index{Binaries!separation} The recent expansion of statistics for the youngest stars lends credence to the notion that binaries form early and thus much of the orbital evolution occurs before the natal cloud disperses (see \S\ref{sec:ProtoStatistics}). Migration can occur either due to interactions with both bound or unbound low density molecular gas, or with a well defined circumbinary disk.\index{Disks!circumbinary}

{\it{Gas-dynamical friction}}:
For widely separated bound binaries, the combined action of gas accretion and the generation of a dense trailing wake provide a torque that reduces the semi-major axis. Numerous works have studied this so-called ``dynamical-friction force" \citep{BateBonnell1997,OstrikerGammie1999,Stahler2010,LeeStahler2011}.  \citet{LeeOffner2019pub} showed that a simple first order differential equation for $\dot{a}$ that includes both accretion and the torque due to dynamical friction with the gas $\dot{L}= \vec{r} \times (-\dot{m} v_*)$, can explain the observed rate of orbital decay in turbulent MHD simulations. The removal of angular momentum by magnetic breaking also helps to drive inward migration \citep{ZhaoLi2013}. Thus, so long as the gas cloud is present and stars continue to accrete, gas can efficiently reduce orbital separations from thousands of au to hundreds or even tens of au in well under a Myr.

Though a simple model can explain the observed migration in simulations, it does not easily translate into a generic prediction for the expected orbital evolution for a population of binaries, as it is a function of the local environment. Nevertheless, phenomenological models have had some success in reproducing observed distributions \citep{TokovininMoe2020}.

{\it{Disk-driven migration:}}
Migration of binaries embedded in a shared circumbinary disk\index{Disks!circumbinary} also crucially shapes the final orbital and mass ratio distributions. The classic theory for migration suggests that inward migration should dominate \citep{LubowArtymowicz1996a}, however this work only includes the torque from a circumbinary disk on point particles and neglects the often substantial torque generated by small individual disks surrounding each star, as well as the contribution from ongoing accretion. Accretion and circumstellar torques in particular push objects in the opposite direction and can lead to substantial outward migration \citep{MunozMiranda2019}. The net torque including circumbinary disks, circumstellar disks, and ongoing accretion is a function of binary mass ratio, eccentricity, effective viscosity, and aspect ratio \citep{SatsukaTsuribe2017, DempseyMunoz2021}. \index{Binaries!mass ratio} Ongoing work will map this space theoretically, which will inform future population synthesis models. However, it is clear that more extreme mass ratios experience inward migration for typical disk properties. This is somewhat consistent with disk fragmentation\index{Disks!fragmentation} simulations that undergo mergers, although numerical viscosity may also play a role \citep{VorobyovBasu2005b, ZhuHartmann2012}.  The production of near equal mass binaries at close separations may also derive from this process, as rapid inward migration at high mass ratios is followed by mass equalizing accretion from the circumbinary disk. This process can lead to stalled inward migration and even reverse the migration direction \citep{TokovininMoe2020}.

\subsubsection{\textbf{Secular evolution}}
\label{sec:secular}
 An alternative model for the origin of very close binaries ($P<7$ days) is Kozai-Lidov (KL) cycles coupled with tidal friction \citep{Kozai1962b,Lidov1962,FabryckyTremaine2007c}. This model posits that binaries with $P<10$ days form from triple star systems with an inclined outer tertiary.\index{Multiple Systems!higher order} Exchange of angular momentum between the orbits can drive long timescale eccentricity-inclination oscillations. When the pericenter of the inner binary is pushed to very close separations, tides begin to circularize the orbit, both shutting off the oscillations and inducing drastic period decay  \citep[e.g.,][]{Dabringhausen2022}. This model relies on the high fraction of triples amongst the closest binaries \citep[][see \S\ref{sec:statisticsMS}]{TokovininThomas2006}. However, such oscillations cannot account for the population of co-planar, close-in triples identified by \cite{BorkovitsHajdu2016}. A population synthesis study by \cite{MoeKratter2018} found that less than half of binaries with $P<10$ days could be plausibly explained by KL cycles. One of the key reasons KL cycles do not reproduce the observed population is that the pre-MS\index{Pre-Main Sequence} close binary fraction is consistent with that of the field (see \S\ref{sec:statisticsPMS}), suggesting that evolutionary mechanisms act on Myr not Gyr timescales: only a tiny fraction of KL susceptible systems have predicted orbital circularization timescales of less than a Gyr. The need for rapid migration  likely requires gas-driven migration to efficiently drain energy and angular momentum from binary and triple star orbits.\index{Angular Momentum|)}



\subsection{Impact of Environment on Multiplicity}\label{sec:enviornment}

\begin{deluxetable}{@{\extracolsep{4pt}}lccccccccccc}
\tablecaption{Star Cluster Calculation Statistics \label{table:StarCluster}}
\tablewidth{0pc}
\setlength{\tabcolsep}{4pt}
\tabletypesize{\scriptsize}
\tablehead
{
\colhead{}& \multicolumn{7}{c}{Properties\tablenotemark{a}} & \multicolumn{4}{c}{Statistics\tablenotemark{b}}
\\
\cline{2-8} \cline{9-12}
\colhead{Reference} & \colhead{B} & \colhead{R} & \colhead{O} & \colhead{C} & \colhead{$L$~(pc)} & \colhead{$M$~($M_\odot$)}
& \colhead{$\Delta x$~(au)}  & \colhead{MF} & \colhead{CF} & \colhead{ $\widetilde{a}$ (au)} & \colhead{$\widetilde {d}_{\rm 2D}$ (au)}  }
\startdata
\citet{Bate2019}    & $...$ & $\checkmark$ &  $...$ & S & 0.808 & 500 & 0.5 & $0.27\pm 0.03$ &  $0.50\pm 0.03$ & 13 &  $...$\\ 
\citet{Bate2019} $Z=0.01Z_\odot$ & $...$ & $\checkmark$ &  $...$ &S & 0.808 & 500 & 0.5 & $0.29\pm 0.05$ & $0.52\pm 0.04$  & 9 & $...$\\
\citet{Bate2019} $Z=0.1Z_\odot$ & $...$ & $\checkmark$ &  $...$ & S & 0.808 & 500 &  0.5 &  $0.24\pm 0.03$ &  $0.47\pm 0.04$  & 8 & $...$\\
\citet{Bate2019} $Z=3Z_\odot$ & $...$ & $\checkmark$ &  $...$ &S & 0.808 & 500 & 0.5 &  $0.22\pm 0.04$ &  $0.57\pm 0.03$ & 25 & $...$ \\
\citet{CunninghamKrumholz2018} $\mu=1.56$ \tablenotemark{c} & $\checkmark$ & $\checkmark$ &  $\checkmark$ & B & 0.65 & 185 & 65 & $0.10\pm 0.03$  &$0.18\pm 0.04$  & 37 & 986 \\
\citet{CunninghamKrumholz2018} $\mu=2.17$ \tablenotemark{c} & $\checkmark$ & $\checkmark$ &  $\checkmark$ & B& 0.65 & 185 & 65 & $0.25\pm 0.05$ & $0.53\pm 0.09$  & 126 & 564\\ 
\citet{CunninghamKrumholz2018} $\mu=23.1$ \tablenotemark{c} & $\checkmark$ & $\checkmark$ &  $\checkmark$ & B& 0.65 & 185 & 32 & $0.20\pm 0.06$ & $0.38\pm 0.10$  & 22 & 411\\
\citet{LeeOffner2019pub} $\mu=2$ & $\checkmark$ & $...$ &  $...$ & B& 2.0 & 601 & 50 & $0.14\pm 0.13$ & $0.43\pm 0.25$ & 449 & 245 \\
\citet{LeeOffner2019pub} $\mu=8$ & $\checkmark$ & $...$ &  $...$ & B& 2.0 & 601 & 50 & $0.38\pm 0.17$ & $1.00\pm 0.35$  & 1146 & 1003\\
\citet{LeeOffner2019pub} $\mu=32$ & $\checkmark$ & $...$ &  $...$ & B& 2.0 & 601 & 50 & $0.08\pm 0.04$  & $0.16\pm 0.06$  &1316 & 295\\
\citet{LiKlein2018} $\mu=1.9$  & $\checkmark$ & $\checkmark$ &  $\checkmark$ & B& 4.2 & 3110 &  28 & $0.40\pm 0.07$  & $0.74\pm 0.13$  & 87 & 578\\ 
\citet{MathewFederrath2021} $\mu=6.6$ & $\checkmark$ & $...$ & $\checkmark$ & B& 2.0 & 775 & 100 & $0.34\pm 0.11$ & $0.70\pm 0.25$ & 291 & 560\\ 
\enddata
\vspace{-0.25in}
\tablecomments{ \index{Binaries!statistics}$^a$Physics included in the simulation: magnetic fields (B), radiative transfer (R), and outflows (O). Column C indicates a spherical (S) or box (B) cloud geometry; $L$, $M$, and $\Delta x$ indicate the cloud diameter, mass, and minimum gas spatial resolution, respectively. The metallicity, $Z$, is noted when it is not solar ($Z_\odot$). Magnetized runs note the initial ratio of the gas mass to critical mass that can be supported by magnetic fields against collapse (``mass-to-flux ratio"), $\mu$. $\mu=M_{\rm gas}/M_{\rm crit} = 2 \pi \sqrt{G}M_{\rm gas}/B_{\rm z}L^2$, where $B_z$ is the initial uniform magnetic field and $L$ is domain length. $^b$ All entries except those of \citet{MathewFederrath2021} assume binomial statistics and Poisson statistics to estimate the MF and CF uncertainties, respectively.  Uncertainties for the \citet{MathewFederrath2021} values adopt the standard deviation of the 10 runs. \citet{Bate2019} uses closest separation to rank pairs and define stellar systems, while the other results use lowest binding energy to define bound systems. This choice has little impact on MF and CF, which differ by less than $1\sigma$. The last two columns list the median semi-major axis, $\widetilde{a}$, and median 2D pair separation computed by assigning system multiplicity following \citet{TobinOffner2021}. $^c$ Statistics represent the final stellar distribution from a run with driven turbulence and one with identical initial conditions in which turbulence naturally decays.}
\vspace{-0.25in}
\end{deluxetable}

In past reviews, the limited number of star-cluster formation simulations and the small statistics of formed systems therein have obscured trends in stellar multiplicity with birth environment.   Over the past few years computational advances have provided new insights into how molecular cloud properties affect stellar multiplicity by modelling more massive clouds, following larger dynamic ranges, and including more physical processes. While the statistics in such calculations remain modest, especially for very low ($<0.1 \msun$) and high mass ($> 10 \msun$) stars, it is now clear that multiplicity exhibits more variation than the stellar IMF, which is surprisingly insensitive to environment \citep{OffnerClark2014}. In this section, we discuss the relationship between environment, physical processes, and multiplicity inferred from current state-of-the art star-cluster calculations. We present a summary of simulation properties and multiplicity statistics in Table \ref{table:StarCluster}.

\subsubsection{\textbf{Radiation}}

Radiative heating from protostars, which predominately influences gas within a few $10^2$~au of the source in the case of low-mass stars, can significantly reduce disk fragmentation (\S\ref{sec:diskfrag}), thereby reducing the formation of brown dwarfs \citep{OffnerKlein2009,Bate2012}.\index{Radiation} \citet{Bate2019} modeled a $500\Msun$ unmagnetized, collapsing cloud with radiative transfer and a model for the diffuse ISM. The MF, separation distributions, and mass ratios at the final time show good agreement with the multiplicity of MS stars (e.g, see Fig.~\ref{FigSeg}), despite the simulated stars being protostellar, i.e., significantly younger. Surprisingly, the same initial conditions evolved with a barotropic equation of state \citep{Bate2012} produced statistically similar multiplicity with only a slightly higher MF for stars with $M>0.2\Msun$. The relatively small number statistics in the radiative transfer calculation, combined with observational uncertainties, prohibit more rigorous conclusions about the impact of radiation on multiplicity. Note that the \citet{Bate2019} calculations do not explicitly include heating due to stellar accretion and nuclear processes, as the magnetized calculations from \citet{CunninghamKrumholz2018}  and \citet{LiKlein2018} (Table \ref{table:StarCluster}), and therefore represent a lower limit on the impact of radiation on multiplicity. 

Models and simulations focusing on multiple formation via turbulent fragmentation\index{Turbulent Fragmentation}, rather than disk fragmentation, find that radiative transfer has little impact on multiplicity. Core and filament fragmentation produce widely separated multiples, which form beyond the sphere of heating, such that calculations with and without protostellar heating give similar multiplicity statistics. However, even without disks, the statistics of very low-mass objects are over-predicted when heating is neglected, since turbulent fragmentation can proceed to smaller scales \citep{GuszejnovHopkins2017}. In simulations with disks, magnetic fields and thermal support due to protostellar heating both act to enhance disk stability and produce little small scale fragmentation \citep{OffnerDunham2016,CunninghamKrumholz2018,GuszejnovGrudic2021}. 

\subsubsection{\textbf{Protostellar outflows}}

Protostellar outflows launched by accreting protostars entrain and expel dense core gas, both reducing the star formation efficiency of dense gas and injecting significant energy and momentum into the natal cloud environment \citep{Bally2016}.\index{Protostellar Outflows} Numerical simulations suggest that protostellar outflows can explain the offset between the stellar IMF and dense core mass function, helping to set the characteristic stellar mass \citep{OffnerChaban2017,GuszejnovGrudic2021,MathewFederrath2021}. \citet{MathewFederrath2021} conducted a suite of MHD turbulent cloud simulations both with and without protostellar outflows. Runs including outflows reproduced the expected trend of increasing multiplicity with stellar mass but under-predicted the multiplicity of low-mass ($M\lesssim 0.03 \msun$) multiples. This deficit is likely due to the minimum calculation spatial resolution of $\sim 10^2$~au, which limits the formation of close separation systems (although disk fragmentation does occur). Comparing the results of magnetized star cluster calculations with and without outflows in Table \ref{table:StarCluster} suggests outflows have little direct impact on CF and MF.

\subsubsection{\textbf{Magnetic fields}}

Magnetic fields provide additional pressure support that reduces gravitational collapse, thereby lowering the star formation rate, while increasing the characteristic stellar mass \citep{PadoanFederrath2014,GuszejnovGrudic2021}.\index{Magnetic Fields} While magnetic fields might therefore naively be expected to decrease stellar multiplicity, they may instead enhance multiplicity. \citet{OffnerDunham2016} carried out a suite of magnetized turbulent collapse calculations and found that because magnetic fields  introduce asymmetry, which inhibits spherical collapse, they enhance turbulent fragmentation.  \citet{CunninghamKrumholz2018} and \citet{LeeOffner2019pub} followed magnetized turbulent clouds with different mass-to-flux ratios
and showed that stronger field calculations produce comparable or  {\it higher} MFs and CFs (see Table \ref{table:StarCluster}). 
Likewise, the MF and CF reported by  \citet{LiKlein2018} and \citet{MathewFederrath2021} are higher than those of \citet{Bate2019}, which do not include magnetic fields.
While strong magnetic fields reduce protostar formation overall, they also produce fewer initial high-order systems. Consequently, fewer dynamical interaction occur, which would otherwise eject members and increase the fraction of single stars. This suggests a more subtle relationship between magnetic fields, stellar density and multiplicity.

Magnetic fields treated in the ideal limit of perfect dust-gas coupling transport angular momentum
efficiently, which can lead to compact ($< 50$~au) accretion disks or suppress disk formation altogether \citep{ZhaoTomida2020}. Spatial resolution of $\lesssim$ a few au is required for disk formation in turbulent, ideal MHD calculations. Since 
no magnetized cluster calculations listed in Table \ref{table:StarCluster} adequately resolve disks, multiples in these calculations formed primarily through turbulent fragmentation. 

Close binaries are under-represented in the resulting separation distributions\index{Binaries!separation}, which hints that disk fragmentation, in addition to turbulent fragmentation and dynamical interactions, is required to produce the observed population of close-separation field star binaries.
However, Figure \ref{FigSeg} shows the resulting separation distributions of stars formed in some MHD calculations are consistent with the observed distribution of wide-separation protostars from \citet{TobinOffner2021}. This comparison is apt because the ages of the simulated stars in these calculations are less than 1 Myr and, like protostars, they are less dynamically evolved than the population of field stars.
Note that both simulations and observations of protostars are incomplete for separations below $\sim 25$ au (see \S\ref{sec:ProtoStatistics}).  Meanwhile the separation distribution from \citet{Bate2019} is similar to the observed field population of solar-type stars  and inconsistent with the protostellar distribution despite simulated stellar ages $\lesssim 0.2$ Myr. 

Non-ideal magnetic processes are important in regulating angular momentum transport and setting disk properties \citep[][see also \S\ref{sec:diskfrag}]{ZhaoTomida2020}, however, most calculations of star-cluster formation have assumed ideal MHD. In one exception, \citet{WursterBate2019}  modeled a suite of 50\msun~ magnetized turbulent clouds, including ambipolar diffusion, Ohmic dissipation, and the Hall effect. 
They find that non-ideal magnetic processes do not inhibit the formation of multiple systems, which form independent of the initial field strength. Nevertheless, small number statistics ($N_{\rm sys}< 20$) prohibit stronger conclusions about the impact of non-ideal effects on multiplicity.

\subsubsection{\textbf{Turbulence}}

Turbulent properties, such as the Mach number, virial parameter, and type of forcing (e.g., compressive versus solenoidal), influence the gas distribution and star formation rate \citep{PadoanFederrath2014,Federrath2015}. 
While the stellar IMF\index{IMF} appears to be relatively insensitive to these 
details 
\citep{OffnerClark2014,GuszejnovGrudic2021}, the impact of turbulence on stellar multiplicity is less clear, since few simulations have explored the impact of varying turbulence properties on multiplicity. \citet{CunninghamKrumholz2018} compared multiplicity in turbulent, magnetized clouds and found those with continuous turbulent driving and weaker fields ($\mu = 1.56, 2.17$) over-predicted the multiplicity of solar-type stars, albeit the statistics were small.  \citet{Mignon-Risse2021} found that the collapse of massive cores\index{Dense Cores!collapse} with sub-Alfv\'enic turbulence ($\mathcal{M}_A equiv v / v_A < 1$), i.e., relatively stronger fields, produced single high-mass systems, while cores with super-Alfv\'enic turbulence produced binary formation via disk fragmentation, which was driven by higher levels of angular momentum. 

\begin{figure*}[h]
\begin{center}
 \epsscale{2.2}
 \plottwo{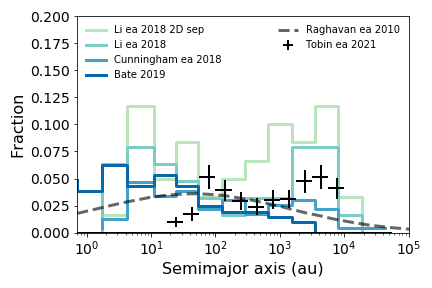}{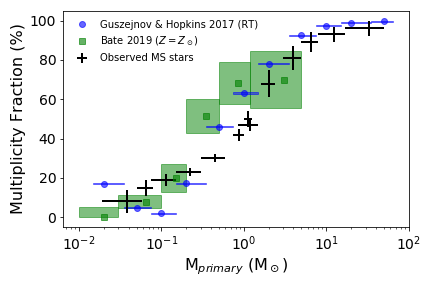}
  \vspace{-0.2in}
 \end{center}
 \caption{\small  \textbf{Left:} Separation distribution for observed protostars (crosses),  
  fit to solar-type field stars (dashed line), 
  and semi-major axis distributions for simulated protostars.
  The number in each bin is normalized by the total number of singles and  separations in each dataset. The ``Li ea 2018 2D sep" curve shows the projected 2D separations of systems identified using the method of \citet[][see Table \ref{table:StarCluster}]{TobinOffner2021}. 
  \textbf{Right:} Multiplicity fraction versus primary system mass. Shading indicates bin ranges and binomial uncertainties. Black points show the bias-corrected MF for field stars (see Table 1).}
 \label{FigSeg}
\end{figure*}

A separate class of semi-analytical models focuses on the fundamental interaction between turbulence\index{Turbulence} and gravity; this approach is simpler and computationally cheaper than performing MHD simulations but cannot capture time dependence or the complex spatial relationship between turbulence and density. 
Despite a variety of approximations, the \citet{GuszejnovHopkins2017} turbulence model (see \S\ref{sec:modes})
successfully predicts increasing multiplicity with stellar mass and the relative frequency of mass-ratios\index{Binaries!mass ratios} for both solar-mass and low-mass stars.  Fig.~\ref{FigSeg} shows that the predicted MF 
agrees reasonably well with that of field stars, although the model over-predicts the MF of solar-type stars. This suggests that supersonic turbulent properties, such as the lognormal density distribution and power spectrum, together with gravity can explain a variety of observed multiplicity properties. However, \citet{GuszejnovHopkins2017}  predict that the initial separation distribution peaks at $\sim 600$~au and has a deficit of close, $d \lesssim 10$~au, binaries. This underscores that fully matching observations requires treatment of disk fragmentation and/or dynamical evolution (see \S\ref{sec:diskfrag})\index{Binaries!dynamical evolution}.



\subsubsection{\textbf{Metallicity}}
\label{sec:MetalsTheory}

It has long been understood that the evolution of metallicity from the first Population III stars to today could substantially impact the IMF \citep{AbelBryan2002}. Gas metallicity influences the efficiency of cooling. Reducing the metallicity of the low density, optically thin gas characteristic of star forming cores inhibits cooling via metal lines. On the contrary, for solar metallicity gas that is optically thick, as in protostellar disks, lowering metallicity can promote cooling by reducing the opacity.
\citet{Bate2019} carried out four calculations of forming star clusters with metallicity, $Z$, ranging from 0.01 to 3 times solar. The calculations included separate treatment for the gas and dust temperatures, but neglected radiative feedback and magnetic fields. While the shape of the stellar IMFs was largely independent of metallicity, the lower metallicity calculations experienced more fragmentation in cores, filaments, and disks due to the higher cooling rate of dense gas. This produced more close binaries ($a\lesssim10$au), which varied from CF=$0.1 \pm 0.03$ for $3Z_\odot$ to $0.53\pm 0.12$  for $0.01Z_\odot$. The close binaries in the lowest metallicity calculation also exhibited a small preference for lower mass ratios, which is likely caused by the enhanced small scale fragmentation. However, Table \ref{table:StarCluster} shows that the overall MF and CF values for the four metallicities fall within $1\sigma$ of one another.

As noted in \S \ref{sec:MultTrends}, observations support a drastic decrease in the close binary fraction as a function of metallicity, with the binary fraction beyond $\sim 200$ au remaining essentially unchanged \citep{MoeKratter2019,El-BadryRix2019}. Analytic calculations suggest that the increase in only the close binary fraction  is consistent with enhanced disk fragmentation, with wide binary formation mechanisms remaining essentially unchanged \citep{MoeKratter2019}. \index{Binaries!fraction}


\bigskip
\noindent

\section{\textbf{Observations of Forming Multiple Systems}}\label{sec:ObsFormMult}

Observations of multiplicity in the youngest stellar systems are key to understanding their origins.  
Unlike the pre-MS and MS systems reviewed in \S\ref{sec:statisticsMS}, they have had little time to dynamically change, and therefore their properties more directly reflect the formation conditions. 

We define the youngest star systems, hereafter protostars, as stars that are still deeply embedded within a dense envelope of dust and gas that obscures them from direct observation (i.e., Class 0\index{Class 0|(} and I\index{Class I|(} sources). Due to high optical depths from the surrounding material, protostellar multiplicity is best observed via high resolution interferometric observations at (sub)millimeter to radio wavelengths where the surrounding envelope is both optically thin and filtered out such that the circumstellar environment can be detected.  At these wavelengths, young stars are primarily detected via thermal dust emission from circumstellar disks , thermal free-free emission from ionizing shocks at the base of jets, or from a combination of these processes \citep[e.g.,][]{Segura-CoxLooney2018,TychoniecTobin2018}. 


In the \ppvi~ review, discussion of protostellar multiplicity centered on infrared observations of Class I systems, since there were few observations of multiplicity in Class 0 systems \citep{ReipurthClarke2014}.  These studies tended to have inhomogenous separation distributions, owing to different region distances, and many were limited to separations $> 100$ au, which does not probe the field star separation peak (see Fig.~\ref{FigSeg}).  Nevertheless, observations provided tantalizing evidence that the multiplicity fraction of protostars is higher than that of pre-MS and MS stars \citep{LooneyMundy2000, ChenArce2013} suggesting that most systems form as multiples.   

In this section, we review recent impactful case studies and large surveys using (sub)millimeter continuum observations from interferometers that probe complete populations of both Class 0 and Class I sources in nearby clouds. 
We describe new emerging trends and
discuss their implications for multiple star formation and evolution.


\subsection{Starless Cores}\label{sec:StarlessCores}

Multiplicity can begin prior to the onset of star formation when starless cores\index{Starless Cores|(} and filaments fragment\index{Dense Cores!fragmentation|(} to form multiple density peaks (see \S \ref{sec:corefrag}). We define {\it starless} cores as those with no detected protostar and independent of whether they are gravitationally bound.
Observations of the internal structure of collapsing starless cores are required to understand how and when this mode of multiple formation occurs.  Detecting multiple density peaks or extended internal structure within a given starless core does not guarantee multiple formation, however, since the objects must also be self-gravitating and undergoing collapse.  Current literature tends to define any compact emission that is detected within a starless core using interferometry as {\it substructure}. The presence of two or more substructures suggests that the core is undergoing the first phase of collapse with possible incipient multiple formation, which is our main focus here.
%

At the time of \ppvi, surveys of starless cores had found few cases of substructure 
\citep{SchneeEnoch2010,SchneeDi-Francesco2012, LeeLooney2013}. The profusion of non-detections placed upper limits on substructure masses of $\sim 0.1 \Msun$.  Deeper studies of a few objects successfully detected substructure: L 183 \citep{KirkCrutcher2009}, R CrA SMM1 A \citep{ChenArce2010b}, Oph A SM1 (which has since been identified as protostellar, see \S \ref{sec:Multigeneration}), and Oph B2-N5 \citep{NakamuraTakakuwa2012}.  These deeper studies estimated substructure masses of $\sim 0.001 - 0.01 \Msun$ on sub-1000 au scales, in agreement with the earlier non-detections.  

ALMA\index{ALMA|(} has enabled large surveys of starless cores on sub-1000 au scales and down to the mass sensitivities of $\sim 0.001 \Msun$ that are necessary to detect substructure \citep{DunhamOffner2016, KirkDunham2017}.
\citet{DunhamOffner2016} found no substructure in 56 starless cores in Chameleon, whereas \cite{KirkDunham2017}  detected substructure in only two out of 37 starless cores in Ophiuchus. The substructures detected by \citet{KirkDunham2017}  include multiple density peaks in Oph A SM1N \citep[see Fig.~\ref{schematic} and][]{FriesenPon2018} and a single substructure in L1689N, a starless core to the east of IRAS 16293-2422 region \citep{LisWootten2016}.  Oph B2-N5 was not detected by \cite{KirkDunham2017}, however, likely due to a combination of lower sensitivities at long wavelengths and loss of emission from spatial filtering compared to \citet{NakamuraTakakuwa2012}. In a targeted study, \citet{CaselliPineda2019} detected multiple density peaks in the starless core L1544. \index{Turbulent Fragmentation}

The relatively low substructure detection rate in starless cores implies that the cores have not reached sufficiently high central densities (possibly because collapse is not occurring) or that the timescale over which the substructure is detectable before protostar formation is short. \citet{KirkDunham2017} reasoned that if the  typical starless core lifetime is comparable to the free-fall time then two detections in 37 starless cores implies a substructure timescale on order of 5 kyr assuming a central density of $10^5$ cm$^{-3}$.   \citet{DunhamOffner2016} used synthetic observations of MHD simulations of turbulent cores and Bonnor-Ebert sphere models to constrain the likelihood of detection. They found that substructure is only detectable at their survey resolution toward the end of the starless stage (after $>0.1$ Myr) when the central density exceeds $\sim 10^7$ cm$^{-3}$, i.e., when the free-fall time itself is less than 10 kyr.     

Assuming typical core properties and that cores fragment on their free-fall timescale, \citet{DunhamOffner2016} expected roughly two cores in the Chameleon sample to show substructure.  Most of these cores, however, are gravitationally unbound, suggesting they are less evolved and 
it is possible some of them will not collapse to form stars at all.   By contrast, the cores in Ophiuchus are denser, likely more evolved and therefore are more likely to contain substructure \citep{KirkDunham2017}.  Given the short timescales during which high densities are present prior to protostar formation and the large fraction of unbound cores in star-forming regions, detecting substructure in starless cores is challenging even with ALMA.  Future studies will require large samples of cores or targeted observations 
with a balance of high resolution and moderate scales. 
\citet{OffnerCapodilupo2012} found that synthetic observations of core fragmentation simulations that combined the main ALMA array with the ACA were key to detecting substructure in starless cores \citep[see also,][]{DunhamOffner2016}.

It is worth noting that of the substructure detections in starless cores, most (4/6) show evidence of multiple density peaks in dust continuum, e.g., as shown in Fig.~\ref{schematic}. 
This occurrence rate is consistent with that of turbulent fragmentation models, which predict that a significant fraction of cores and filaments that do collapse to form stars undergo fragmentation. This sample suggests that when substructure is detected there is a high probability that it shows evidence of multiple formation. 

\begin{figure}[h!]
\begin{center}
\includegraphics[width=0.45\textwidth]{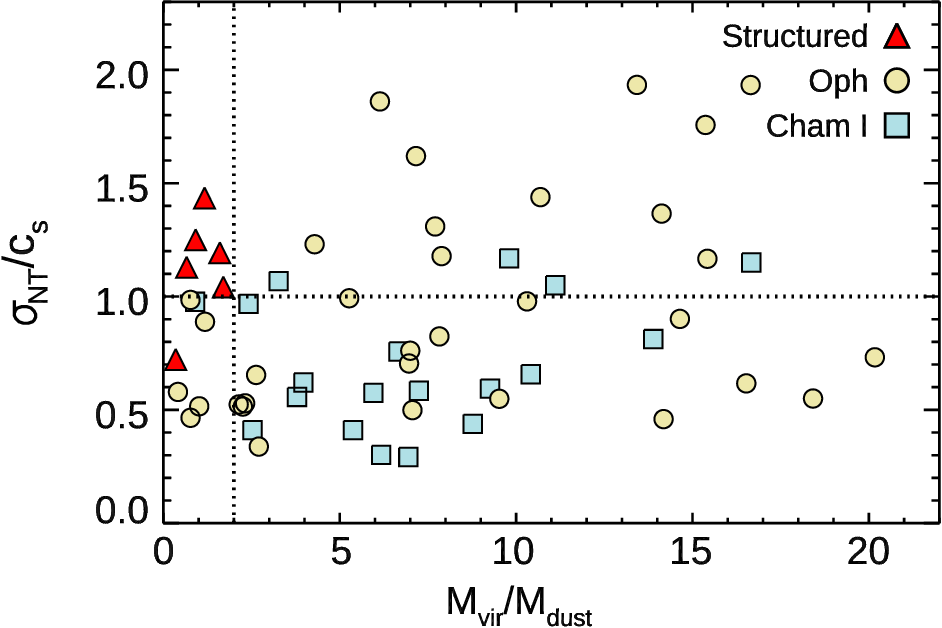}
\end{center}
\vspace{-0.15in}
 \caption{\small Starless core properties for L1688 in Ophiuchus (circles) and Chameleon (squares) with known structured starless cores as triangles.  Data are primarily from single-dish telescopes and are calculated using different tracers, sensitivities, and techniques.  We assume the core gas is isothermal to calculate the mass (measured from dust continuum data) and sound speed. The L1688 core properties are from \citet{KerrKirk2019} and use NH$_3$ observations.  The Chameleon core properties use the N$_2$H$^+$ observations  from \citet{TsitaliBelloche2015}, with the temperature scaled to 12 K to match the dust masses from \citet{BellocheSchuller2011}. The structured cores properties use data from the following references: 
 L183  \citep[][]{LattanziBizzocchi2020, KarolySoam2020}, 
 R CrA SMM1 A  \citep[][]{HarjuHaikala1993, BresnahanWard-Thompson2018},
 Oph A SM1N  \citep[][]{PattleWard-Thompson2015, KerrKirk2019}, 
 Oph B2-N5 \citep[][]{MotteAndre1998, FriesenDi-Francesco2009}, 
 L1689N  \citep[][]{PattleWard-Thompson2015, PanLi2017}, 
 and L1544 \citep[][]{Chacon-TanarroPineda2019, KoumpiaEvans2020}. 
 }
 \label{fig:starless_core_dynamics}
\end{figure}

The amount of core turbulence may be an important signpost for targeted studies searching for fragmentation. However, there has been no rigorous study comparing the turbulent properties of structured and unstructured cores.
Figure \ref{fig:starless_core_dynamics} compares the non-thermal velocity dispersion, $\sigma_{\rm NT}$, to the virial parameter ($\alpha = M_{\rm vir}/M$) 
of starless objects that have detected substructure (black triangles) with the larger starless core populations in Ophiuchus and Chameleon (grey symbols).  To first order, a core has supersonic turbulence if $\sigma_{\rm NT}/c_{\rm s} > 1$ and subsonic turbulence if $\sigma_{\rm NT}/c_{\rm s} < 1$; it is bound if $\alpha = M_{\rm vir}/M < 2$ and unbound if $\alpha = M_{\rm vir}/M > 2$ \citep{McKee1999}.  We calculate a virial mass assuming $M_{\rm vir} = 5\sigma_{\rm v}R/G$, where $\sigma_{\rm v}$ is the total velocity dispersion, $R$ is the size of the core, and we calculate $c_s$ assuming $\mu = 2.37$ and the gas is isothermal.  We note that all values are taken from the literature where the turbulent properties, temperature, and mass estimates are all calculated differently and thus a thorough consistent study of starless core properties is still needed. 

Figure \ref{fig:starless_core_dynamics} shows that the general starless core\index{Starless Cores|)} populations in Ophiuchus and Chameleon exhibit wide ranges of velocity dispersion and degree of gravitational boundedness, whereas the few cores that have detected internal structures are all bound and most (5/6) have supersonic turbulence.  The only subsonic core is L1544, which shows evidence of infall \citep{KetoCaselli2015}, indicating that it is collapsing.  While cores without detections may still have structure below the observed sensitivity limits, the structured cores have density peaks that are more substantial and more likely to result in multiple star formation \citep[e.g.,][]{DunhamOffner2016}.   This distinction suggests that bound, turbulent, and collapsing cores are the best candidates for substructure searches.  \index{Dense Cores!fragmentation|)}

\subsection{Protostellar Multiplicity Statistics}\label{sec:ProtoStatistics}\index{Binaries!statistics}\index{Multiple Systems!embedded|(}

Prior to \ppvi\, the lack of large, unbiased protostellar samples led
to overly broad conclusions based on small samples
\citep{LooneyMundy2000,MauryAndre2010}.
One of the first attempts to create a large sample was by \citet{ChenArce2013}, with an 
archival study of submillimeter observations toward protostars. 
This greatly increased the number of characterized systems, but the archival data had non-uniform sensitivity and resolution. 

The advent of ALMA and the upgraded VLA\index{VLA} have now enabled truly uniform and unbiased surveys of
multiplicity in nearby star-forming regions, where global statistics can be determined with high accuracy. Thus far, the protostellar populations of Perseus \citep{TobinLooney2016a},
Ophiuchus \citep{EncaladaLooney2021}, and Orion \citep{TobinSheehan2020,TobinOffner2021}
have been observed at uniform sensitivities
and resolutions. 
This section reviews the statistical results of these censuses.

\subsubsection{\textbf{Multiplicity and companion frequencies}}
\label{sec:ProtostarMult}

The MF and CF are key metrics for comparison to more evolved populations, but 
for a given sample they depend on separation range and protostellar Class, 
as well as any sample bias.
For instance, the archival study of \citet{ChenArce2013} found MF=0.64$\pm$0.08 and CF=0.91$\pm$0.05
for protostars with a separation range between 50 to 5000~au. However, these values likely missed
companions since the typical spatial resolution was 100s of au rather than 50 au. 

Recent surveys of Perseus and Orion have well-defined angular resolutions
and sensitivities \citep{TobinOffner2021}.  
From 20 to 10000~au, the MFs and CFs for the full protostellar samples (Class 0, Class I and Flat Spectrum) are 0.30$\pm$0.03 and 0.44$\pm$0.03, respectively, in Orion and 
0.36$\pm$0.06 and 0.52$\pm$0.06, respectively, in Perseus. Both results are consistent within their
uncertainties and both surveys have comparable minimum separations set by 
the angular resolution of the observations. 
For Class 0 protostars with separations between 20 to 10000~au 
the MFs and CFs are 0.38$\pm$0.05 and 0.62$\pm$0.05 for Orion and 
 0.47$\pm$0.09 and 0.74$\pm$0.08 for Perseus, which are consistent despite environmental differences. These results are lower overall
 with respect to \citet{ChenArce2013}, despite the fact that \citet{TobinOffner2021} considers a wider
 range of separations.
While the VANDAM surveys discovered a number of new companions with separations $<$200~au and wider companions at $>$5,000~au, these were offset by
the detection of many additional single systems, thereby reducing the total MF and CF. Previous archival studies
were limited to data biased toward the brightest millimeter
sources, which also tend to have higher luminosities; \citet{TobinOffner2021} found that 
multiple protostar systems, regardless of separation, had systematically
higher luminosities than single systems.

Multiplicity statistics change between protostellar classes within the same
20 to 10$^4$~au range of separations
\citep{TobinLooney2016a,TobinOffner2021}. 
In Orion, the MFs and CFs of Class I protostars are 
0.23$\pm$0.04 and 0.32 $\pm$ 0.05, respectively, while for Perseus they
are 0.27$\pm$0.09 0.35$\pm$0.09, respectively. Thus, Class I protostars
have lower multiplicity than Class 0 protostars for the same separation range.
The relative MFs and CFs are $\sim$40\% and $\sim$50\% lower for both regions, respectively,
having $>$ 3$\sigma$ differences.

The MF and CF for Flat Spectrum protostars in Orion are 
consistent with the Class
I statistics. Flat Spectrum protostars were not classified separately for Perseus but are combined with the Class I 
sample.
These differences 
indicate multiplicity evolution between Classes 
(see \S \ref{sec:densities}).

Other comprehensive studies of multiplicity have been carried out in Orion
at infrared wavelengths, but these studies were limited to $<$1000~au separations due to contamination on larger scales. \citet{KounkelMegeath2016}
measured a CF of 0.14$^{+0.011}_{-0.013}$, which includes mostly Class I and Flat Spectrum sources, demonstrating the large variation
in MF and CF for different separation ranges and evolutionary stages (see \S\ref{sec:obsprotosep}).
In this range of separations for a comparable sample of Class I
and Flat Spectrum protostars, Orion 
has a MF and CF of 0.10$\pm$0.02 and 0.11$\pm$0.02 \citep{TobinOffner2021}, respectively. 
These are slightly lower but still 
consistent with the statistics from \citet{KounkelMegeath2016}, which is 
likely because the centimeter/millimeter 
observations are more incomplete toward Class I/Flat Spectrum protostars than
near-infrared observations \citep{TobinOffner2021}.

Finally, \citet{EncaladaLooney2021} characterized the Class I and Flat Spectrum protostellar multiplicity in
Ophiuchus for
the separation range between 15~au and 1600~au. They found a combined Class I and Flat Spectrum MF and CF of 0.25$\pm$0.09 and CF=0.33$\pm$0.10, respectively, which is consistent with those of Orion and Perseus.

\subsubsection{\textbf{Separation distributions}}\label{sec:obsprotosep}

High spatial resolution, uniform surveys now measure companion separation\index{Binaries!separation}
distributions down to $\lesssim$50~au, similar to studies done for MS
stars (see \S\ref{sec:ObsStellarMultiplicity}). 
The large sample sizes enable studies as a function of
protostellar class, which have distributions that appear distinct from those of MS stars (Figures \ref{fig:Floga} \& \ref{FigSeg}). 

\citet{TobinLooney2016a} detected an apparent bimodal separation distribution
for the full sample of Perseus protostars with peaks at $\sim$75~au and 
$\sim$3000~au. The positions of the two peaks indicate both disk and turbulent fragmentation operate. However, the wide-separation feature is mainly produced by the 
Class 0 sources, and a
reanalysis of the Perseus data does not rule-out a log-flat distribution.
Orion also exhibits a bimodal separation distribution with similar peak locations
\citep{TobinOffner2021}.
The larger samples for Orion alone and the combined Orion and Perseus samples
strongly disfavor a log-flat distribution for both Class 0 and Class I protostars.

\citep{EncaladaLooney2021} measured the separation distribution from 15 to 1600~au for Class I and Flat spectrum protostars in Ophiuchus. Despite small numbers, this distribution
is quite similar to that of Perseus and Orion and has
an average separation of 183~au. 
The similarity of 
separation distributions in different star-forming environments suggests
multiple systems form via similar physical processes.


\subsection{\textbf{Markers of Binary Formation}}

\subsubsection{\textbf{Multi-generational systems}}\label{sec:Multigeneration}

MS stars in multiple systems are generally expected to have formed together and have similar ages.  Nevertheless,
theoretical models predict age differences up to a few Myr between stars in multiple systems. While a few Myr age difference is difficult to identify in MS multiples, it can potentially be observed in younger systems.  In this section, we review forming multiple systems that appear to have distinct star formation epochs and discuss the degree of age difference.  We use the term {\it non-coeval} to describe apparent protostellar age differences.  

\citet{Murillovan-Dishoeck2016} examined the spectral energy distributions (SEDs) of 16 protostellar multiples in Perseus with data from micron to millimeter wavelengths and found that roughly a third of the wider ($\gtrsim 2000$ au) multiple systems contained members with different Classes. Assuming Class is an accurate proxy for age, the implied age differences are consistent with those predicted ($\Delta t \lesssim 3 \times 10^5$~Myr) by numerical simulations \citep[e.g.,][]{LeeOffner2019pub}.

Recent ALMA observations have identified starless substructures near protostars that may point to core fragmentation after star formation is well underway. Two
instances are Oph A SM1 and Oph A N6 \citep{FriesenPon2018}.  Oph A N6 exhibits two peaks, one of which appears protostellar \citep{FriesenPon2018}.
Recent ALMA observations show that one of the peaks in Oph A SM1 is also likely protostellar \citep{FriesenDi-Francesco2014}. Oph SM1 appears connected to the structured starless core SM1N to the north (see Fig.~\ref{schematic}) and a chain of other protostars to the south \citep{FriesenDi-Francesco2014, KirkDunham2017}, suggesting that the entire region may be undergoing filamentary fragmentation with an age gradient from north to south.  

Another example is B5, a bound elongated core that contains two starless filaments that extend away from a central Class I protostar \citep{PinedaGoodman2011}.  
 \citet{PinedaOffner2015} found three dense condensations within these filaments as shown in Fig.~\ref{schematic}, which are incipient sites of star formation that could ultimately produce a non-coeval quadruple system.   \citet{SchmiedekePineda2021} showed that the mass per unit length ($M/L$)
exceeds the expected critical value required for stability (see \ref{sec:corefrag}), in agreement with the presence of the embedded condensations.  Thus, core and filament gas surrounding young protostars may still fragment, producing age differences comparable to the duration of the protostellar lifetime, i.e., $\sim 0.5$ Myr.



\subsubsection{\textbf{Protostellar disk stability}}\label{sec:DiskStability}

Observations of massive circumbinary and circum-multiple disks\index{Disks!circumbinary|(} provide evidence for the occurrence of 
disk fragmentation.  A number of studies have found extended structures around already formed, tight multiple systems \citep[e.g.,][]{TobinLooney2016a, TobinKratter2016, LimYeung2016, BrinchJorgensen2016a, TakakuwaSaigo2017, KrausKluska2017, HarrisCox2018, Arturdela-VillarmoisKristensen2018, AlvesCaselli2019,  LeeLi2020, TobinSheehan2020}, some of which exhibit Keplerian rotation  signifying that they are circumbinary disks \citep[e.g.,][]{Murillovan-Dishoeck2016, Arturdela-VillarmoisKristensen2019, ReynoldsTobin2021}.  Direct confirmation of disk fragmentation, however, is difficult since protostars formed by turbulent fragmentation combined with orbital evolution can drive wide companions to closer separations (see \S\ref{sec:capture}).  The Toomre Q parameter\index{Toomre Instability} provides insight into the actual degree of disk stability  (\S \ref{sec:diskfrag}). In this section, we adopt a critical Toomre Q of $Q_{crit} = 1$; while this threshold for fragmentation is not universal, the variation is far smaller than the observational error bars \citep{KratterLodato2016}.

The number of disk stability measurements using the Toomre Q parameter are increasing, although 
all calculations of Toomre Q should be considered first-order estimates due to uncertain temperatures, dust opacities, and dust-to-gas conversion factors that are necessary to obtain surface densities.  Additional diagnostics,  such as evidence of disk structure, are therefore necessary to verify instability. The circum-multiple disk of L1448 IRS3B in Perseus (Fig.~\ref{schematic}) is a prime candidate for disk fragmentation. The disk geometry and rotation is centered on a close binary system, but a third object is detected in the outer spiral arm ($< 300$ au from the binary).  This tertiary feature also appears protostellar, with a compact source that is driving an outflow \citep{ReynoldsTobin2021}.  At the radius of the tertiary source the disk has $Q < 1$ \citep{TobinKratter2016, ReynoldsTobin2021}, supporting the theory that this protostar companion was produced by disk fragmentation. 
Similarly, the Class I protostar HH111 MMS has spiral structure, $Q < 1$ \citep{LeeLi2020} and a wide companion.  Although no fragments are detected toward this disk yet, the similarities to L1448 IRS3B suggest it is a prime candidate for fragmentation. 

 Nevertheless, most protostellar disks with measured Toomre Q values appear to be stable\index{Disk Fragmentation}.
\citet{TobinSheehan2020} measured the Toomre Q parameter for 259 disks in Orion that are in either single star systems or in wide binaries, finding that only six were consistent with being gravitationally unstable \citep[see also,][]{SharmaTobin2020}.  Moreover,
some structured circumbinary disks  have large Toomre Q values indicating that they are stable. \citet{ReynoldsTobin2021} found hints of spiral structure toward the disk of L1448 IRS3A, a more distant companion to L1448 IRS3B \citep{TobinKratter2016}, with $Q \approx 5$. Similarly, \citet{AlvesCaselli2019} and \citet{Diaz-RodriguezAnglada2022} each found prominent spiral structure in a Class I circumbinary disk\index{Disks!circumbinary|)} and $Q \gtrsim 4$.  These cases show that spiral structure alone does not indicate gravitational instability.  Alternatively, spiral structure may be caused by dynamical processes such as interactions between the embedded protostars or it may be a remnant of a previously more massive disk from which the stars formed. 
We note that protostellar disks can accrete material from the surrounding envelope.  Thus, this simple picture of gravitational instability is more complicated, since the disk mass and surface density change over time.

The lack of marginally-unstable disks suggests disk instability is either short-lived or rare. In the former case, disks fragment quickly such leaving behind a compact multiple system.  In the latter case, if few disks ever become gravitationally unstable then most protostellar companions at separations $< 300$ au must be produced by migration. Further observations tracking fragmentation and disk stability in different cloud environments across multiple evolutionary stages are necessary to distinguish the fractions of close multiples produced in situ within disks versus those produced by migration. 

\subsubsection{\textbf{Protostellar disk properties}}\label{sec:DiskProperties}

\begin{figure*}[h]
 \epsscale{2.2}
 \begin{center}
 \plottwo{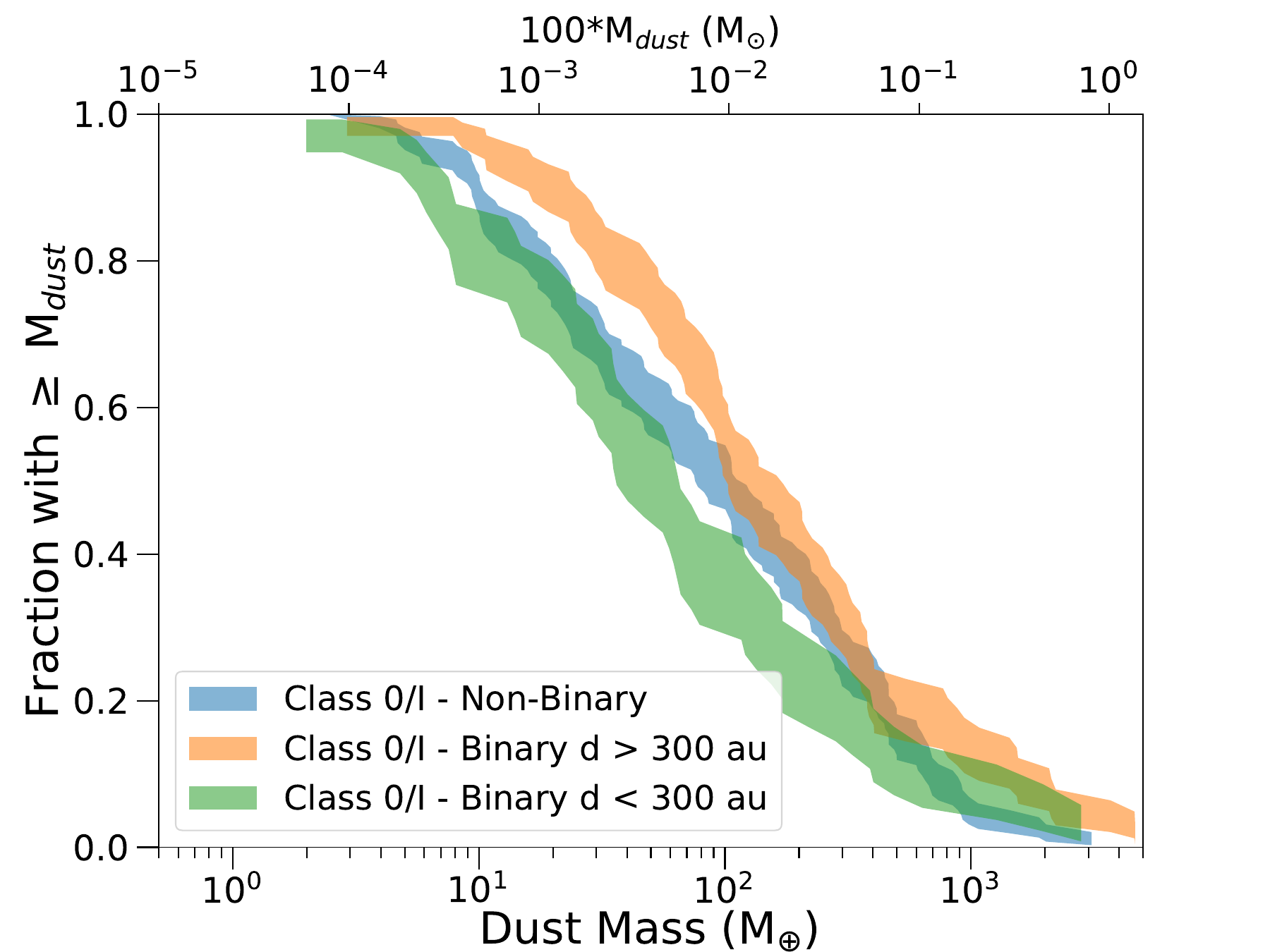}{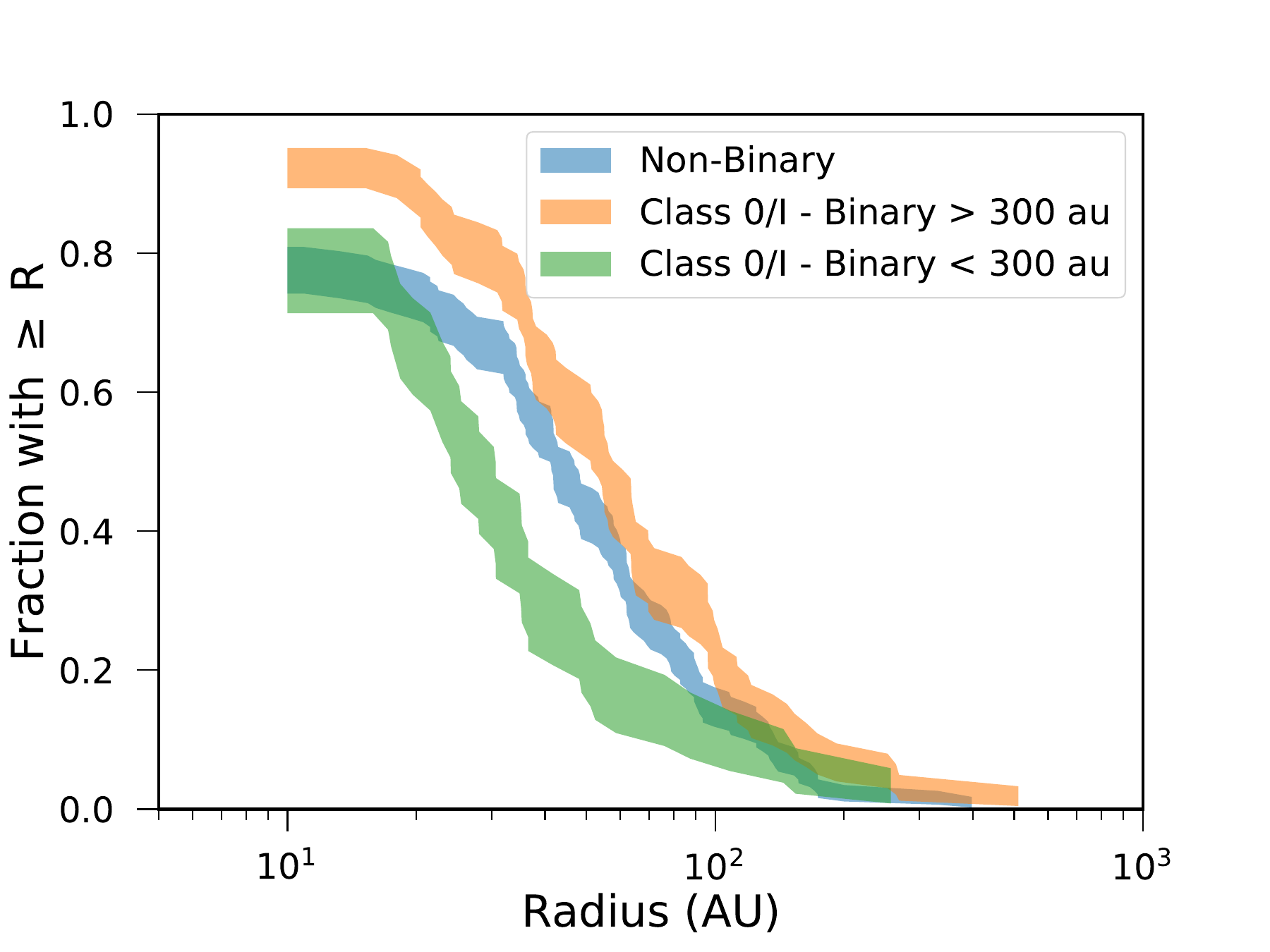}
 \end{center}
 \vspace{-0.16in}
 \caption{\small Comparison of circumstellar disk mass ({\bf left}) and radius ({\bf right}) for single and multiple Class 0 and I protostellar systems in Orion
 \citet{TobinSheehan2020}. 
 Masses are calculated assuming optically thin dust emission at 870 $\mu$m and an assumed dust temperature of 20 K and disk radii are derived from the 2$\sigma$ extent of Gaussian fits to dust continuum images. 
The division at 300~au
 contrasts the effect of intermediate and wide separation companions on disk masses and extents.
 Surprisingly, systems with wide companions tend to have larger and more massive disks than those without companions. The maximum fraction is $<$1.0 primarily due to unresolved disks. }
 \label{fig:Orion_disk_comparison}
\end{figure*}

In this section, we compare protostellar disk properties in systems with multiple and single stars. 
For simplicity, we exclude systems that have confused disks that are blended with circumbinary structures (see \S\ref{sec:circumbinary}) or very close companions.  Fig.~\ref{fig:Orion_disk_comparison} shows cumulative mass and size distributions for protostellar disks in Orion, which is the largest available sample 
\citep{TobinSheehan2020}.  
The disks are separated into three populations: single-star systems, close and intermediate multiple
systems, and wide multiple systems. 

Disks in close and intermediate multiple systems are statistically smaller in mass\index{Disks!masses} and size\index{Disks!radii} than those in wide multiples with p$<$0.01 from a Kolmogorov-Smirnoff test, whereas disks in single and wide-multiple systems tend to be consistent in both mass and radius. These trends indicate that the proximity of companions in close and intermediate multiple systems affects young disk properties.  Wider companions, however, have little to no effect, and disks in those systems appear similar to those of single stars (see \S \ref{sec:MaterialforPlanets}).  


\textbf{
}




\subsubsection{\textbf{Protostellar system orientations}}\label{sec:Orientations}\index{Disks!in binary systems|(}

Stars that form via turbulent fragmentation should have uncorrelated angular momenta, since these protostars form in distinct, well-separated core substructures, whereas companions produced by disk fragmentation, form in a single plane and should have preferentially aligned angular momentum vectors \citep{OffnerDunham2016,Bate2018}. Early observations of field star rotation appeared to support this picture \citep{Hale1994a}.  However,  a rigorous statistical reanalysis of the \citet{Hale1994a} data concludes that these data are insufficient to make any claims about spin-alignment \citep{JustesenAlbrecht2020}. A handful of individual measurements of eclipsing binaries reveal heterogeneous orientations, but with a preference towards alignment \citep{SybilskiPawlaszek2018}.
To complicate matters, a myriad of mechanisms can torque multiple star systems into or out of alignment, such as turbulent accretion, magnetized accretion, ejections, inclination oscillations, and damping from the disk  \citep{LubowOgilvie2000, ValtonenKarttunen2006, BatyginAdams2013, FieldingMcKee2015, LeeHull2017, JenningsChiang2021}. 

Observations of forming systems provide a stronger test of the natal spin orientations of multiple systems before dynamical processes affect them.  A variety of recent surveys have explored the angular momentum of young multiple systems by using the outflow or disk orientations as a proxy for angular momentum direction. We note that even in these studies, where an angular momentum direction can be robustly measured and assigned to each protostar, such proxies only reflect the {\it current} angular moment direction of gas on $\sim 1-100$ au radii from the stars and do not directly probe the (proto)stellar spin or system orbit.

Using observations of outflows\index{Protostellar Outflows}, \citet{LeeDunham2016} studied the orientations of nine multiple systems in the Perseus cloud with separations $>2000$ au using data from the MASSES survey \citep{StephensBourke2019}. They found that the relative outflow orientations in these systems were consistent with either random or preferentially perpendicular alignments. For companions with separations less than a few hundred au, however, individual outflows can overlap in projection \citep[e.g.,][]{StephensDunham2017}.  Blended outflows are difficult to disentangle and can bias the angular  differences toward misalignment \citep{OffnerDunham2016}. 

To date, there are no systematic studies of protostellar disk alignments. 
Many protostellar binaries with separations less than a few hundred au have disks with similar inclinations and position angles \citep[see][]{TobinLooney2016a, TobinSheehan2020}.  This similarity in disk geometries implies aligned angular momentum vectors, in agreement with the simple picture of disk fragmentation. Similarly, several wider multiples have misaligned disks in agreement with turbulent core fragmentation producing random angular momentum vectors, i.e., NGC 1333 IRAS2 \citep[][]{TobinDunham2015}, IRAS 04191+1523 \citep[][]{LeeLee2017},  BHR 71 \citep[][]{TobinBourke2019}, GSS 30 \citep{SadavoyStephens2019}, HOPS 182 \citep{TobinSheehan2020}, and HH111 MMS \citep{LeeLi2020}.  

Nevertheless,  some protobinary systems with separations $\lesssim 100$ au seem misaligned.  Oph-IRS~43 and Oph-IRS 67 are close binaries that appear misaligned with respect to circumbinary disks, whereas the circumstellar disks for the tight binary in L1551 IRS 5 are inclined moderately ($\sim 30$\arcdeg) relative to the circumbinary disk and the circumstellar disks in a wider binary, L1551NE, are inclined relative to each other \citep{BrinchJorgensen2016a, TakakuwaSaigo2017, Arturdela-VillarmoisKristensen2018,Cruz-SaenzdeMieraKospal2019}.  Recent high resolution ALMA observations of the outflow of VLA 1623A\index{VLA 1623A} show that its tight  binary ($< 30$ au) has misaligned outflows and disks \citep{HaraKawabe2021}.   Similarly, a disk resolved in the triple system IRAS 16293-2422 exhibits a $\sim 90\arcdeg$ misalignment with the surrounding circumbinary disk \citep{MaureiraPindea2020}. Such misaligned systems may have formed at larger separations via turbulent fragmentation and then migrated inward (see \S \ref{sec:TheoreticalModels}) or the system may have been perturbed from an originally aligned orientation by dynamical interactions. \index{Disks!in binary systems|)} 

\subsection{\textbf{Environment Around Forming Binaries}}

\subsubsection{\textbf{Core and envelope properties}}


Proto-multiple systems are embedded in dense cores and envelopes, which likely provide clues about multiple formation. However, to date there has been few systematic studies of the surrounding environments of such systems.  \citet{SadavoyStahler2017} compared the separations of multiple protostellar systems in the Perseus cloud with the properties of their host cores.   They found that the projected distance vectors of wide binaries with separations $\gtrsim 500$ au are preferentially aligned with the long axes of their host cores, whereas tighter systems are more randomly orientated.  This correlation suggests core fragmentation proceeds within a preferred plane, possibly due to rotation, magnetic fields, or both (see \S \ref{sec:TheoreticalModels}), whereas disk fragmentation or dynamics randomizes the orbit orientation relative to the host core.  \citet{LuoLiu2022} examined the binarity and core properties of 43 systems in the Orion cloud. They found that the cores hosting multiple systems tended to have higher column and volume densities and higher Mach numbers than those hosting single stars.  Higher densities and Mach numbers promote thermal Jeans fragmentation and turbulent fragmentation, respectively. This result suggests that multiple stars form from cores that are both turbulent and dense, in agreement with the structured detections for starless cores (see Fig.~\ref{fig:starless_core_dynamics}). 

Magnetic fields can greatly influence the formation of multiple systems, particularly if the field is aligned with the system angular momentum vector (see \S \ref{sec:TheoreticalModels}).  However, magnetic field observations exist for only a few protobinary system envelopes  ($\lesssim 1000$ au). L1448 IRS 2 is a Class 0 protobinary system with an inferred envelope magnetic field resembles an hourglass and is aligned with the system rotation axis  \citep{KwonStephens2019}.   
Models indicate this field geometry enhances magnetic breaking and might produce more compact disks. Nevertheless, L1448 IRS 2 has a circumbinary disk \citep{TobinLooney2018}.  Similar polarization morphologies are seen toward the envelopes or circumbinary material of the protobinaries Per-emb-2 \citep{CoxHarris2018}, BHB07-11 \citep{AlvesGirart2018} VLA 1623A \citep{SadavoyMyers2018}, IRAS 16293A \citep{SadavoyMyers2018a}, and BHR 71 \citep{MyersStephens2020}, although it is unclear whether the polarization uniquely traces pinched fields versus outflow cavities.\index{ALMA|)} 

Some protobinaries also have complex circum-multiple material. Theoretical studies predict that bridges of dust and gas can connect binary stars, thereby affecting their formation and accretion properties (\S \ref{sec:signatures}).  At present, only the IRAS 16293-2422 protobinary system shows such a filament of dust and gas connecting young low-mass stars  \citep[separated by $\sim 700$ au][]{Jorgensenvan-der-Wiel2016}.  This dust bridge has relatively uniform polarization indicating a structured magnetic field \citep{SadavoyMyers2018a}, allowing gas to flow freely along the bridge.
MHD simulations by \citet{KuffmeierCalcutt2019} show that such features occur when binaries form at wider separations ($\gtrsim 1000$au) and migrate inward. If bridge features are transient, large high-resolution surveys are needed to constrain their properties and lifetimes. Binary interactions may also enhance the degree of bursty accretion and affect gas chemistry \citep{JorgensenKuruwita2022}.

\subsubsection{\textbf{Stellar densities}}\label{sec:densities}

The local density of stars 
influences the evolution of multiple systems, such that wider members may be dynamically stripped from systems located in dense cluster environments.
This process is illustrated by the deficit of Class II and III companions with separations greater than 250~au 
in the ONC, where the stellar surface density is 770~pc$^{-2}$ \citep{ReipurthGuimaraes2007}. The companion frequency is higher for field stars and pre-MS stars in Taurus, which have stellar densities down to 35~pc$^{-2}$ (see right panel of Fig.\ref{fig:Floga}).

\citet{KounkelMegeath2016} completed a large 
\textit{Hubble Space Telescope} and NASA Infrared Telescope Facility Survey of Class I and Flat Spectrum multiplicity toward protostars in the \textit{Herschel} Orion Protostar Survey 
\citep[HOPS,][]{FurlanFischer2016}.
They estimated the YSO density around each protostar by combining the \textit{HST} 
data with complementary \textit{Spitzer} and X-ray data.
Adopting 45~pc$^{-2}$ as the boundary between high and low YSO surface density, they find the CF for the 100-1000~au separation range is larger by $\sim$50\% (0.05 to 0.11) in higher surface density regions.

Building on these results, \citet{TobinOffner2021} used the ALMA and VLA surveys of Orion to examine the influence of YSO surface density on protostellar multiplicity. 
They find higher MF and CFs in regions with YSO surface densities above 30~pc$^{-2}$, confirming
the results of \citet{KounkelMegeath2016}.
They also demonstrated a relationship between stellar density,
separation range, and protostellar Class.
For a separation range of 20 to 10$^4$~au,
all protostellar Classes have higher MFs and CFs in higher YSO density regions. 

This trend holds for Class I\index{Class I|)} and Flat Spectrum sources with smaller separation ranges of 20 to 10$^3$~au and 100 to 10$^3$~au. 
Class 0 protostars, however, have MFs and CFs that are statistically consistent in low and high YSO surface densities over all separation ranges.

These results indicate that YSO surface density impacts multiplicity on scales $<$~1000~au, but that impact is only apparent after the Class 0 phase.  \citet{TobinOffner2021} suggested that the Class 0\index{Class 0|)} results reflect the primordial multiplicity, which does not have a strong dependence on YSO density.
In contrast, the elevated MFs for more-evolved protostars are likely produced by the migration of companions from
$>$1000~au to $<$1000~au (see \S \ref{sec:capture}). The higher CFs from 20 to 10$^4$~au for all Classes at higher 
YSO density
naturally provide a reservoir of companions that can migrate to smaller separations.
This is consistent with the turbulent fragmentation model, which predicts companions
form at wide separations and then migrate inward (\S\ref{sec:corefrag}).

Thus, higher YSO densities may 
{\it enhance} rather than reduce
multiplicity. These high YSO densities are likely produced by higher initial gas densities, which may favor fragmentation \citep{GutermuthPipher2011}. We note, however, that the YSO surface densities in the Orion molecular clouds
are lower than the stellar densities in the ONC, which exceed 100~pc$^{-2}$. Above some threshold, high YSO density likely hinders multiple star formation and the ability of such systems to retain wide companions. 

\index{Multiple Systems!embedded|)}

\section{\textbf{IMPACT OF MULTIPLICITY ON PLANET FORMATION AND SYSTEM ARCHITECTURES}}
\label{sec:PlanetFormation}

The high frequency of multiple systems has the potential to have a strong impact on planet formation, both in terms of the overall frequency of planetary systems and on the architecture of those systems.  In this section, we review the connections between multiplicity and planet formation, but we do not attempt a comprehensive review of all observations of protoplanetary disks in multiple systems. Instead, we focus on some specific areas where the observations place particular constraints on planetary system properties, either current or future. After discussing disks in binaries, we summarize the final population statistics for planets in binary systems among field stars, again focusing on aspects that shed light on formation models.

\subsection{\textbf{Raw Material for Planet Formation: Disk Sizes and Masses in Multiple Systems}}\label{sec:MaterialforPlanets}
\index{Disks!protoplanetary}
\index{Disks!in binary systems|(}

\subsubsection{\textbf{Disk masses and millimeter fluxes}}
\label{sec:diskmasses}\index{Disks!masses|(}

It has long been known that stellar companions with separations on the order of disk radii ($\sim$\,10 to 100 au) affect disk structure, reducing disk masses compared to disks around single stars \citep{BeckwithSargent1990a, OsterlohBeckwith1995, JensenMathieu1996, HarrisAndrews2012a}.  Indeed, surveys that start from a sample of young stars without pre-selecting them based on whether or not they host disks find that the overall disk frequency is lower in close binaries than in wider binaries or single stars \citep[e.g.,][]{CiezaPadgett2009, CheethamKraus2015}.

Since \ppvi, ALMA\index{ALMA} observations have added nuance to such studies due to the combination of ALMA's high spatial resolution and sensitivity, particularly focusing on stars that still retain disks, i.e., the Class II phase, and measuring the properties of those disks that still survive at this age.  ALMA has resolved many more binaries at millimeter wavelengths, allowing comparative studies of disk radii, and determining the flux distribution in circumprimary and  circumsecondary disks. Increased sensitivity has allowed millimeter-wavelength detection of many more disks around secondary stars in particular, enabling comparisons of disk properties around primary and secondary stars within binary systems.  Such studies also extend to much lower stellar masses for both binary and single stars, thus including disks that are arguably more typical of the overall population of young stars.

\begin{figure*}[th]
\begin{center}
 \includegraphics[width=0.95\textwidth]{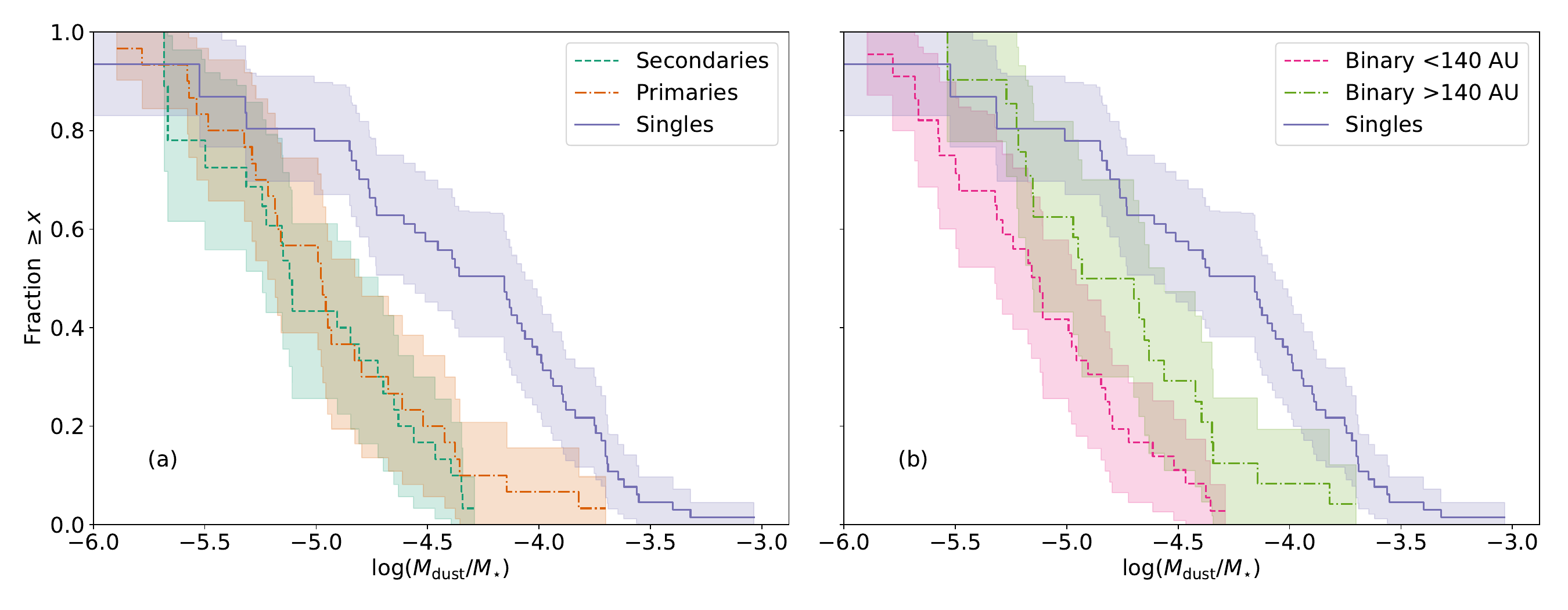}
 \end{center}
 \vspace{-0.26in}
 \caption{{\small Disk mass (assuming optically thin emission) for Taurus Class II binaries and single stars normalized by stellar mass.  {\bf Left:} After normalizing by stellar mass, secondary and primary stars in binaries have similar disk mass distributions, but both are lower than the distribution of single star disk masses. {\bf Right:} Smaller-separation binaries have lower-mass disks on average (relative to their stellar masses) than wider binaries, which in turn are lower than that of single stars. (Figure adapted from \citealt{AkesonJensen2019}.)}}
 \label{fig:disk-mass-binaries}
 \vspace{-0.1in}
\end{figure*}

\citet{AkesonJensen2019} combined new ALMA observations with literature data to assemble a sample of Class II\index{Class II} binaries in Taurus\index{Clusters!Taurus} wider than 30 au and with spectral types of M6 and earlier and compared these to a sample of single stars. To isolate the effects of binarity as cleanly as possible, the single-star sample was required to have high spatial resolution in the optical or IR, and the binary sample was restricted to double systems only, removing the complicating effects of hierarchical systems.  As has been seen previously, binaries with a companion closer than 140 au show reduced mm flux, while the mm flux distribution in wider binaries is indistinguishable from that of single stars.   However, some interesting new results emerge when the individual stellar masses are taken into account. 
It is well established that disk mass is correlated with stellar mass \citep[e.g.,][]{AndrewsRosenfeld2013, PascucciTesti2016}. 
To account for this, \citet{AkesonJensen2019} compared the distributions of $M_{\rm disk}/M_*$ in binaries and single stars and found that both primaries and secondaries, in both close {\it and\/} wide binaries, have reduced disk masses (on average) compared to single stars when corrected for stellar mass (Fig.\  \ref{fig:disk-mass-binaries}a). While the difference between wide binaries and single stars is less than that for close binaries vs.\ singles (Fig.\ \ref{fig:disk-mass-binaries}b), this result suggests that even the disks in wider binaries are influenced by membership in a binary system. In addition, after correcting for stellar mass, primaries do not dominate the disk mass in a given binary system; the ratio $M_{disk}/M_*$ has about the same distribution for primaries and secondaries.  We note here that this and other similar studies assume that the mm continuum emission is largely optically thin, so that the emission traces the total dust mass in the disks. 

\citet{ZurloCieza2020, ZurloCieza2021} studied disks in binaries in $\rho$ Oph\index{Clusters!rho Oph} and Lupus\index{Clusters!Lupus}, combining new adaptive optics observations to search for companions in their sample with previous ALMA observations of disks from the ODISEA survey \citep{CiezaRuiz-Rodriguez2019, WilliamsCieza2019}.  They found that the most massive disks are present only around single stars, and they argue that for more typical disks (those with $M_{\rm dust} < 50 M_\oplus$) the dust mass distributions are similar for single stars and multiple systems.

While the studies discussed above focus on Class II systems in 1--3 Myr old star-forming regions, \citet{BarenfeldCarpenter2019} find somewhat different results for Class II systems with ages of $\sim$5--11 Myr in Upper Sco\index{Clusters!Upper Sco}.  Unsurprisingly the 0.88 mm flux density distributions of all groups (close binaries, wide binaries, and single stars) in Upper Sco are shifted to lower values compared to the younger Taurus\index{Clusters!Taurus} association.  In contrast to Taurus, Upper Sco exhibits no significant difference in the mm flux distributions of these three groups; in particular, close binaries have the same flux distribution as single stars.  The stellar mass distributions of the three groups are similar as well, so this appears to be a real difference in the disk properties and not an artifact of different stellar masses.
\index{Disks!masses|)}

\subsubsection{\textbf{Disk radii}}\label{section:disk-radii} 
\index{Disks!radii|(}

One notable change in the ALMA era, compared to work with previous millimeter facilities, is the ability to measure (or at least constrain much better) disk radii, not only for the largest and brightest disks, but for more typical disks across a range of stellar masses.  \citet{CoxHarris2017}, studying a sample of mostly Class II stars in $\rho$ Oph\index{Clusters!rho Oph}, found that dust disk radii among the binaries in their sample are smaller on average than those of single stars.  \citet{ManaraTazzari2019} found a similar result in Taurus\index{Clusters!Taurus}, with the added information that their single-star and binary samples have similar distributions of stellar masses, showing that this result is not a side effect of the single stars in the sample being preferentially more massive. 
In contrast, \citet{BarenfeldCarpenter2017} found that dust disks in Upper Sco {\it in general}, i.e., in both singles and binaries, 
are smaller by a factor of $\sim$3 than those in similar-stellar-mass samples in Lupus\index{Clusters!Lupus} or Taurus\index{Clusters!Taurus} and Ophiuchus\index{Clusters!rho Oph} \citep[][see also \citealt{HendlerPascucci2020}]{TazzariTesti2017,TripathiAndrews2017}. 
\citet{BarenfeldCarpenter2019} suggested that much of the difference between the single and multiple-star samples may be due to differing evolution of disk radii; tidal truncation may limit young disk sizes in binaries, 
but single star disks later ``catch up'' in reducing their sizes, perhaps by radial drift of dust, photoevaporation, dust fragmentation, and/or viscous accretion \citep[e.g., ][]{GortiHollenbach2015}.

Both \citet{CoxHarris2017} and \citet{ManaraTazzari2019} found that the measured dust disk sizes in binary systems are smaller than would be expected from tidal truncation alone, unless all systems had very large eccentricities. However, dust and gas disk sizes need not be the same; \citet{ZagariaRosotti2021b} showed that once the faster radial drift of dust expected in binary disks is taken into account, the observed dust disk sizes in binaries are consistent with the expected tidal truncation radii for modest eccentricities. \citet{RotaManara2022} confirmed this observationally, showing that most of these systems have gas disks that are larger than their dust disks, consistent with radial migration of the dust.  There is no discrepancy between the observed {\em gas\/} disk radii and expectations from tidal truncation models, assuming eccentricities consistent with the field eccentricity distribution. 

For the older disks in Upper Sco\index{Clusters!Upper Sco}, it is not only curious why the disks in binaries have fluxes similar to those for single stars, but why (at least for the binaries) the disks are there at all. 
As noted above, at 1--2 Myr ages the binary disks were already small, and both gas dispersal and radial drift of the dust should be faster in binaries, with models predicting that disks in binaries with separations less than 100 au should be depleted completely in just a few Myr \citep{RosottiClarke2018,ZagariaRosotti2021}. There is also no evidence of a circumbinary reservoir\index{Disks!circumbinary} that is replenishing these disks (see \S\ref{sec:circumbinary}).
Here it is important to note that the studies of \citet{BarenfeldCarpenter2017} and \citet{BarenfeldCarpenter2019} (as well as most of the other studies cited above) adopt a target sample containing only Class II systems 
in order to study why systems with ages similar to Upper Sco\index{Clusters!Upper Sco} still retain some disk material. Thus, the similarity of the disk properties in binaries and singles may reflect similar mechanisms for {\it retaining\/} disks on timescales of 10 Myr in any type of system. If the remaining disks are those that have developed large pressure bumps and/or dust traps, the similar disk properties might suggest that the dust-retention mechanism, rather than multiplicity, is now the dominant force driving disk evolution. 
\index{Disks!radii|)}

\subsection{\textbf{Circumbinary Disks}}\label{sec:circumbinary}
\index{Circumbinary Disks|(}\index{Disks!circumbinary|(}
Most of the above discussion concerns disks that are circumstellar with a given disk centered around one specific star in the system, but orbital dynamics also allow for circumbinary disks, the presumed birthplace of circumbinary planets, roughly a dozen of which have now been discovered by {\it Kepler\/} (summarized in \citealt{Martin2019}) and {\it TESS} \citep{KostovOrosz2020, KostovPowell2021}.\index{Kepler}\index{TESS}

While there are some notable examples of Class II\index{Class II} circumbinary disks, they are surprisingly rare.
Stable circumbinary orbits exist for disk material starting at 2--3 times the binary semi-major axis, depending on eccentricity \citep{ArtymowiczLubow1994a}. 
If close binaries only impacted a parent disk via inner tidal truncation, we would easily detect circumbinary disks extended out to $\sim 100-200$ au around systems with separations of several tens of au, i.e.\ those near the peak of the binary period distribution (\S\ref{sec:statisticsPMS}). 
In fact, such disks are uncommon.  \citet{CzekalaChiang2019} compiled a list of known circumbinary disks around young stars; among the stars with ages of 1--20 Myr, only 4 of the 30 circumbinary disks are around binaries with semi-major axes wider than 10 au\null. 
Given the completeness and high-sensitivity of ALMA surveys in nearby star-forming regions,  the lack of circumbinary disks around multiples with separations $\gtrsim$10--20 au suggests they are truly scarce.  Anecdotally, at younger ages there are a number of known circumbinary structures around systems in this separation range (\S\ref{sec:DiskStability}), but it is not clear that the difference in disk frequency is statistically significant.  If the difference is real, it could imply the structures seen in the embedded phase are not stable Keplerian disks or that circumbinary disks dissipate rapidly during the final phases of binary formation. As we discuss further in \S\ref{sec:cbpdisk}, the reappearance of circumbinary debris disks only deepens this puzzle.  In higher order multiples\index{Multiple Systems!higher order}
, it may be that confinement of the circumbinary disk due to both the internal binary and an external companion slows its viscous evolution and contributes to a longer lifetime (e.g., HD 98800,\index[obj]{HD 98800} \citealt{RibasMacias2018, RoncoGuilera2021}; TWA 3,\index[obj]{TWA 3} \citealt{CzekalaRibas2021}).

Among closer binaries, there may be an anti-correlation between the binary mass ratio and the likelihood of retaining a circumbinary disk.  As noted in \S\ref{statisticsTTS}, in a large sample of young stars, \citet{KounkelCovey2019} find that sources with disks are much less likely to be double-lined spectroscopic binaries\index{Binaries!spectroscopic} than those without disks, but that the deficit of binaries is recovered when considering single-lined systems with RV variability.  Thus, perhaps the formation of unequal-mass binaries allows or is aided by the presence of a circumbinary disk.
\index{Circumbinary Disks|)}\index{Disks!circumbinary|)}

\subsection{\textbf{Orbit-Disk Alignment}}\label{sec:OrbitDiskAlignment}
\index{Angular Momentum}
\index{Disks!alignment|(}\index{Disks!angular momentum}
Within a given young binary system, the relative orientation of the binary orbital plane and any disks in the system provides constraints on binary formation mechanisms (see \S\ref{sec:Orientations}). Even if the birth alignment is substantially altered by the time of the Class II phase, disk alignment also provides insight into the architecture of future planetary systems, since the planets will (at least initially) lie near the disk plane.  Resolved disk images trace out potential planetary orbits on the sky and thus 
provide insights into the initial orientation of those orbits, informing our understanding of future (mis)alignment.  As in the previous sections, we separate the discussions of circumbinary and circumstellar disks, since their observational constraints differ. 

\subsubsection{\textbf{Alignment of circumbinary disks}}\index{Circumbinary Disks}
The only systems for which the binary orbital orientation can be meaningfully constrained on an individual (rather than statistical) basis are those that complete a non-negligible fraction of their orbits on human timescales, i.e., those with periods less than a few hundred years.  
For low-mass stars, this implies semi-major axes less than tens of au and thus, as discussed above, any remaining disks (and certainly most {\it resolvable\/} disks) tend to be circumbinary rather than circumstellar.  Despite their rarity, known circumbinary disk systems are valuable since they allow direct comparison of orbital and disk orientations. 

For systems with double-lined spectroscopic orbits and circumbinary disks, the binary orbital inclination $i_{\rm binary}$ can be measured by determining $(M_1 + M_2) \sin^3 i_{\rm binary}$ from the orbit and the sum of the stellar masses from the rotation of the disk \citep{PratoSimon2002, RosenfeldAndrews2012, CzekalaAndrews2016, CzekalaAndrews2017, PratoRuiz-Rodriguez2018}.  For short-period binaries, measurements of $i_{\rm binary}$ generally agree 
with the observed disk inclination $i_{\rm disk}$, strongly suggesting coplanarity of disk and binary orbital planes.  Formally,  $i_{\rm binary} =  i_{\rm disk}$ does not require coplanarity since $\Omega_{\rm binary}$, the position angle of the ascending node of the binary orbit, is unobserved and not guaranteed to be the same as the position angle of the disk.  Nonetheless, \citet{CzekalaChiang2019} showed
observations can constrain the {\it distribution\/} of relative orientation angles $\theta$.  They found that the data are consistent with binaries with $P < 40$ days being aligned with their disks to within $\theta < 5\arcdeg$. Most of these systems also have circular orbits.  In one case (V4046 Sgr\index[obj]{V4046 Sgr}), coplanarity is confirmed by the detection of eclipse-like shadows cast on the disk surface as the stars orbit \citep{D'OraziGratton2019}.

For longer-period binaries with circumbinary disks, the situation is dramatically different; most have eccentric orbits and $\theta > 20\arcdeg$ \citep{CzekalaChiang2019}. While some recent discoveries follow these patterns \citep{CzekalaRibas2021, RagusaFasano2021}, notable counterexamples exist, as well; WW Cha\index[obj]{WW Cha} ($P = 207$ days) and V892 Tau\index[obj]{V892 Tau} ($P = 7.7$ yr) are each aligned with their circumbinary disks to within $\sim8\arcdeg$ \citep{Gravity2021, LongAndrews2021}.
Two systems, 99 Her\index[obj]{99 Her} \citep{KennedyWyatt2012} and HD 98800B\index[obj]{HD 98800} \citep{KennedyMatra2019}, have disks with a polar orientation to the binary orbital plane. 

These two populations agree well with predictions from the theory of disk-binary alignment.  Short-period systems, particularly those that accrete from a circumbinary disk, are expected to efficiently align the disk and binary orbital planes on timescales shorter than a typical disk lifetime \citep{FoucartLai2013}.  Alignment is faster for short-period systems both because it scales with orbital period and because these systems circularize their orbits quickly \citep{Mathieu1992}; a circular orbit means that the disk inner edge can be closer to the binary \citep[e.g.,][]{ArtymowiczLubow1994a}, increasing the alignment torque \citep{FoucartLai2014}.  In contrast, in the presence of sufficient initial eccentricity and misalignment, a variant of the Lidov-Kozai mechanism \citep{Lidov1962, Kozai1962b} can cause an outer low-mass body (in this case, a disk gas or dust parcel) orbiting a massive inner pair to exchange eccentricity for angular momentum, causing its inclination to librate around $\theta = 90\arcdeg$ \citep{VerrierEvans2009, FaragoLaskar2010}.  In a dissipative disk, librations are damped and the disk can settle into a polar configuration \citep{MartinLubow2017}. 
The higher the eccentricity, the lower the initial misalignment needed for the disk to reach polar alignment \citep{MartinLubow2018}.  Some misalignment is required, though, providing a constraint on binary formation theories; the presence of these polar-aligned disks shows that at least some close binaries form with misaligned circumbinary disks. 
 
Finally, we note that a circumbinary disk may not be a single, planar structure.  Theory suggests that the disk can be warped or torn into separate rings \citep{NixonKing2013, FacchiniLodato2013, FacchiniJuhasz2018}, and
these effects are observed in V892 Tau\index[obj]{V892 Tau} \citep{LongAndrews2021} and the triple system GW Ori\index[obj]{GW Ori} \citep{KrausKreplin2020}, respectively.  

In some cases the inner circumstellar disk is misaligned with the surrounding circumbinary disk \index[obj]{HD 142507}\citep[HD 142507;][]{PriceCuello2018, ClaudiMaire2019}, causing shadowing of the outer disk \citep{FukagawaTamura2006, AvenhausQuanz2014}.  However, it is not clear if all cases of inner/outer disk misalignment are caused by stellar companions since shadows are also present in some systems that lack known companions \citep{BenistyJuhasz2018, PinillaBenisty2018}. 

\subsubsection{\textbf{Alignment of circumstellar disks}}

Wider young binary systems, especially those wide enough to host individually-resolvable circumstellar disks, typically have orbital periods that are too long for their orbits to be meaningfully constrained on an individual basis.  In such systems the orbital inclination is unknown and cannot be compared to the disk inclination.  However, in systems where more than one star has a disk, the orientations of the disks can be compared.  If the disk inclinations are similar, there is a high likelihood that they also lie near the binary orbital plane, since the binary orbit has most of the system's angular momentum, and thus disk orientations tend to be influenced more by the stellar companion than by the other star's disk \citep{BateBonnell2000, LubowOgilvie2000, ZanazziLai2018}. 

It is clear that disks in young binaries are not all aligned, as numerous examples exist of individual systems with differing disk orientations \citep{StapelfeldtKrist1998, Koresko1998, RoccatagliataRatzka2011, JensenAkeson2014b, WilliamsMann2014, Fernandez-LopezZapata2017}. The eponymous T Tauri system\index[obj]{T Tauri} itself is a clear case of misalignment in a triple system \citep[and references therein]{BeckSchaefer2020}.  For constraining binary formation models, however, 
systematic studies can inform the extent to which disks are aligned, misaligned, or even randomly oriented.  

Some early surveys used polarization of unresolved disks to trace disk orientation in binary systems, finding that the two stars in most systems have polarization position angles that are more similar than would be expected from a random distribution \citep{MoninMenard1998, JensenMathieu2004, MoninMenard2006}.  This suggests a tendency toward disk alignment, though interpretation of the data is complicated by the relatively small polarizations in the unresolved light and possible contamination by interstellar polarization.  Studies with resolved disks are clearly the way forward, and the advent of ALMA has made it possible to resolve disks in many more systems.  However, the small sizes of disks in binaries (\S\ref{section:disk-radii}) and the need to resolve both disks in a given system means that it is still challenging to assemble large samples to study this question. 

\citet{JensenAkeson2020} combined new ALMA observations with data from the literature to assemble a sample of 8 binary systems (projected separations 100--2000 au) in which both disks are resolved and disk rotation is detected in CO for both components.  The latter criterion makes it possible to determine whether disks with similar position angles and inclinations also rotate in the same direction, allowing three-dimensional constraints on the disks' angular momentum.  Indeed, \citet{JensenAkeson2020}  found there is a strong tendency for both disks in most systems to have similar orientations, and despite the modest sample size the overall distribution is inconsistent with random relative orientations at a high level of significance.  There is no obvious correlation of misalignment angle with projected separation. 

\subsubsection{\textbf{Implications for binary formation models}}\index{Binaries!formation}

One challenge in connecting observed disk alignment to binary formation models is 
the uncertainty in how much disk orientations evolve between formation and the 1--2 Myr ages at which they are observed. Provided there is some dissipation in the disk, circumstellar disks should be driven toward alignment with the binary orbit \citep{LubowOgilvie2000, Bate2000a, ZanazziLai2018}. This evolution means that observed Class II samples are likely to be more co-aligned on average than the systems were at formation.   The presence of some systems with misalignments of many tens of degrees, and few systems that are perfectly aligned, thus is consistent with binary formation models (e.g.\ turbulent fragmentation; \S \ref{sec:corefrag}) that preferentially produce systems with disks that are misaligned with the binary orbit, at least for systems wider than a few hundred au.   At the same time, theoretical calculations of alignment timescales \citep[e.g.,][]{ZanazziLai2018} predict significant tilt evolution in 2 Myr for systems with semi-major axes less than $\sim$200 au\null. Thus, the presence of well-aligned systems with separations of order 1000 au and the lack of systems with misalignments $> 90\arcdeg$ at any separation suggest that disk alignment is much faster than currently predicted and/or there is some tendency for binary formation to favor systems with non-random disk orientations. Rapid alignment may be more likely given observations of misaligned outflows and disks during the protostellar phase (\S \ref{sec:Orientations}).  Preliminary evidence of a tendency toward alignment between binary companions and the orbits of close-in planets around one star \citep{ChristianVanderburg2022, DupuyKraus2022} is consistent with the binary-disk alignment\index{Disks!alignment}
seen at young ages, though there may be some difference in alignment statistics for close-in terrestrial planets and hot Jupiters \citep{BehmardDai2022}.

Finally, we note that the tendency toward disk-binary alignment means that most planetary systems in binaries will not have a stellar companion with a sufficient relative tilt for Kozai-Lidov oscillations\index{Kozai-Lidov mechanism} to be effective in driving high-eccentricity migration, at least not for very long.
\index{Disks!alignment|)}


\subsection{\textbf{Constraints From Planets in Multiple Systems}}
\index{Planet Formation!in binary systems}
Theoretical investigations of planet formation and survival in binary systems predate their discovery 
\citep{Harrington1977,SzebehelyMcKenzie1981,InnanenZheng1997,HolmanWiegert1999,QuintanaLissauer2002,Nelson2003,QuintanaLissauer2006}. 
Early RV surveys were strongly biased against the detection of planets in intermediate and close separation binaries. Nevertheless a $2M_{\rm jup}$ mass planet was discovered in the $\gamma$ Cephei binary in the early 2000s \citep{HatzesCochran2003}.  With a semi-major axis of only 2 au in a binary with semi-major axis of 20 au, it remains among the tightest binaries with a planet in a circumstellar orbit. The first circumbinary planet, Kepler 16b, was discovered by the transit method nearly a decade later \citep{DoyleCarter2011a}. 

Since these first discoveries, there has been a deluge of new detections of planets in binaries. Some have derived from targeted RV surveys \citep{Eggenberger2010}, coupling RV with adaptive optics \citep{WangFischer2014,WangFischer2015a}, Kepler and TESS follow-up \citep{KrausIreland2016b,WangFischer2015b,ZieglerTokovinin2020}, and direct imaging \citep{WangFischer2015c}. 
Here, we focus specifically on the statistics and properties of planets in multiple systems in so far as they constrain the dominant formation mechanism of the host stars. 

\subsubsection{\textbf{Planet occurrence rates in circumstellar disks}}

For circumstellar planets, we have robust data sets demonstrating that intermediate-wide separation binaries reduce disk masses and truncate disks, thereby substantially reducing the total mass reservoir available for planet formation (see \S\ref{sec:diskmasses},\ref{section:disk-radii}). Planet occurrence rates in intermediate separation binaries are correspondingly depleted compared to single stars. \citet{MoeKratter2021} estimated that between $\sim 10-200$au, planet occurrence rates rise from $15\%$ of the single star rate to $100\%$.  It is noteworthy that both the disk properties and planet occurrence rates seem to converge to the singleton result around $200$ au. While there are claims in the literature that Hot Jupiters reside predominantly in intermediate separation binaries \citep{NgoKnutson2016}, these claims are not borne out with revised statistical analyses \citep{MoeKratter2019b}.

The connection between disk properties and planets is unsurprising. Binary eccentricity, disk truncation, and overall mass reduction should correlate with a reduced population of observable planets in intermediate separation binaries. Eccentric binaries especially inhibit planet formation by exciting collisional velocities between pebbles and planetesimals, favoring fragmentation over agglomeration \citep{SilsbeeRafikov2015}. Small disks can hinder planet formation even at disk radii well below the truncation radius; outer disks may serve as a vital reservoir for solids that are subject to rapid radial drift \citep{YoudinChiang2004,NajitaKenyon2014}. Starving the inner disks of solids could also yield planets so small that they are mostly undetectable \citep{DupuyKratter2016a}. These results imply that planets in binaries may have distinct mass and semi-major axis distributions. There is only tentative evidence for such a shift, data sets only support an overall reduction \citep{MoeKratter2021, HirschRosenthal2021}, motivating future dedicated surveys.

Extracting constraints on formation channels from intermediate separation binaries is not straightforward. Disk and planet properties both return to their nominal single star values at the same separations where we expect disk fragmentation to be inefficient and turbulent fragmentation to dominate. It is then tempting to conclude that turbulent fragmentation does not substantially suppress planet formation. The coupled disk and planet trends also suggest that disk material beyond $\sim 50$ au contributes little to the planetary systems that dominate our data sets, i.e., those with orbits $<10$ au. Future {\it Gaia} studies of slightly wider planets from e.g., astrometric detections,  will confirm or refute this claim.

It is unclear whether a deficit of planets at smaller orbital separations indicates a transition between binary formation mechanisms.
Both turbulent fragmentation with rapid gas-driven migration and disk fragmentation could create inhospitable environments for planets.  Thus we believe the current data is insufficient to conclude that either a majority of closer separation binaries derive from disk fragmentation or that disk fragmentation is inherently less favorable to subsequent planet formation. We nevertheless put forward several hypotheses as to why disk fragmentation may hamper planet formation. Binaries formed via disk fragmentation might consume more of the remnant disk material than single stars -- both because the instability drives rapid accretion prior to fragmentation and because of the introduction of a second mass sink. Moreover, as reviewed in other PPVII chapters current disk mass measurements in single stars point towards the very early growth of solids to planetesimal sizes. If gravitational instability prior to fragmentation reduces the survival rate of large grains, these systems might never 
quickly agglomerate dust into planets before it is lost to the star. Note that the presence of spiral arms due to gravitational instability could enhance solid growth rates in some cases \citep{RiceLodato2004,BaehrZhu2021}. Finally, 
disk substructure around single stars due to pressure bumps may be crucial for retaining disk material and growing planets. The maintenance of these delicate structures might similarly be inhibited by both an instability preceding fragmentation and by the subsequent stirring of an embedded, migrating binary. This latter concern would be equally valid for binaries that derive from turbulent fragmentation and migrate to intermediate separations.

\subsubsection{\textbf{Planet occurrence rates in circumbinary disks}}
\label{sec:cbpdisk}
\index{Circumbinary Disks}
The relative rarity of circumbinary disk detections seems at odds with the discoveries from {\it Kepler} and {\it TESS} \citep{CzekalaChiang2019}. The dearth of disks,  especially around binaries with $a>10$ au is (at least naively) unsurprising, given the myriad hurdles to survival posed by close binary formation models. Dynamical instability, capture, or secular oscillations like KL cycles would all be very destructive to circumbinary disks \citep{ReipurthClarke2001,ClarkePringle1991a,MunozLai2015,HamersPerets2016,MartinMazeh2015}. Even disk fragmentation and subsequent migration might significantly deplete the circumbinary mass reservoir. Circumbinary planet statistics are improving if still limited \citep{ArmstrongOsborn2014, Martin2019,KostovOrosz2020,KostovPowell2021}. The uncertainty is driven not only by the ad-hoc nature of most discoveries, e.g., by eye rather than via pipeline, though see \citet{MartinFabrycky2021}, but due to unknown inclination distributions. As circumbinaries are identified via transit surveys, only those well aligned with an eclipsing binary will be easily discovered. Note that in principle misaligned systems could show single transit events. Thus estimates for the occurrence rates of planets around close binaries range from $10-50\%$ \citep{ArmstrongOsborn2014}. 

We cannot easily reconcile these statistics with those of circumbinary disks, but ancillary evidence suggests that disk detections underestimate the true underlying reservoir of mass for forming such planets.  Both historical and recent surveys of debris disks paint a different picture \citep{TrillingStansberry2007,YelvertonKennedy2019}. These studies mirror the dearth of bright disks in intermediate separation binaries but find debris disks in close binaries
have a frequency consistent with that of single stars. One  interpretation of the dichotomy between protoplanetary disks and debris disks is that the lifetime of the bright, gas-rich phase is very short, but not so short as to preclude planet formation. 

Whether circumbinary disks provide a net boost or hindrance to planet formation is an area of active inquiry. Numerous studies conclude that planetesimal formation is more challenging in circumbinary disks due to higher collisional velocities and nodal precession driven by both the disks and binaries \citep{MoriwakiNakagawa2004,MarzariThebault2013c,SilsbeeRafikov2015a}. \citet{BromleyKenyon2015}  have argued that beyond an inner critical radius that depends on binary properties, planet formation proceeds much like it does around single stars. Formation of planetesimals by the streaming instability could ameliorate the bottleneck even at closer separations \citep{YoudinGoodman2005,SilsbeeRafikov2021}. If formation is inhibited in-situ, inward migration must occur. State-of-the-art migration models for more massive planets can often reproduce the observed semi-major axis distribution 
of the known planets \citep{PenzlinKley2021}. This corroborates the view that planet scattering and dynamical instability are not responsible for the abundance of planets near the stability boundary set by the binary orbit \citep{LiHolman2016,SmullenKratter2016}. Though small planets remain undetected as of yet, terrestrial planet formation is also expected to proceed quickly, at least beyond some inner critical radius \citep{QuintanaLissauer2006}. Recent models by \citet{ChildsMartin2021} found that terrestrial planet formation can occur on an accelerated timescale in circumbinary systems, although the final masses are typically smaller than those in single star systems. This finding is roughly consistent with the dearth of circumbinary disks. At present all known circumbinary planets are above the terrestrial boundary $6R_{\rm earth}< R$. While smaller planets are harder to find, the detection limit is thought to be a factor of two smaller \citep{MartinFabrycky2021}. Future surveys will clarify whether the dearth of small planets is genuine or due to selection bias.
\index{Disks!in binary systems|)}

\section{\textbf{FUTURE OBSERVATIONS AND OUTLOOK}}

\subsection{\textbf{Summary of Open Questions}}

While observations and theory have made great progress advancing our understanding of the origin of multiplicity over the last decade, a variety of open questions remain.

{\it What portion of multiples form via core versus disk fragmentation? } Star cluster calculations have  demonstrated that a single formation mechanism cannot reproduce observed multiplicity statistics. Obtaining statistics of forming multiples --  catching formation sufficiently early that evolution cannot blur the lines between disk and core fragmentation -- is an observational challenge for the next decade. This entails making high-resolution observations of Class 0 disks at the earliest possible times and measuring their stability, as well as observing cores while collapse is in progress. More predictive models of gas-driven migration are also necessary to explain the final period distribution, especially for the closest binaries.
Only a handful of starless cores have confirmed substructure in either dust continuum or a dense gas tracer, indicating incipient multiple formation \citep[e.g.,][]{KirkCrutcher2009,ChenArce2010, NakamuraTakakuwa2012, LeeLooney2013, KirkDunham2017}. 
Future surveys of both line and dust emission are needed to identify and confirm gas substructures and to ascertain whether these peaks represent collapsing regions. 

{\it What are the underlying mechanisms responsible for multiple formation in complex numerical simulations?} We have reviewed the manner in which multi-physics star-formation simulations do and do not match observed multiplicity statistics. However, the literature often lacks a critical analysis of {\it why}. Why did the included physics lead to the observed population of multiples? What channels of formation dominate on different scales or environments? Answering these questions 
requires tracing the flow of mass and angular momentum for each star formed over the course of the simulation. While attempts at such analysis have been made \citep{SmithClark2009,SmullenKratter2020}, it will require methodologies, like tracer particles, that follow the time-evolution of gas parcels and capture all important physical processes \citep[e.g.,][]{GrudicGuszejnov2021}. Machine learning algorithms could be leveraged to identify underlying evolutionary trends or mechanisms that depart from the classical analytic constraints highlighted in \S3. 

{\it How do initial conditions, particularly those in extreme environments, impact multiplicity? } Observations of local star-forming regions tentatively find differences in multiplicity properties. Early dynamical evolution, rather than innate differences in the gas conditions, appear to drive some of these differences \citep{TobinOffner2021}. Simulations suggest that both metallicity and mean magnetic field impact the primordial multiplicity, however, order of magnitude variations in these conditions may be required to detect robust statistical variation. Local star-forming regions, from Taurus to Orion, exhibit relatively modest physical differences. Future studies must push the resolution limits of statistical surveys or develop new techniques to probe multiplicity in more extreme environments. Such observations are required to disentangle the influence of nature (initial conditions) versus nurture (dynamical evolution).

{\it Do high-mass stellar multiples form similarly or differently from low-mass stars?} 
High-mass protostars ($M_*~\gtrsim 8\ \Msun$) are more challenging to observe due to larger distances, higher optical depths, and complex chemistry \citep{RosenOffner2020}. Only a few high-mass protobinaries have been observed to date \citep[e.g.,][]{BeltranCesaroni2016, BeutherLinz2017,BeutherWalsh2017,ZhangTan2019,ZapataGaray2019}.  Models suggest that high-mass protostellar disks are more likely to be gravitationally unstable \citep{KratterMatzner2006}.  Observations, however, are inconclusive \citep{Cesaroni2005a,ChiniHoffmeister2004a,PatelCuriel2005,Koumpia2021,BoscoBeuther2019,MotogiHirota2019}.  Robustly linking O-star multiplicity with high-mass star formation requires more high-resolution observations of
high-mass protobinaries and their disks properties.  

{\it How do the properties and lifetimes of disks in multiple systems vary from those in single star systems? }
Observations demonstrate that disk properties vary with stellar mass, age, and environment as well as multiplicity. Larger statistical samples are needed to separate these effects and determine exactly how disks form and evolve differently in multiple systems. 
To date, star cluster calculations have not addressed differences in disk properties in multiple systems. Non-ideal magnetic processes and au-scale resolution are required for future calculations to adequately follow disk formation and evolution.

{\it What are the separation limits and properties of very wide binaries?}  Previous studies of multiplicity (and multiple-system properties) have been limited at wide separations by the challenges of confirming whether wide systems are bound or only chance projections. With the advent of {\it Gaia}, it is much more feasible to identify and characterize the widest bound systems \citep[e.g.,][]{El-BadryRix2021}.  Work to date has only begun to mine the available data.

\subsection{\textbf{Next Generation Instruments}} \label{sec:nextgen}

A variety of new surveys and observational advances over the coming decade hold 
significant promise for multiplicity studies. 
The next generation Very Large Array (ngVLA) is a proposed single-configuration
interferometer with $\sim$10 times the resolution and sensitivity of the current VLA, 
which will excel in characterizing close-separation protostellar multiplicity in the youngest
systems. It is currently in the design phase with construction tentatively 
planned to start in 2025. The ngVLA 
will simultaneously offer $\sim$au-scale spatial resolution and high
sensitivity, enabling searches for closely separated protostellar multiples
in nearby star-forming regions at wavelengths of $\sim$7~mm and $\sim$3~mm. While 
ALMA\index{ALMA} can provide comparable spatial resolution, the
 time available for the highest resolution studies is quite limited. Meanwhile, the high
dust opacity of protostellar disks at $\lambda$~$\sim$~1~mm reduces the utility 
of ALMA for characterizing close-separation multiplicity.
ALMA will soon be able to observe at 8~mm wavelengths, though the finest spatial resolution 
will be less than the VLA currently provides. The longer wavelengths of the ngVLA will
enable it to probe multiplicity within protostellar disks at wavelengths where the dust is optically thin. 

The higher resolution capabilities of the ngVLA will provide insight into multiplicity in more distant massive star-forming regions and across more diverse environments.  
The ngVLA sensitivity and resolution in regions a few kpc away will be comparable to that of current studies in nearby clouds.
Since most stars in fact form in high-mass environments \citep[e.g., the total number of Class 0 and I protostars within 500~pc is $\sim$500][]{DunhamStutz2014}, the general conclusions that can be drawn from studies of nearby regions is limited in scope and statistical robustness.

More distant regions, however, do not currently have well-characterized solar-type protostar populations. 
Accurate classification is critical to identify YSOs and study multiplicity evolution.   
The {\it James Webb Space Telescope (JWST)} will provide some of the needed characterization 
for wavelengths less than 24~\micron. However, photometry from
the near-infrared to the submillimeter is also required and is essential
for the youngest sources. Consequently, robust measurements of protostellar multiplicity in distant star-forming regions will require combining data from next
generation space-based far-infrared observatories, like the proposed {\it Origins} Space Telescope,
with data from large ground-based (sub)millimeter telescopes such as the forthcoming Large
Millimeter Telescope (LMT) and Fred Young Submillimeter Telescope (FYST) or the proposed Atacama Large Aperture Submillimeter Telescope (AtLAST). 

Surveys with large ground-based (sub)millimeter telescopes will enable studies of
the environments around proto-multiples. The LMT, FYST, and AtLAST
telescopes will achieve resolutions of $< 5000$ au in dust continuum and dust polarization 
in nearby clouds ($\lesssim 500$ pc). Planned surveys with LMT and FYST will resolve large ($>10^3$) core
populations, 
enabling robust comparative studies of cores hosting multiple and single systems.
Moreover, these facilities will better resolve the material surrounding wide binaries, and in particular, measure  magnetic 
field properties on $\sim 10^3$\,au scales, constraining
the role of magnetic fields in fragmentation and
dynamical evolution. While angular resolution will be lower compared to JWST, ALMA, and (ng)VLA making confusion and
crowding an issue for characterizing multiple systems, they will be able to study 
YSOs in nearby star-forming regions with unprecedented 
angular resolution and observe more distant regions near the current spatial resolution.

Upcoming optical and infrared facilities will probe uncharted regions of the parameter space of MS and pre-MS binaries \citep{Price-Whelan2019,Schaefer2019,Rix2019}. {\it JWST}, the {\it Roman Space Telescope},  GRAVITY+, and the Giant Magellan Telescope will better characterize cool, faint BD/late-M binaries and the disk properties of T Tauri binaries \citep{FontaniveBardalez-Gagluiffi2021pub}. Ground-based Extremely Large Telescopes will provide the highest resolution NIR imaging of star-forming regions and thus push multiplicity studies significantly ahead from Class II through MS phases.
The third full data release of {\it Gaia} expected in mid-2022 will include multi-epoch spectroscopy, photometry, and astrometry of billions of stars, yielding a catalog of millions of spectroscopic, eclipsing, astrometric, and visual binaries with full orbital solutions \citep{EyerHoll2013,EyerRimoldini2015}. This catalog will likely reveal hidden trends in multiplicity statistics, such as the relationship between stellar mass and the distributions of triple star periods, mass ratios, and orientations. Finally, the Large Synoptic Survey Telescope at the Vera Rubin Observatory will discover tens of millions of eclipsing binaries \citep{PrsaBatalha2011,Geller2021b}, providing insight into how close binary properties vary with stellar properties and environment.

\subsection{\textbf{Challenges for Theoretical Models}}

While current state-of-the-art simulations reproduce a number of key observables, such as the strong dependence of multiplicity on stellar mass (\S\ref{sec:enviornment}), most do not reproduce the observed separation distribution or its evolution over time.
In addition, models must also reproduce the observed triple fraction and companion frequency as shown in Fig.~\ref{fig:MultFrac}.
Published comparisons often benchmark numerical results using MS field star population statistics, which 
are inappropriate for simulated sources $\lesssim$~0.5~Myr years old. 
It is critical to make equivalent comparisons by accounting for observational constraints, including sensitivity  limits and projected  separation  (rather than 3D separation or binding energy). Finally, public release of numerical methods and simulation data will increase transparency and encourage broader community participation.

To obtain robust stellar statistics, future numerical studies must capture a dynamic range approaching seven orders of magnitude in spatial scale, while 
also including all important physical processes. This requires exploiting computational advances in clock speeds and system architectures and developing novel, efficient methods to model star cluster formation.  
Future star-cluster calculations must also incorporate non-ideal magnetic effects, which are important for modelling disk evolution.
Including these processes, together with sub-au resolution, may mitigate current discrepancies with observation and point the way to a fully realized theory of multiple formation.

\section{\textbf{CONCLUSIONS}} 


We have reviewed the formation and evolution of stellar multiplicity from protostellar birth through the main sequence lives of stars. Observations and theory agree that most stars are born in small-order multiple systems. Multiplicity arises primarily from the fragmentation of filaments, dense cores and massive accretion disks. Young stellar systems evolve through dynamical interactions such that multiplicity declines with time. System statistics, such as the fraction of multiple systems, frequency of higher order multiples and median separation, are strong functions of primary stellar mass. These also vary with environment density and metallicity. 

While angular momentum misalignment is common during the embedded phase, as indicated by protostellar outflow and disk orientations, older systems more often exhibit orbit-spin alignment and have aligned disks. 
Protoplanetary disks in multiple systems are on average less massive and more compact than those around single stars, while circumbinary protoplanetary disks are rare. These trends suggest planet formation and planetary architectures differ in multiple systems. 
Additional high-resolution surveys with existing facilities, such as ALMA and {\it Gaia}, as well as data from future facilities such as the JVLA, extended ALMA and {\it JWST}, promise significant strides towards addressing the remaining mysteries of multiple star systems.

\bigskip

\noindent\textbf{Acknowledgments} SSRO acknowledges support from NSF CAREER grant 1748571 and NSF AST-1812747.  JJT acknowledges  support from  NSF AST-1814762. MM and KMK acknowledge financial support from NASA ATP via grant 80NSSC18K0726. The National Radio
Astronomy Observatory is a facility of the National Science Foundation 
operated under cooperative agreement by Associated Universities, Inc.  SIS acknowledges support from the Natural Sciences and Engineering Research Council of Canada (NSERC) Discovery Grant (RGPIN-2020-03981).

\bigskip

\bibliographystyle{pp7}
\bibliography{thecleanPPVIIsuperduperbib.bib,newcites.bib}

 
\end{document}